\documentclass[12pt]{article}

\usepackage{graphicx,psfrag,epsf,color}
\usepackage{amsmath,amssymb,amsfonts}
\usepackage{array}
\usepackage{cite}
\usepackage{multirow}
\usepackage{rotating}
\usepackage{slashed}
\usepackage{bm}
\usepackage{graphbox}

\usepackage{enumerate}

\usepackage{tensor}
\usepackage{mathtools}
\usepackage[colorlinks]{hyperref}
\usepackage{slashed}
\usepackage[dvipsnames]{xcolor}
\usepackage[capitalise]{cleveref}
\usepackage{dsfont}
\usepackage{simpler-wick}
\hypersetup{pageanchor=false,linkcolor=NavyBlue,citecolor=NavyBlue,urlcolor=RoyalPurple}

\newcounter{MBQ}

\newcounter{RSQ}

\newcounter{PHQ}

\newcounter{PHT}

\newcounter{PHA}

\newcommand{\be}{\begin{equation}}
\newcommand{\ee}{\end{equation}}
\newcommand{\bea}{\begin{eqnarray}}
\newcommand{\eea}{\end{eqnarray}}
\newcommand{\bi}{\begin{itemize}}
\newcommand{\ei}{\end{itemize}}
\newcommand{\ben}{\begin{enumerate}}
\newcommand{\een}{\end{enumerate}}
\newcommand{\bt}{\begin{tabular}}
\newcommand{\et}{\end{tabular}}
\newcommand{\lc}{\left[}
\newcommand{\rc}{\right]}
\newcommand{\lp}{\left(}
\newcommand{\rp}{\right)}

\newcommand{\np}{n_+}
\newcommand{\nm}{n_-}

\newcommand{\brac}[1]{\left[#1\right]}

\newcommand{\nip}{n_{i+}}
\newcommand{\nim}{n_{i-}}
\definecolor{navy}{rgb}{0.0,0.0,0.5}

\newcommand{\nn}{\nonumber}

\numberwithin{equation}{section}

\usepackage{scalerel,stackengine,amsmath}
\newcommand\equalhat{\mathrel{\stackon[2pt]{=}{\stretchto{%
    \scalerel*[\widthof{=}]{\wedge}{\rule{0.9ex}{2.9ex}}}{0.6ex}}}}

\newcommand{\innfwhat}[2]{%
  \styletofont{#1}%
  \dimen0 \fontcharic\next1 \skewchar\next1
  \advance\dimen0 -\fontcharic\next1`#2%
  \makebox[0pt][l]{$#1#2$}%
  \makebox[\widthof{$#1#2$}]{$#1\kern.5\dimen0 \widehat{\vphantom{#2}}$}}

\begin{document}
\allowdisplaybreaks

\begin{titlepage}

\begin{flushright}
{\small
TUM-HEP-1366/21\\
February 16, 2022 \\
arXiv:2110.02969 [hep-th]
}
\end{flushright}

\vskip1cm
\begin{center}
{\Large \bf Gravitational soft theorem from 
emergent\\[0.2cm] soft gauge symmetries}
\end{center}
  \vspace{0.5cm}
\begin{center}
{\sc Martin~Beneke,$^{a}$ \sc Patrick~Hager,$^{a}$ and Robert~Szafron$^{b}$} 
\\[6mm]
{\it ${}^a$Physik Department T31,\\
James-Franck-Stra\ss e~1, 
Technische Universit\"at M\"unchen,\\
D--85748 Garching, Germany}\\[6mm]
{\it ${}^b$Department of Physics, Brookhaven National Laboratory,\\ 
Upton, N.Y., 11973, U.S.A.}
\end{center}
\vskip1cm

\begin{abstract}
\noindent 
We consider and derive the gravitational soft theorem 
up to the sub-subleading power from the perspective 
of effective Lagrangians. The emergent 
soft gauge symmetries of the effective Lagrangian provide 
a transparent explanation of why soft graviton emission 
is universal to sub-subleading power, but gauge boson 
emission is not. They also suggest a physical 
interpretation of the form of the soft factors in terms 
of the charges related to the soft transformations and 
the kinematics of the multipole expansion. The derivation 
is done directly at Lagrangian level, resulting in 
an operatorial form of the soft theorems. In order to 
highlight the differences and similarities of the 
gauge-theory and gravitational soft theorems, we include 
an extensive discussion of soft gauge-boson emission 
from scalar, fermionic and vector matter at 
subleading power.
\end{abstract}

\end{titlepage}


\section{Introduction}

Soft or low energy theorems play a crucial role in understanding quantum field theory. They provide a connection to classical field theory, allow performing resummation of large infrared logarithms, and constrain scattering amplitudes. Soft theorems for the emission of single gauge bosons hold to next-to-leading power in the soft expansion \cite{Low:1958sn, Burnett:1967km}. Gauge symmetry constrains the form of the next-to-soft terms and guarantees the universality of the soft limit. The soft theorem has been generalised to graviton emission by Weinberg \cite{Weinberg:1965nx} and subsequently by Gross and Jackiw \cite{Jackiw:1968zza,Gross:1968in}. Recent developments of spinor-helicity amplitude methods renewed interest in the gravitational soft theorem and led to the discovery \cite{Cachazo:2014fwa,Bern:2014vva,Schwab:2014xua,Broedel:2014fsa,Zlotnikov:2014sva} that the 
universality of soft graviton emission extends to the first three 
terms in the soft expansion, that is, to sub-subleading power. 
A relation between the asymptotic symmetries of
\cite{Bondi:1962px,Sachs:1962wk} and the soft theorems has been 
uncovered \cite{Strominger:2013lka,He:2014laa}, which extends 
to subleading power \cite{Kapec:2014opa,Campiglia:2014yka} as well.

Why are there three universal terms in the soft theorem for gravity but only two for gauge theories? What role do the local symmetries underlying gravity and gauge theory play? And is there a more direct way of deriving the soft theorem from the underlying Lagrangians? In the spinor-helicity formalism, the existence of a third term in the gravitational soft theorem is a consequence of little-group scaling of the spinor-helicity amplitude. Compared to gauge-boson amplitudes, the helicity-two nature of the graviton leads to an additional singular term after soft rescaling of the amplitude, which can be related to the non-radiative amplitude through a recurrence relation. In relativistic quantum field theory, the helicity of the emitted particle is closely related to the gauge symmetry of the theory, and its coupling to the conserved currents. However, the gauge symmetries -- non-abelian and diffeomorphism invariance -- of the full relativistic Lagrangians are not suited to make the soft theorem manifest. In this work, we approach the above questions from the notion of soft-collinear effective Lagrangians for gauge theory \cite{Bauer:2000yr,Bauer:2001yt, Beneke:2002ph,Beneke:2002ni} and gravity \cite{Beneke:2012xa,Beneke:2021aip} and show that the soft theorem is essentially dictated by the powerful constraints imposed by the emergent {\em soft} gauge symmetry on the Lagrangian. This approach allows us to derive the soft theorem in an operatorial form that makes the appearance of the angular momentum operator transparent. It also explains the number of universal terms as a consequence of soft gauge invariance without any calculation. While the result is of course the well-known soft theorem, our approach provides an interesting new perspective on the structure and interpretation of the various terms in the soft theorem, especially for gravity, for which the sub-subleading term was uncovered only using the spinor-helicity formalism.

To be more specific, for gauge theory, the Low-Burnett-Kroll (LBK) theorem \cite{Low:1958sn, Burnett:1967km} relates the amplitude $\mathcal{A}_{\rm rad}$, with an additional single soft gauge boson emitted, to the non-radiative amplitude $\mathcal{A}$, stripped of its external polarisation vectors, through the formula 
\begin{equation}\label{eq:LBK-QCD}
\mathcal{A}_{\rm rad}=-g_s\sum_{i=1}^n t^a_i \,\overline{u}(p_i)\lp\frac{p_i\cdot\varepsilon^a(k)}{p_i\cdot k}+\frac{k_{\nu}\varepsilon^a_{\mu}(k)J^{\mu\nu}_i}{p_i\cdot k}\rp\mathcal{A}\,,
\end{equation}
where $p_i$ is the momentum of the emitter, 
$t_i^a$ its non-abelian ``colour'' charge and 
$\overline{u}(p_i)$ its polarisation vector (spinor). 
The momentum $k$ and $\varepsilon^a(k)$ refer to the momentum and polarisation vector of the emitted soft gauge boson, respectively.  The first term is gauge-invariant after one sums over all charged external particles and imposes charge conservation $\sum_i t_i^a=0$.\footnote{In QCD this condition is referred to as colour-neutrality of the source of the emitter particles.} The second term is manifestly gauge-invariant due to the anti-symmetry of the angular momentum operator 
\begin{align}\label{eq:J}
J^{\mu\nu}_i = L^{\mu\nu}_i+\Sigma^{\mu\nu}_i= p_{i}^{\mu} \frac{\partial}{\partial p_{i \nu}}-p_{i}^{\nu} \frac{\partial}{\partial p_{i \mu}} + \Sigma^{\mu\nu}_i\,,
\end{align}
where $L^{\mu\nu}_i$ is the orbital angular momentum operator of particle $i$, and $\Sigma^{\mu\nu}_i$ the spin operator. 

In contrast, the single-graviton radiative amplitude is related to the non-radiative amplitude as
    \begin{equation}\label{eq:SoftTheorem}
        \mathcal{A}_{\mathrm{rad}} = \frac{\kappa}{2}\sum_i \overline{u}(p_i) \left( \frac{\varepsilon_{\mu\nu}(k)p_i^\mu p_i^\nu}{p_i\cdot k} + \frac{\varepsilon_{\mu\nu}(k)p_i^\mu k_\rho J_i^{\nu\rho}}{p_i\cdot k} 
        +\frac{1}{2}\frac{\varepsilon_{\mu\nu}(k) k_\rho k_\sigma J_i^{\rho\mu}J_i^{\sigma\nu}}{p_i\cdot k}
        \right)\mathcal{A}\,.
\end{equation}
The appearance of an additional universal term suggests that the gauge symmetry of gravity provides stronger constraints than in the ordinary gauge-theory case.
The first two terms resemble (\ref{eq:LBK-QCD}) if one replaces the gauge charge by momentum, $t_i^a\to -p_i^\mu$,  the gauge-boson polarisation vector by the graviton-polarisation tensor $\varepsilon_\mu^a(k) \to \varepsilon_{\mu\nu}(k)$, and adjusts the coupling constant $g\to \kappa/2$, very suggestive of the gauge-gravity double copy \cite{Bern:2008qj,Bern:2010yg}, for a comprehensive review, see \cite{Bern:2019prr}.
Indeed, the first two terms of the soft theorem can be constructed this way from the LBK amplitude already at the Lagrangian level in the soft-collinear effective theory \cite{Beneke:2021ilf}.

In this work, we suggest an alternative perspective on the terms appearing in the soft theorem for gauge and graviton emission, which emphasises the underlying local symmetries in these two cases.
This manifests itself already in the structure of the theorems themselves. In the gauge-theory soft theorem, the next-to-soft term involving the angular momentum operator is manifestly gauge-invariant for every $i$, while 
in the case of gravity the first two terms are gauge-invariant only after imposing total momentum and angular momentum conservation,  $\sum_i p_i^\mu=0$ and ${\sum_i J_i^{\mu \nu}=0}$, respectively, of the 
source. Only the third term is manifestly gauge-invariant for every $i$.
This difference points out that the second term for gravity has a different origin than the corresponding term for gauge theory, despite their similarity in form. 
Our main objective is to investigate these differences and shed some light on the connection of different terms to gauge symmetry, providing a novel interpretation of the origin of the subleading terms.  From this perspective the next-to-next-to-soft term in the gravitational soft theorem should be viewed as the analogue of the next-to-soft term in gauge theory, while the first two terms in gravity are related to the soft gravitational gauge symmetries. We explore and rederive the soft theorem using the effective field theory (EFT) formalism and demonstrate how the soft theorem follows directly from the structure of the subleading Lagrangian and its soft gauge symmetry. For gravity, 
the soft gauge symmetry consists 
of local translations and Lorentz transformations of a soft background field, which lives on the light-cones defined by the emitter 
particles.  
More importantly, we show that these theorems can be recast into an operatorial statement within the EFT formalism.

Soft theorems encapsulate factorisation between universal long-distance soft radiation and short-distance hard scattering. 
The separation of long- and short-distance effects is most conveniently formulated in the modern EFT language. A fascinating and advanced EFT construction is known as soft-collinear effective field theory (SCET) \cite{Bauer:2000yr,Bauer:2001yt,Beneke:2002ph,Beneke:2002ni}. SCET constructs an expansion of scattering amplitudes around the light-cone, describing energetic particles, collinear modes, and their 
interactions with soft modes. The fundamental role which gauge symmetry plays in the construction of the SCET Lagrangian has been recognized early on \cite{Beneke:2002ni,Bauer:2003mga}. However, although 
the subleading soft theorem for gauge theories has already been thoroughly discussed in the SCET context in \cite{Larkoski:2014bxa} (see also \cite{Beneke:2017mmf}), the powerful implications of soft gauge 
symmetry have not yet been fully exploited. Beyond the leading 
term, the soft theorem in gravity has not yet been considered in 
SCET, since SCET gravity was not constructed beyond the leading 
power at all. The present work uses the construction of \cite{Beneke:2021aip}, 
simplified to tree-level, single-emission of soft gravitons, and 
highlights the many similarities and differences between 
soft-collinear gauge theory and gravity.

The paper is organised as follows: 
first, in Section~\ref{sec:sQCD}, we provide an introduction to the basic formalism and notation in SCET. Although the main topic of this work is gravity, we 
then review the derivation of the gauge-theory soft theorem (``LBK theorem'') for the emission from scalar, spinor, and vector 
particles in Sections~\ref{sec:SoftTheoremSQCD} to \ref{sec:vQCD}, which serves to illustrate the main ideas. Rather than 
calculating the amplitude, we perform the derivation at the operatorial level by manipulating the Lagrangian until the 
angular momentum operator represented on fields becomes manifest. 
In Section \ref{sec:GR} we derive the soft theorem in perturbative gravity. We show that the universality of the first three terms is a direct consequence of the SCET gravity soft gauge symmetry. In 
particular, the absence of soft source operators up to the next-to-next-to-soft order immediately implies the existence of three universal terms in the 
gravitational soft theorem. 


\section{Preliminaries}
\label{sec:sQCD}

\subsection{Invitation}
\label{sec:invitation}

In this section, we set up the effective field theory notation, introduce the notion of the soft symmetry and discuss the matching of the non-radiative amplitude to its SCET representation. These preliminary constructions, in particular the non-radiative matching, are valid for both gauge theory and gravity. For the soft gauge symmetry and soft emission, there are differences, which we point out once relevant, and focus here on gauge theory.

It is important to note that SCET enters merely as a framework to separate the soft physics from the energetic, collinear physics of the particles generated by the hard process. SCET captures the respective soft and collinear limit of the underlying full theory, QCD or perturbative gravity, at the Lagrangian level, and its Feynman rules precisely reproduce the full-theory amplitudes in these limits.
While the framework may appear very technical at first, it 
pays off due to the conceptual clarity of the field theory representation of the physics underlying soft and collinear processes. From the perspective of SCET, soft theorems are tree-level computations within an EFT that has much broader applicability.
This means that once we understand the complicated notation, the soft theorem follows almost immediately from the effective soft gauge symmetry and the allowed currents and Lagrangian interactions by a simple computation.

To illustrate this point, we note that an energetic 
fermion with its large 
momentum $p^\mu$ directed along the light-like vector 
$n_-^\mu$ interacts with a soft gauge boson 
through the effective Lagrangian 
\cite{Dugan:1990de,Bauer:2000yr,Beneke:2002ph}
\begin{equation}
{\cal L}_{\rm gauge}^{(0)}=\overline{\chi}_c\,
g_s n_-^\mu A_{s\mu}(x_{-})\frac{\slashed n _{+}}{2}\,\chi_c
\label{eq:softL0gauge}
\end{equation}
at leading order in the soft expansion. Similarly, 
for the soft graviton \cite{Beneke:2012xa},
\begin{equation}
{\cal L}_{\rm grav}^{(0)}=\overline{\chi}_c
\left[-\frac{\kappa}{4} n_-^\mu n_-^\nu 
s_{\mu\nu}(x_{-})
\,i n_+\partial
\right]  \frac{\slashed n _{+}}{2}\,\chi_c\,. 
\label{eq:softL0grav}
\end{equation}
The graviton coupling is given by 
$\kappa=\sqrt{32\pi G_N}$ in terms of Newton's constant.
The structure of the leading term in the soft theorems 
\eqref{eq:LBK-QCD}, \eqref{eq:SoftTheorem} is already 
manifest in the Lagrangians, which couples the energetic 
particle to the soft gauge field and graviton only 
proportional to the large momentum 
$p^\mu\propto n_-^\mu$. The content of the leading term in the soft 
theorem can now be stated in operatorial form as 
\begin{equation}
\sum_ii\int d^{4}x\;T\left\{\overline{\chi}_{c_i}(0), \mathcal{L}_i^{(0)}(x)\right\}\Big\rvert_{\rm tree}\,,
\end{equation}
where the sum over $i$ runs over the energetic particles 
created in the hard process. At this point, it is essential  
that soft gauge bosons and gravitons cannot be emitted 
directly from the hard vertex at this order in the 
soft expansion, since there are no source operators 
containing soft fields that would be invariant under the soft gauge symmetry. 
The entire radiative amplitude originates from the time-ordered product with the universal Lagrangian interaction.
This guarantees the universality of the soft theorem, 
that is, its form is independent of the non-radiative, 
hard process. We shall show  from the general building principles of the 
soft-collinear effective Lagrangians for gauge fields and 
gravitons that these considerations 
extend to the next-to-soft order for gauge-boson emission 
and to next-to-next-to-soft order for graviton 
emission. 

In the following, we introduce a simplified and decluttered 
SCET notation that allows the non-expert to follow the discussion and makes our derivation more transparent. This notation is chosen specifically to work with tree-level single-emission processes in the context of soft radiation, and, in general, more care has to be taken. We refer to the literature for the general definitions \cite{Beneke:2002ni,Beneke:2017mmf,Beneke:2018rbh,Beneke:2021aip}.

\subsection{Notation and structure of the Lagrangian}

We consider processes involving a number of energetic particles, described by collinear fields $\psi_i$, moving in different, well-separated directions, and low-energy radiation described by soft modes $\psi_s$. The different collinear directions are specified by light-like vectors $n_{i-}$.  One also introduces a corresponding light-like reference vector $n_{i+}$, such that $n_{i\pm}^2=0$ and $n_{i-}n_{i+}=2$.
With these vectors, we decompose the collinear momentum 
$p_i^\mu$ into 
\begin{equation}
    p^\mu_i =n_{i+}p_i\, \frac{n_{i-}^\mu}{2} +p_{i\perp}^\mu + n_{i-}p_i\, \frac{n_{i+}^\mu}{2}\,.
\end{equation}
It is implicitly understood that the transverse component 
for an $i$-collinear momentum is defined with respect to 
the $n_{i\pm}$ reference vectors. By construction, 
$n_{i+}p_i$ is the large component of the energetic 
particle's momentum. Thus, collinear momentum scales as 
\begin{equation}
    (\nip p_i\,, p_{i\perp}\,, \nim p_i) \sim Q (1,\lambda,\lambda^2)\,,
\label{eq:collinearscaling}
\end{equation}
where $Q$ is some generic hard scale (often omitted in the 
following), and 
\begin{equation}
    \lambda \sim \frac{\lvert p_{i\perp}\rvert}{\nip p_i}\ll 1
\end{equation}
is the SCET power-counting parameter. With this counting, 
soft momentum components scale as $k^\mu\sim \lambda^2$ 
uniformly. A next-to-soft term therefore corresponds 
to a subleading $\mathcal{O}(\lambda^2)$ term, and 
next-to-next-to-soft to $\mathcal{O}(\lambda^4)$.

Computations of scattering amplitudes in SCET require two classes of objects, the Lagrangian interactions, and sources -- the $N$-jet operators. The SCET Lagrangian takes the form
\begin{equation}
    \mathcal{L}_\text{SCET} = \sum_{i=1}^N \mathcal{L}_i(\psi_i,\psi_s) + \mathcal{L}_s(\psi_s)\,,
\end{equation}
where $i$ denotes the different collinear sectors, and $s$ denotes soft fields.
Notably, the Lagrangian only contains interactions within the same collinear sector, interactions between collinear particles and the soft background, as well as purely-soft self-interactions. There are no direct interactions between fields from different collinear sectors in the SCET Lagrangian.
Such terms, where a hard vertex creates multiple particles belonging to different collinear sectors, are encapsulated in the so-called $N$-jet operators. Broadly speaking, these $N$-jet operators correspond to the non-radiative amplitude, while the Lagrangian interactions represent the soft emission from the external legs. However, in SCET this separation is organised in a manifestly gauge-invariant way.

Indeed, an important consequence of this structure is that each collinear sector transforms under its own collinear gauge symmetry, and all fields transform under the soft gauge symmetry with transformations \cite{Beneke:2002ni}
\begin{equation}
    \begin{aligned}
    &\text{$i$-collinear:} & A_{c_i} &\to U_{c_i} A_{c_i} U_{c_i}^\dagger + \frac{i}{g} U_{c_i}\lc D_s\,, U_{c_i}^\dagger \rc, \; & \phi_{c_i} &\to U_{c_i}\phi_{c_i}\,,\\
    &\text{soft:} & A_{c_i} &\to U_s(x_{i-}) A_{c_i} U_s^\dagger(x_{i-})\,, & \phi_{c_i} &\to U_s(x_{i-})\phi_{c_i}\,,
    \end{aligned}
\label{eq:gaugetrafos}
\end{equation}
where we introduced
\begin{equation}
\label{eq:prelim:CovariantDerivative}
D_{s\mu} = \partial_\mu - ig \nim A_s(x_{i-})\,\frac{n_{i+\mu}}{2}\,,
\end{equation}
 the soft-covariant derivative with homogeneous $\lambda$-scaling, and defined 
\begin{equation}
x_{i-}^\mu = \nip x \,\frac{\nim^\mu}{2}\,.
\label{eq:xminusdef}
\end{equation}

\subsection{Soft gauge symmetry}
\label{sec:prelim:soft}

We now take a closer look at the above gauge symmetries, 
which play an important role in our construction, and even 
more so in gravity. For the following discussion, we 
focus on the interactions of soft gauge fields with 
collinear matter fields, i.e. we do not discuss soft 
matter and collinear gauge fields, which are also present 
in the theory.

The transformations \eqref{eq:gaugetrafos} do not take the form of the standard gauge transformation in non-abelian gauge theory. We 
note the following two important properties:
\begin{itemize}
\item Only the $\nm A_s$ component of the soft gauge field 
appears in the covariant derivative (as is the case for the 
leading Lagrangian \eqref{eq:softL0gauge}). Hence, 
for collinear fields, 
only this component acts as a gauge field, but not the 
transverse and $\np A_s$ components. Moreover,  
$\nm A_s$ appears as a background field in the collinear 
gauge transformation.
\item The soft gauge field and soft gauge transformations 
in collinear interactions ``live'' on the light-cone 
or classical trajectory $x_-^\mu$ of the energetic particle. This important property of the soft gauge symmetry stems directly from the expansion in the parameter $\lambda$, 
since for a systematic and homogeneous power expansion 
in $\lambda$, the soft fields at space-time point $x$ must be 
multipole-expanded as 
\begin{equation} 
\psi_s(x) = \psi_s(x_-) + x_\perp\cdot \partial\psi_s(x_-) + 
\ldots
\end{equation}
about $x_-^\mu$
in interactions with collinear fields. 
This expansion can be performed in a straightforward and systematic fashion \cite{Beneke:2002ni}.
\end{itemize} 
The effective theory allows us to make the soft gauge symmetry manifest, which in turn provides an understanding and an interpretation for the individual terms in the soft theorems. Remarkably, a soft gauge symmetry with very similar properties but a generalised soft-covariant derivative also arises in the gravitational soft-collinear effective theory \cite{Beneke:2021aip}, as explained in \cref{sec:GR}, even though in full Einstein gravity, covariant derivatives are absent for scalar fields, which highlights the different nature of the soft gauge symmetry relative to the original one. For this reason, the soft gauge symmetry is an emergent one
in the infrared.

To illustrate the different nature of the components of the soft gauge field, let $\chi_c$ be a fermionic collinear matter field. Its interaction Lagrangian with soft fields is constructed as a power-series in $\lambda$,
\begin{equation}
    \mathcal{L} = \mathcal{L}^{(0)} + \mathcal{L}^{(1)} + \mathcal{L}^{(2)} + \mathcal{O}(\lambda^3)\,.
\end{equation}
The leading piece takes the form
\begin{equation}
\mathcal{L}^{(0)}=\overline{\chi}_{c}\left(
in_{-}D_s(x_-)+ i\slashed{\partial}_\perp\frac{1}{i\np\partial} 
i\slashed{\partial}_\perp\right)\frac{\slashed n _{+}}{2}
\chi_c\,,
\label{eq:fermionicLO}
\end{equation}
and it contains the soft field only in the covariant 
derivative $\nm D_s$, as it must be, as this is the only soft gauge invariant term at this order. 
In fact, the soft covariant derivative {\em only} appears 
in this term, while all subleading soft-collinear interactions are expressed as couplings of the field-strength tensor $F^s_{\mu\nu}$ and its covariant derivatives to the multipole moments of the fluctuations around the classical trajectory. 
For example, the first subleading interaction takes the form
\begin{equation}
    \mathcal{L}^{(1)} = \frac 12 x_\perp^\mu \np j^a \nm^\nu g_s F_{\mu\nu}^{s \;a}\,,
\end{equation}
where $j^a$ denotes the Noether current, and $a$ denotes an adjoint representation index. We can identify $x_\perp^\mu \np j^a$ as part of the dipole moment. Both the leading and the subleading term are universal and appear in similar form for matter fields of different spin. The dipole term is relevant for the second, next-to-soft term in the soft 
theorem \eqref{eq:LBK-QCD}.

\subsection{$N$-jet operators}

\begin{figure}
    \centering
    \includegraphics[width=0.42\textwidth]{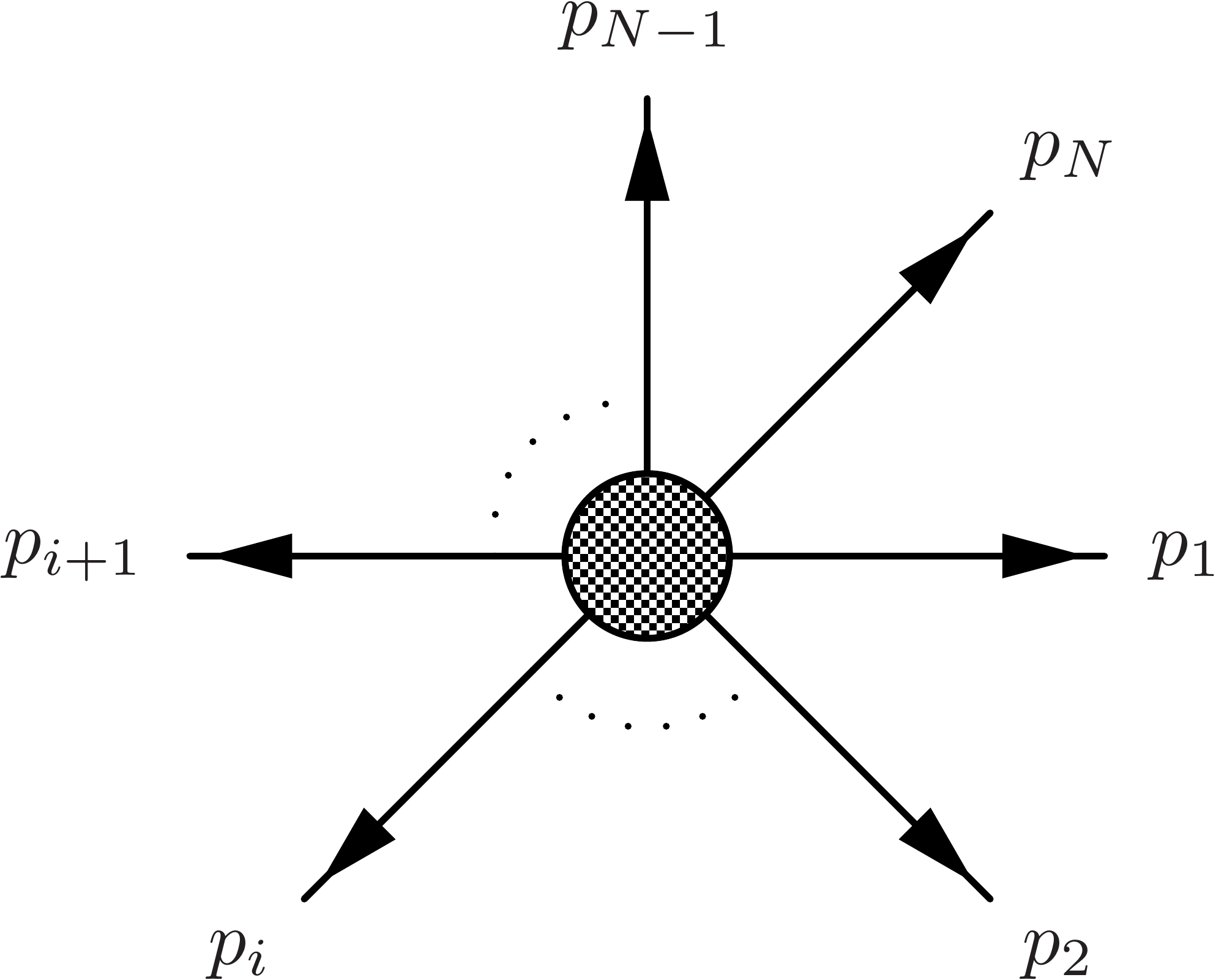}
    \caption{Non-radiative $N$-jet amplitude. }
    \label{fig:Njet}
\end{figure}

In SCET, the source that generates the energetic particles, 
or ``$N$-jet operator'', depicted in \cref{fig:Njet}, is represented by an operator that is non-local along the light-cones of the energetic particles.  
Schematically, we have
\begin{align}\label{eq:GenericNJet}
 \hat{ \mathcal{A}} = \sum_X \int [dt]_N\: \widetilde{C}^X({t_1,\ldots, t_N}) \left(\prod_{i=1}^N J^X_i(t_i)\right)J^X_s(0) \,,
  \end{align}
  where $[dt]_N = \prod_{i=1}^N dt_i$, $t_i\in [0,\infty)$ and $X$ enumerates different possible operators. 
The collinear currents  $J_i(t_i)$ are displaced along their respective light-cone direction, and constructed from collinear gauge-invariant fields, convoluted with the hard matching coefficients $\widetilde{C}^X$. $J_s(0)$ is a soft gauge-covariant product of soft fields. However, the soft gluon and graviton fields cannot appear in 
these $N$-jet operators up to order $\mathcal{O}(\lambda^4)$ and $\mathcal{O}(\lambda^6)$, respectively \cite{Beneke:2017ztn, Beneke:2021aip}.\footnote{Note that soft {\em matter} fields can appear already at $\mathcal{O}(\lambda^2)$ for the 
scalar case. However, these are not relevant to the soft gluon/graviton emission processes we consider here, and hence we can ignore them.} This follows from the fact that the soft 
covariant derivative can be eliminated by applying the 
equation of motion for the collinear fields, and further 
because the field strength tensor $F^s_{\mu\nu}$ (Riemann tensor in gravity) scales as $\mathcal{O}(\lambda^4)$ ($\mathcal{O}(\lambda^6)$ for the Riemann tensor). {\em It is this simple consequence of soft gauge symmetry which implies that there is 
some form of universal soft theorem including a next-to-soft term in the gauge theory soft theorem, and a further universal next-to-next-to-soft term for gravity, as any soft emission up these orders must arise from universal Lagrangian terms, independent of the source for the energetic particles.}

To derive the specific form of the soft theorem requires 
the subleading soft-collinear interaction Lagrangians, 
which will be discussed in subsequent sections, and 
the construction of a complete and minimal operator basis, using soft and collinear gauge symmetry \cite{Beneke:2017ztn,Beneke:2017mmf,Beneke:2018rbh}.
For our purpose -- the derivation of the LBK amplitude or soft theorem, respectively -- we can restrict ourselves to
interaction Lagrangians that describe tree-level single soft-emission processes, and to $N$-jet operators with a single energetic particle in a given direction.
This means that each $J_i$ contains a single collinear field
\begin{align}\label{eq:JA0Blocks}
 J_i^{A0}(t_i) \in \left\{\chi_i(t_i n_{i+}),\chi^\dagger_i(t_i n_{i+}),\mathcal{A}_{i\perp }(t_i n_{i+}) \right\}   \,.
\end{align}
Here, $\chi_i$ is a collinear gauge-invariant matter (scalar, spinor, or vector) field, and $\mathcal{A}_{i\perp}$ is a collinear gauge-invariant gluon field.
Subleading currents are constructed from this using the $\partial_\perp$ derivative. Thus, the leading-power operator reads
\begin{align}
   \hat{\mathcal{A}}^{(0)} =  \int [dt]_N\: \widetilde{C}^{A0}(t_1,\ldots, t_N)  \prod_{i} J_i^{A0}(t_i) \,.
\label{eq:defAhat0}
\end{align}
At subleading power, we further need 
 the collinear building blocks 
\begin{equation}\label{eq::NJet::RelevantOperators}
\begin{aligned}
&\mathcal{O}(\lambda^1): & &J^{A1\mu}_{\partial\chi_{i}^\dagger}(t_i) = i\partial^{\mu}_{i\perp}\chi_{i}^\dagger(t_in_{i+})\,,\\
&\mathcal{O}(\lambda^2): & &J^{A2\mu\nu}_{\partial^2\chi_{i}^\dagger}(t_i) = i\partial^{\mu}_{i\perp}i\partial^{\nu}_{i\perp}\chi_{i}^\dagger(t_in_{i+})\,,
\end{aligned}
\end{equation}
and so on, up to $J^{A4\mu_1\mu_2\mu_3\mu_4}_{\partial^4\chi_{i}^\dagger}(t_i)$ for the case of gravity. We use $\chi^\dagger_i$ in the building blocks as we employ all-outgoing particle convention. The soft emission is then 
generated from the time-ordered products
\begin{equation}
i\int d^4x\: T\{ J^{Ak}(t_i),
\mathcal{L}^{(n)}_{i}(x)\} \sim \mathcal{O}(\lambda^{k+n})
\end{equation}
of the collinear currents with the soft-collinear Lagrangian interactions with $n=0,1,2$ (up to $n=4$ in gravity).

\subsection{Non-radiative matching}
\label{sec:prelim:nonradmatching}

In this section, we clarify how the non-radiative amplitude $\mathcal{A}$ appearing in the soft theorem \eqref{eq:LBK-QCD} and \eqref{eq:SoftTheorem} is related to the coefficients of the above $N$-jet operators. Although the main results, \eqref{eq:CA0} and \eqref{eq:CA1} below, are rather obvious, their precise formulation appears somewhat cumbersome. This is reminiscent of imposing momentum conservation in the traditional derivation \cite{Low:1958sn,Burnett:1967km}.
The technical nature of these expressions and the derivation is unavoidable in the EFT framework, and the reader should not get distracted by the technical details.
The main result is that, roughly speaking, the $N$-jet operator corresponds to the non-radiative amplitude and its momentum derivatives order-by-order in the $\lambda$-expansion.
We first focus on a non-radiative amplitude of energetic spin-0 particles.
The generalisation to the fermionic and vectorial matter particle amplitudes is straightforward and explained in the respective later sections.

With only a single particle in a given collinear direction, 
it is always possible to align the reference vectors 
$\nim^\mu$ with the collinear particle momentum $p_i^\mu$, 
such that 
$p_i^\mu = (\nip p_i) \,\nim^\mu/2$. We will adopt this 
choice for the {\em radiative} amplitude. If the emitted 
soft particle carries away momentum $k$, one of the lines 
connecting to the hard vertex will have momentum 
$p_i+k$, see Fig.~\ref{fig:Njet_rad} below, which is 
{\em not} aligned with $\nim$. We therefore have to 
perform the non-radiative matching for the general case 
where the transverse momenta 
$p_{i\perp}\sim\mathcal{O}(\lambda)$ at the source are 
non-vanishing.

To connect the SCET operator to the full non-radiative 
amplitude, we perform tree-level matching, that is, we 
adjust the coefficients $\widetilde{C}^X({t_1,\ldots t_N})$ of the $N$-jet operator \eqref{eq:GenericNJet}
order by order in $\lambda$ to reproduce the full amplitude. The latter depends on 
the scalar products  $p_i\cdot p_j$. 
Given the collinear scaling \eqref{eq:collinearscaling}, 
these are expanded as
\begin{equation}
    p_i\cdot p_j = \frac{n_{i-}n_{j-}}{4}\,\nip p_i\, n_{j+}p_j + \frac{n_{i-}p_{j\perp}}{2} \,\nip p_i + \frac{n_{j-}p_{i\perp}}{2}\, n_{j+} p_j + \mathcal{O}(\lambda^2)\,.
\label{eq:scpexp}
\end{equation}
Thus, we Taylor-expand the full amplitude in $\lambda$,
\begin{align}
\mathcal{A} = \mathcal{A}^{(0)}+\mathcal{A}^{(1)}+\mathcal{O}(\lambda^2)\,,
\end{align}
with 
\begin{align}
 \mathcal{A}^{(0)}&= \mathcal{A}\Bigr\rvert_{p_i^\mu = n_{i+}p_i\, n^\mu_{i-}/2 }\,,\\
 \mathcal{A}^{(1)}&= 
 p_{i \perp }^\mu
 \left.\lp\frac{\partial}{\partial p^{\mu}_{i\perp }} \mathcal{A}\rp\right\rvert_{p_i^\mu = n_{i+}p_i\, n^\mu_{i-}/2 } \,. \label{eq:dA}
\end{align}
The leading term in this expansion must equal the matrix 
element of the leading-power SCET amplitude operator 
$\hat{\mathcal{A}}^{(0)}$ defined in \eqref{eq:defAhat0}, 
hence
\begin{align}\label{eq:CA0}
   \mathcal{A}^{\left(0\right)} &=
 \langle p_1,\ldots, p_N  |\hat{ \mathcal{A}}^{(0)} | 0 \rangle  \nn \\ &=
  \int [dt]_N\:e^{i\sum_i  n_{i+}p_i \; t_i} \widetilde{C}^{A0}({t_1,\ldots t_N}) \equiv {C}^{A0}(n_{1+}p_1,\ldots, n_{N+}p_N)\,.
\end{align}
In the last step, we introduced the Fourier transform of the position-space matching coefficient $\widetilde{C}^{A0}(t_1,\ldots,t_N)$. 
 
As mentioned before, there are multiple ways to generate 
next-to-leading power (NLP) currents. The only one relevant 
here are transverse derivatives acting on the collinear fields in the $A0$ operator. At $\mathcal{O}(\lambda)$, we have 
\begin{align}
 J_i^{A1\,\mu}(t_i) \in \left\{i\partial_\perp^\mu \chi_i(t_i n_{i+}),i\partial_\perp^\mu\chi^\dagger_i(t_i n_{i+}),i\partial_\perp^\mu\mathcal{A}_{i\perp i}(t_i n_{i+}) \right\}  \,. 
\end{align}
The relevant power-suppressed $N$-jet amplitude operator is 
\begin{align}\label{eq:A1operator}
   \hat{\mathcal{A}}^{(1)} = \sum_j\, [dt]_N\: \widetilde{C}^{A1\,\mu}_j(t_1,\ldots, t_N)\,J_{j\,\mu}^{A1}(t_j)\, \bigg( \prod_{i\neq j} J_i^{A0}(t_i)\bigg) \,.
\end{align}
The coefficient $\widetilde{C}^{A1\,\mu}_j(t_1,\ldots, t_N)$ can be obtained as before by matching $\mathcal{A}^{(1)}$ with the matrix element of $\hat{\mathcal{A}}^{(1)}$:
\begin{eqnarray}
       \mathcal{A}^{\left(1\right)} &=&
 \langle p_1,\ldots, p_N  | \hat{\mathcal{A}}^{(1)} | 0 \rangle  \nn =
 -p_{j\perp }^\mu \int [dt]_N\:e^{i\sum_i  n_{i+}p_i \; t_i} \widetilde{C}_{j\,\mu}^{A1}({t_1,\ldots t_N}) 
\nonumber\\
 &=&  -p_{j\perp }^\mu {C}_{j\,\mu}^{A1}(n_{1+}p_1,\ldots, n_{N+}p_N)\,.
\label{eq:CA1}
\end{eqnarray}
While \eqref{eq:dA} and (\ref{eq:CA1}) completely determine the subleading $A1$ matching coefficient in terms of the full-theory amplitude, it is useful to note the following relation.  Using (\ref{eq:dA}) and (\ref{eq:CA0}), as well as the fact that the amplitude depends only on the scalar products $p_i\cdot p_j$ and their $\lambda$-expansion \eqref{eq:scpexp}, one finds  
\cite{Beneke:2019kgv,BGS}
\begin{align}\label{eq:RPIs}
   C^{A1\,\mu}_j(n_{1+}p_1,\ldots,n_{N+}p_N) = -\sum_{k\neq j} \frac{2 n_{k-}^\mu}{n_{k-} n_{j-}} \frac{\partial}{\partial n_{i+}p_i}C^{A0}(n_{1+}p_1,\ldots,n_{N+}p_N)\,.
\end{align}
With the help of reparametrisation invariance (RPI) \cite{Manohar:2002fd,Marcantonini:2008qn} of SCET or Lorentz invariance of the full amplitude, one can prove that this relation between the matching coefficients holds to all orders. Thus, the subleading contributions to the non-radiative amplitude are in principle completely determined already from the leading-power matching.
To summarise, \eqref{eq:CA0} and \eqref{eq:CA1} state that, without soft radiation, the matrix element of the $N$-jet operators $\hat{\mathcal{A}}^{(i)}$ corresponds order-by-order to the $\lambda$-expanded full-theory amplitude $\mathcal{A}^{(i)}$.


\section{Soft theorem in scalar QCD}
\label{sec:SoftTheoremSQCD}
In this section, we derive the soft theorem for emission from scalar particles within the SCET of scalar QCD.
We further show that the essence of the theorem can be recast in an operatorial statement in the form of a time-ordered product of the non-radiative amplitude and a soft-emission vertex containing the angular momentum operator.
To obtain this result, we identify the Lagrangian terms that contribute to the single soft-emission process, and manipulate them directly under the assumption that these terms are evaluated inside a matrix element.
In this way, we can cast the interaction vertex in a form that directly yields the soft theorem.
Crucially, this simple tree-level computation already demonstrates the universality of the soft theorem, i.e., the radiative amplitude $\mathcal{A}_{\mathrm{rad}}$ expressed entirely in terms of the non-radiative one, $\mathcal{A}$, without any further calculations.

\subsection{Set-up}
\label{sec:sQCD:setup}

\begin{figure}
    \centering
    \includegraphics[width=0.28\textwidth]{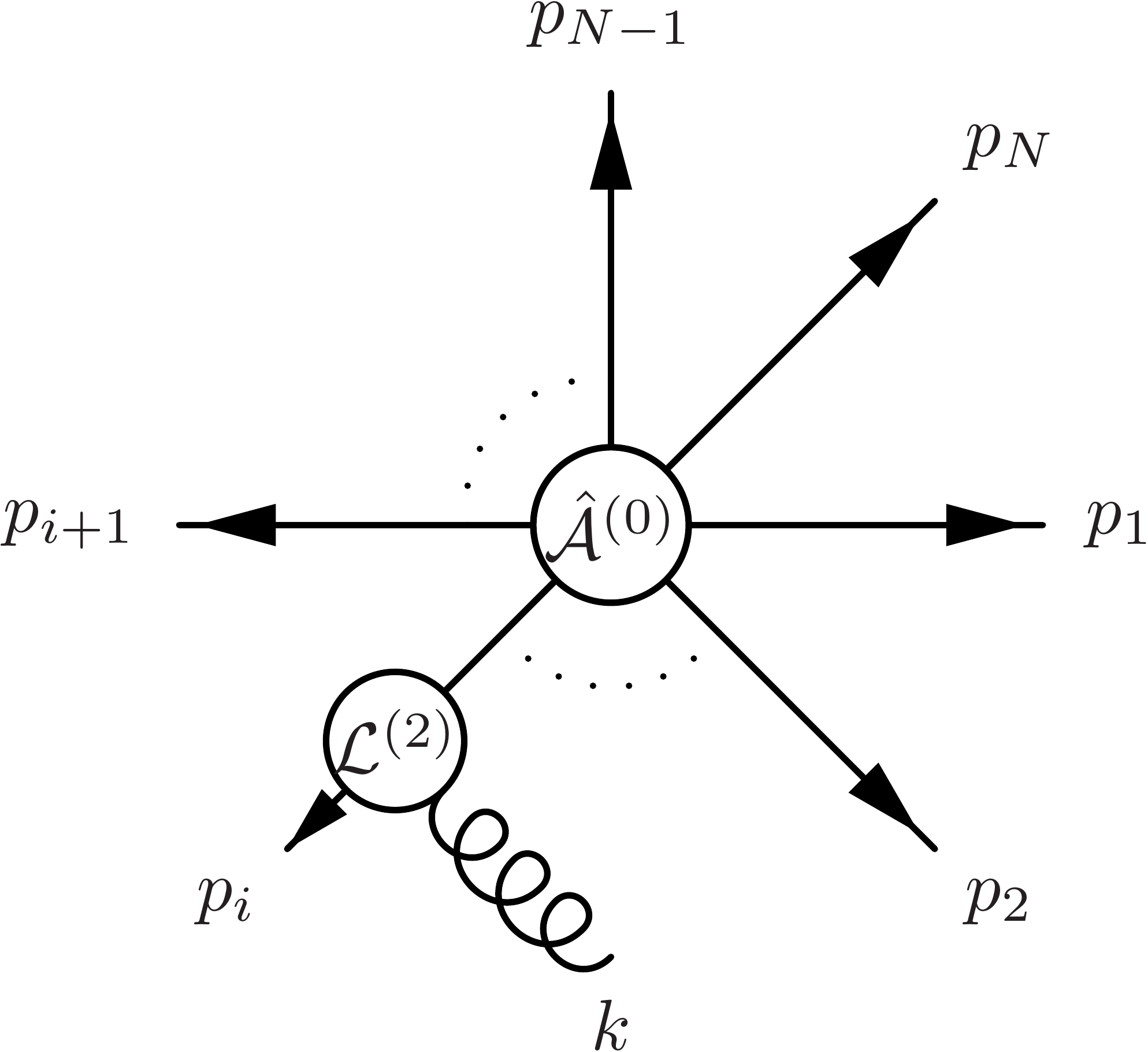}
    \quad
    \includegraphics[width=0.28\textwidth]{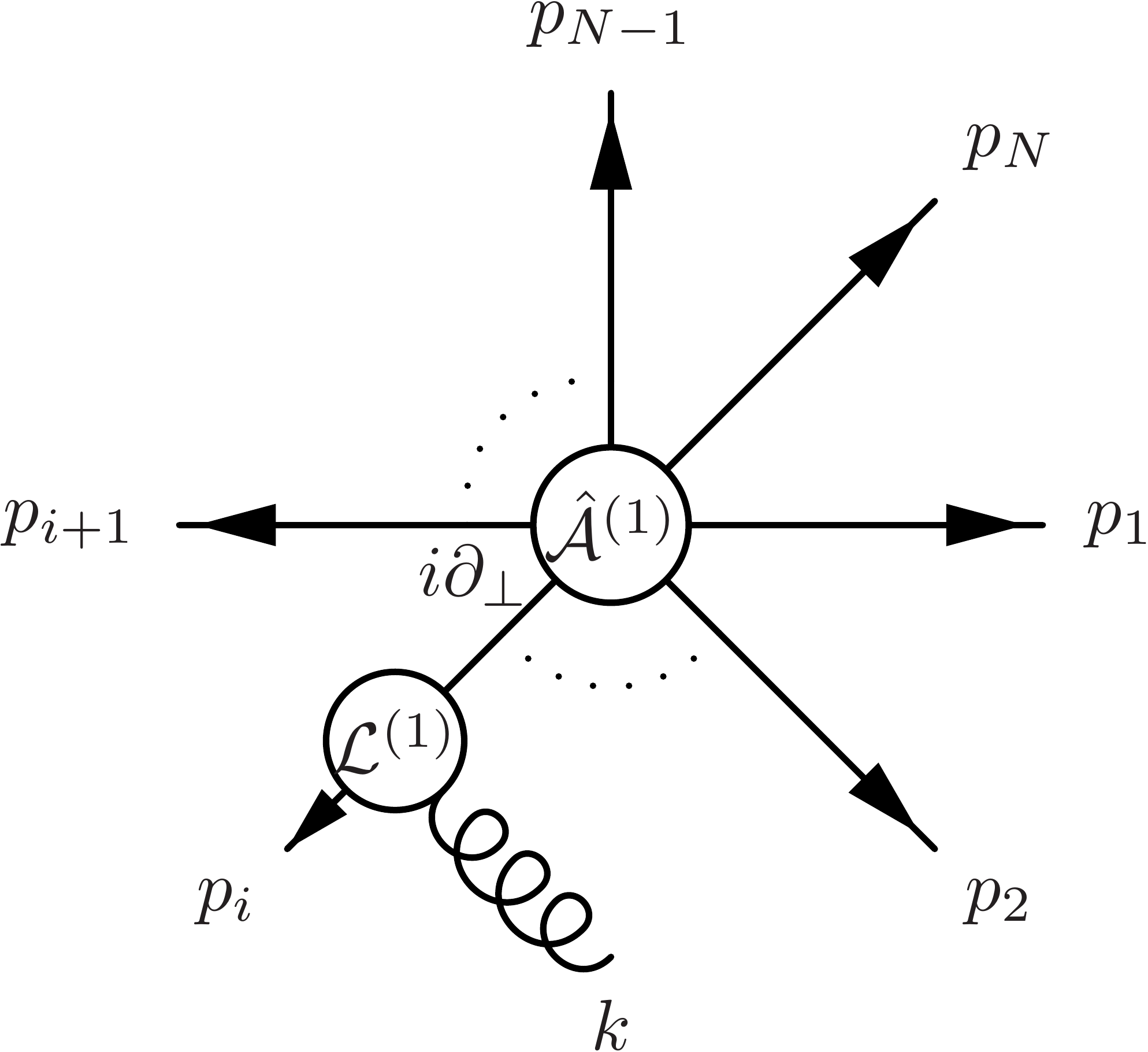}
    \quad
    \includegraphics[width=0.28\textwidth]{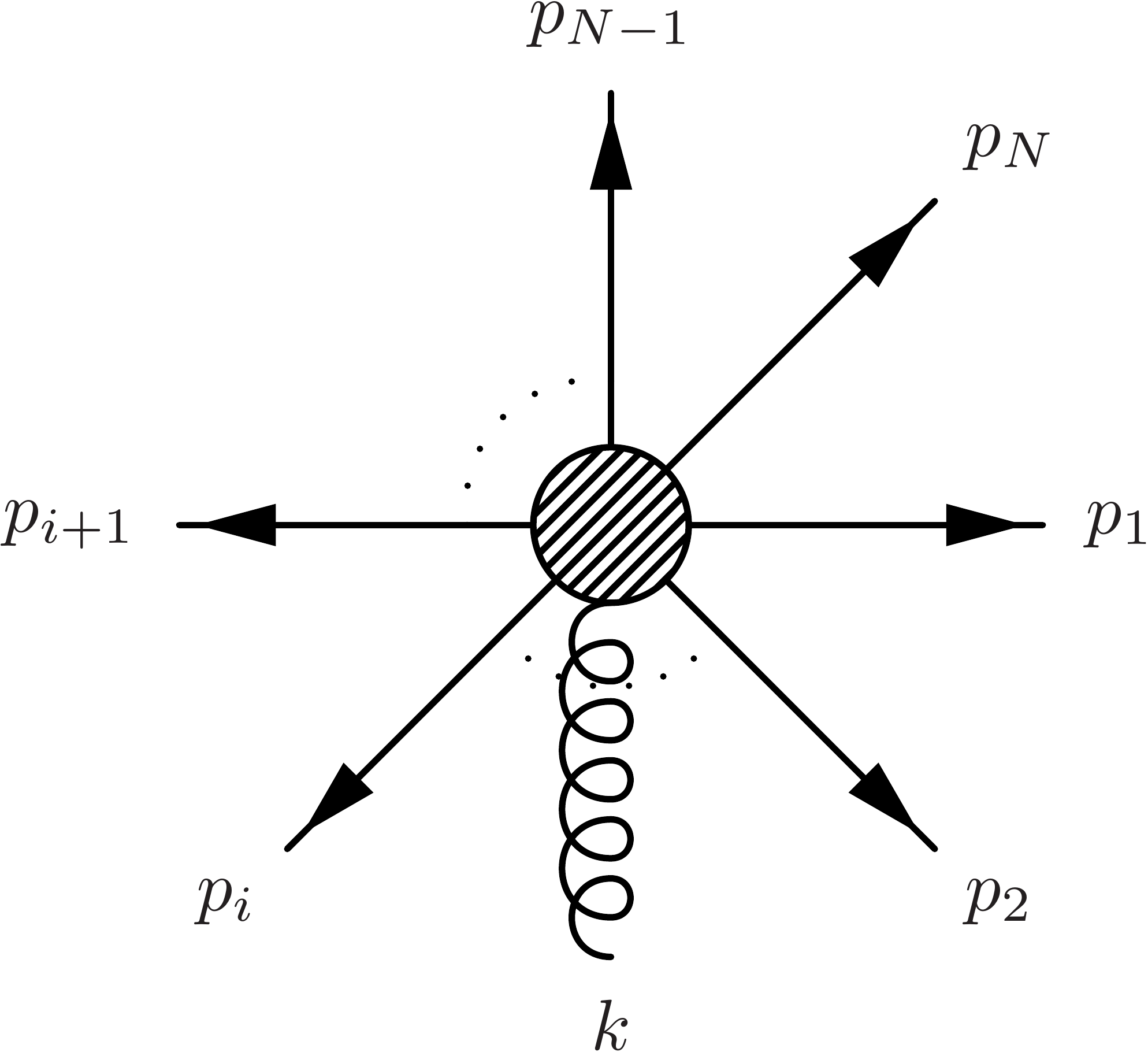}
    \caption{Three possible contributions to the radiative amplitude in SCET at NLP. The first diagram represents the time-ordered product of the leading power current and the $\lambda^2$ suppressed Lagrangian. In the second diagram, both current and Lagrangian are suppressed by a single power of $\lambda$. There are no soft building blocks at order $\lambda^2$; hence, the third diagram does not contribute to the LBK amplitude in SCET. The leading-power emission would correspond to the first diagram with an $\mathcal{L}^{(0)}$ insertion.
\label{fig:Njet_rad}}
\end{figure}

The process we consider consists of $N$ energetic particles 
moving in well-separated directions, each described by their own collinear Lagrangian, and one soft gluon.
The collinear particles are described by complex scalar fields in some representation of $SU(N)$.
As we do not consider additional collinear emission, we can drop collinear Wilson lines $W_c$ and directly work with the manifestly gauge-invariant scalar fields $\chi_c = W_c^\dagger \phi_c$ introduced in \eqref{eq:JA0Blocks}.
Soft gauge symmetry imposes that there are no soft-gluon building blocks in the $N$-jet operator basis until $\mathcal{O}(\lambda^4)$, where the soft field-strength tensor enters.
All building blocks appearing in the $N$-jet operator \eqref{eq:GenericNJet} are thus the collinear fields $\chi_c^\dagger$, considering the situation of outgoing particles.
This means that the radiative amplitude is obtained entirely from time-ordered products of the non-radiative $N$-jet operator with Lagrangian interactions, visualised as emission off the legs in the first two diagrams in Fig.~\ref{fig:Njet_rad}.
Notably, diagrams of the third type, containing a soft building block from which the gauge boson can be emitted, are not possible at $\mathcal{O}(\lambda^2)$. Such diagrams would introduce process-dependence, since the corresponding operators have short-distance coefficients unrelated to those of the non-radiative amplitude. The absence of such operators at 
leading and first subleading order in $\lambda^2$ 
allows us to focus only on a single collinear direction and the relevant Lagrangian insertion.
The final result is then given by the sum over all collinear sectors, already matching the form of \eqref{eq:LBK-QCD}.

To declutter the notation, we omit the indices numbering the external legs, as well as irrelevant variables, e.g., we write ${C}^{A0}(n_{1+}p_1,\ldots, n_{N+}p_N)={C}^{A0}(n_+p)$.  
As explained in \cref{sec:prelim:nonradmatching}, we align 
the collinear momentum with the corresponding reference vector, i.e. we have $p_i^\mu = \nip p_i\frac{\nim^\mu}{2}$ for the external momenta. This choice can always be made without loss of generality.

The Lagrangian is defined as a power series in $\lambda$
\begin{equation}
    \mathcal{L}_\chi = \mathcal{L}^{(0)}_\chi + \mathcal{L}^{(1)}_\chi + \mathcal{L}^{(2)}_\chi + \mathcal{O}(\lambda^3)\,,
\end{equation}
and the leading piece is given by
\begin{equation}\label{eq::sQCD::LeadingLagrangian}
\mathcal{L}^{(0)}_\chi=\frac{1}{2}\lc i \np\partial\chi_{c}\rc^{\dagger} i\nm D_s\chi_{c}
+\frac{1}{2}\lc i\nm D_s \chi_{c}\rc^{\dagger}i\np \partial \chi_c +\lc i\partial_{\perp\mu}\chi_{c}\rc^{\dagger}i \partial_{\perp}^{\mu}\chi_{c}\,,
\end{equation}
where $n_-D_s$ is given in \eqref{eq:prelim:CovariantDerivative}.
The relevant parts of the subleading Lagrangian can be expressed in terms of the Noether current\footnote{See also  
\cite{Beneke:2021aip} for a more extensive discussion of scalar SCET.}
\begin{align}\label{eq:NoetherScalar}
    j^\mu_a = \chi^\dagger_c \,  t^a \, i \partial^\mu \chi_c +\lc i\partial^\mu\chi_c\rc^\dagger \, t^a \, \chi_c\,,
\end{align}
and soft field-strength tensors $F^s_{\mu\nu}$, such that
\begin{align}\label{eq:LSQCD}
    \mathcal{L}_{\chi}^{\left(1\right)}&	= 
\frac{1}{2} x_{\perp}^{\mu}\, n_{+}j_a\, n_{-}^{\nu}g_sF_{\mu\nu}^{s \;a}\,, \\
\mathcal{L}_{\chi}^{\left(2\right)}	&\equalhat \mathcal{L}_{\chi}^{\left(2a\right)}+\mathcal{L}_{\chi}^{\left(2b\right)}+\mathcal{L}_{\chi}^{\left(2c\right)}\,,
\label{eq:L2SQCD}
\end{align}
with
\begin{align}
\label{eq:sQCD2a}
\mathcal{L}_{\chi}^{\left(2a\right)} &=
\frac{1}{2} x_{\perp}^{\nu}j^{\mu_\perp}_a g_sF_{\nu\mu}^{s\; a}\,,  \\
\mathcal{L}_{\chi}^{\left(2b\right)}&=
\frac{1}{4}
n_{-}x\,n_{+}^{\mu}\, n_{+} j_a \, n_{-}^{\nu}g_sF_{\mu\nu}^{s\;a}\,, \\
\label{eq:sQCD2c}
\mathcal{L}_{\chi}^{\left(2c\right)}&=
\frac{1}{4}  x_{\perp}^{\mu}x_{\perp\rho}\,n_{+} j_a\,n_{-}^{\nu}\,{\rm tr}\left(\left[D_{s}^{\rho},g_sF_{\mu\nu}^{s}\right] t^a \right)
\,.
\end{align}
Here, we introduced the symbol $\equalhat$, which indicates that we omitted terms that do not contribute to the specific tree-level matrix elements considered here, i.e.\ with single soft emission, no collinear emissions, and $\perp$-component of the external collinear particle momenta set to zero. In particular, this implies that we can drop the collinear Wilson lines $W_c$ and do not need to distinguish between the gauge-invariant building block $\chi_c = W_c^\dagger \phi_c$ and the collinear scalar field $\phi_c$. The form of the subleading Lagrangian, which contains $x^\mu$ and $\partial_\mu$ explicitly due to the multipole expansion,  already resembles the angular momentum operators appearing in the LBK theorem. There are, however, subtleties related to the action of the derivatives, which we explain in the following.

The actual derivation of the soft theorem now reduces to the computation and manipulation of three contributions, namely
\begin{align}
    \left\langle p,k\right\rvert i \int d^4x\: &T\left\{ \hat{\mathcal{A}}^{(0)},\mathcal{L}^{(0)}_\chi\right\} \left\lvert 0 \right\rangle\,,\label{eq:TA00}\\
    \left\langle p,k\right\rvert i \int d^4x\: &T\left\{ \hat{\mathcal{A}}^{(0)},\mathcal{L}^{(2)}_\chi\right\} \left\lvert 0 \right\rangle\,,\label{eq:TA02}\\
    \left\langle p,k\right\rvert i \int d^4x\: &T\left\{ \hat{\mathcal{A}}^{(1)},\mathcal{L}^{(1)}_\chi\right\} \left\lvert 0 \right\rangle\,,\label{eq:TA11}
\end{align}
where $k$ denotes the soft gluon momentum. All other time-ordered products vanish. These three contributions correspond to the insertion of leading-power and subleading power soft gluon emissions in the external legs, as depicted in \cref{fig:Njet_rad}.

In these computations, the collinear matrix element is proportional to the universal contraction 
\begin{align}\label{eq:eikonal}
 \wick{
\int d^4x\: e^{i\frac{1}{2}n_- k\, n_+ x }  \langle \c2 p \vert \;  \c1 \chi^\dagger_c(0),\; \c2 \chi_c^\dagger(x)\; [in_+ \partial \c1 \chi_c (x)]\;   \vert 0 \rangle= \frac{i n_+ p }{2p\cdot k} = \frac{i}{n_-k} \,,
  }
\end{align}
which reproduces the eikonal propagator in the LBK theorem. We included the factor $in_+ \partial$ acting on the $\chi_c(x)$ field to ensure that, for the specific kinematics chosen here, both $n_-x$ and $x_\perp^\mu$ vanish for the universal contraction term:
\begin{eqnarray}
\label{eq:nmx}
&&\int d^4x \,e^{i\frac{1}{2}n_- k\, n_+ x } \,n_-x \,\wick{\langle \c2 p \vert \;  \c1 \chi^\dagger_c(0),\;  \c2\chi_c^\dagger(x)\; [in_+ \partial  \c1 \chi_c (x)]}\;   \vert 0 \rangle = 0,\\
\label{eq:xperp}
&&\int d^4x \,e^{i\frac{1}{2}n_- k\, n_+ x } \,x_\perp^\mu \,\wick{\langle  \c2 p \vert \;   \c1 \chi^\dagger_c(0),\;  \c2 \chi_c^\dagger(x)\; [in_+ \partial \c1 \chi_c (x)]}\;   \vert 0 \rangle 
\nn\\
&&\hspace*{1cm} \,= \int d^4x \,e^{i\frac{1}{2}n_- k\, n_+ x+i\frac{1}{2}n_+p n_-x } \,x_\perp^\mu \int \frac{d^4q}{(2\pi)^4}e^{-iqx}\frac{in_+q}{q^2}
\nn\\
&&\hspace*{1cm} \,=2\int \frac{d^4q}{(2\pi)^4} \,\delta(n_-q -n_-k)\delta(n_+q-n_+p)\,i\frac{\partial \delta^{(2)}(q_\perp)}{\partial q_{\perp\,\mu}}\,\frac{in_+q}{q^2}= 0\,,
\end{eqnarray}
where the last line vanishes by virtue of $\delta^{(2)}(q_\perp)$ after integration by parts.

In momentum space, explicit factors of $x$ turn into derivatives with respect to the momentum, and (\ref{eq:nmx}), (\ref{eq:xperp}) ensure that these derivatives do not act on the eikonal propagator but only on the hard matching coefficients. This is in line with (\ref{eq:LBK-QCD}), where the angular momentum operator only acts on the non-radiative amplitude.  
Further, we note that our choice of external $p^\mu_\perp=0$ allows us to neglect any terms with $\perp$-derivative acting on the $\chi_c^\dagger(x)$ field (but not when acting on $\chi_c(x)$), since we adopt the outgoing-particle convention, such that $\chi_c^\dagger(x)$ is contracted with the external state, whereas $\chi_c(x)$ is contracted with the $N$-jet operator. In summary, the choice of contraction \eqref{eq:eikonal} guarantees that collinear particles have an eikonal propagator. Thus, the momentum derivatives corresponding to the explicit $\nm x$ and $x_\perp$ appearing in the subleading interactions can be moved past these propagators, and only act on the hard matching coefficient, i.e. on the non-radiative amplitude.

With $p_{i\perp}^\mu =0$ and $p_i^2=0$, which implies 
$n_{i-}p_i=0$, the orbital angular momentum 
appearing in \eqref{eq:LBK-QCD} simplifies to 
\begin{equation}\label{eq:sQCD:AMmomspace}
    L^{\mu\nu}_i = \frac 14 \nip^{[\mu} \nim^{\nu]}\nip p_i \frac{\partial}{\nip p_i} + \frac 12 \nip p_i \nim^{[\nu}\frac{\partial}{\partial p_{i\perp\mu]}}\,,
\end{equation}
where we defined the anti-symmetrisation $\np^{[\mu}\nm^{\nu]} \equiv \np^\mu \nm^\nu - \np^\nu\nm^\mu$.
The LBK theorem \eqref{eq:LBK-QCD} then takes the form\footnote{For a scalar field, we have $J^{\mu\nu}=L^{\mu\nu}$, as there is no spin term, and $\overline{u}_i(p_i)=1$ as there is no polarisation.}
\begin{align}\label{eq:lbk-explicit}
    \mathcal{A}_{\rm rad}&= -g_s\sum_{i=1}^n t^a_i \lp\frac{p_i\cdot\varepsilon^a(k)}{p_i\cdot k}+\frac{k_{\nu}\varepsilon^a_{\mu}(k)L^{\mu\nu}_i}{p_{i}\cdot k}\rp\mathcal{A}\nn\\
    &=-g_s\sum_{i=1}^n t^a_i \lp\frac{\nim \varepsilon^a(k)}{\nim k}\mathcal{A}^{(0)}
    +\frac{k_\mu \varepsilon^a_\nu(k)}{\nim k}\lp\frac 12 \nip^{[\mu} \nim^{\nu]}\frac{\partial}{\nip p_i}\mathcal{A}^{(0)}
    + 
    \nim^{[\nu}\frac{\partial}{\partial p_{i\perp\mu]}}\mathcal{A}^{(1)}\rp\rp\,.
\end{align}
\vskip-0.2cm
\noindent We note that $\mathcal{A}^{(0)}$ depends only on $n_{i+}p_i$, hence the derivative with respect to $p_{i\perp\mu}$ gives the first non-vanishing contribution only when acting on $\mathcal{A}^{(1)}$. 
In the following, we derive the operatorial version of these terms directly from the Lagrangian, and show the equivalence to \eqref{eq:lbk-explicit} subsequently.

\subsection{Leading-power term}\label{sec:sQCD:leading}

The leading term is given by a time-ordered product \eqref{eq:TA00} of the leading current $\hat{\mathcal{A}}^{(0)}$ with the leading-power Lagrangian \eqref{eq::sQCD::LeadingLagrangian}
After integration by parts, we identify the eikonal interaction
\begin{equation}\label{eq:sQCD:eikonal}
    \mathcal{L}^{(0)}_{\rm eikonal} = g_s\nm A_s^a \chi^\dagger_c t^a i\np\partial\chi_c\,.
\end{equation}
The radiative amplitude is then obtained from the time-ordered product 
\begin{align}\label{eq:OP-LP}
      \mathcal{A}_{\rm rad}^{(0)} \equalhat i\int d^{4}x\;T\left\{\mathcal{\hat{A}}^{(0)}, \mathcal{L}_{\rm eikonal}^{\left(0\right)}\right\}\,.
 \end{align}
 We see that the leading eikonal term in \eqref{eq:lbk-explicit} is simply due to the fact that in the eikonal interaction \eqref{eq:sQCD:eikonal}, only $n_-A_s$ appears in combination with $n_+p$, consistent with the soft gauge symmetry.

\subsection{Subleading-power / next-to-soft term}
\label{sec:sQCD:subleading}

Next, we consider the subleading terms and rearrange each term to show the form of the LBK theorem manifestly. For $\mathcal{L}_{\chi}^{\left(1\right)}$, a single integration by parts shifts the $n_+\partial$ from the $\chi_c^\dagger$ to the $\chi_c$ field such that 
\begin{align}\label{eq::SQCD::L1}
\mathcal{L}_{\chi}^{\left(1\right)}&	= \chi_{c}^\dagger\, x_{\perp}^{\mu}n_{-}^{\nu}g_sF_{\mu\nu}^{s}in_{+}\partial\chi_{c}\,,
\end{align}
and (\ref{eq:eikonal}) and (\ref{eq:xperp}) can be applied. 
This ensures that 
\begin{align}\label{eq:A0L1}
    i\int d^{4}x\;T\left\{\hat{\mathcal{A}}^{(0)}, \mathcal{L}_{\chi}^{\left(1\right)}\right\} \equalhat 0\,,
\end{align}
so there is no contribution at $\mathcal{O}(\lambda)$, consistent with \eqref{eq:lbk-explicit}.

At $\mathcal{O}(\lambda^2)$, there are two contributions -- one from the time-ordered product $T\{\hat{\mathcal{A}}^{(0)}\,, \mathcal{L}^{(2)}_\chi\}$ and one from $T\{\hat{\mathcal{A}}^{(1)}\,,\mathcal{L}_\chi^{(1)}\}$, see \eqref{eq:TA02} and \eqref{eq:TA11} and Fig.~\ref{fig:Njet_rad}, corresponding to the terms proportional to $\mathcal{A}^{(0)}$ and $\mathcal{A}^{(1)}$, respectively.
First, we discuss the contribution from $\mathcal{L}_\chi^{(2)}$ given in \eqref{eq:sQCD2a}--\eqref{eq:sQCD2c}.
Let us start with $\mathcal{L}_{\chi}^{\left(2a\right)}$. We integrate-by-parts the $\perp$-derivative acting on the $\chi_c$ field and find 
\begin{align}
    \mathcal{L}_{\chi}^{\left(2a\right)} = [i\partial_{\perp\mu} \chi_{c}]^\dagger x_{\perp}^{\nu}g_sF_{\nu\mu}^{s}\chi_{c}+\frac{1}{2}i\chi^\dagger_{c}\eta_\perp^{\mu\nu}g_sF_{\nu\mu}^{s}\chi_{c} \,.
\end{align}
The second term vanishes due to the anti-symmetry of $F_{\nu\mu}^{s}$. We can drop the terms proportional to $\partial_\perp \chi_c^\dagger(x)$, which vanish with our choice $p_{\perp}^\mu=0$. Hence 
\begin{align}
 \mathcal{L}_{\chi}^{\left(2a\right)}(x)  \equalhat 0\,.
 \end{align}
Similarly, we get 
\begin{align}\label{eq:L2b}
\mathcal{L}_{\chi}^{\left(2b\right)}&=
\frac{1}{2}
\chi^\dagger_{c}\,n_{-}xn_{+}^{\mu}n_{-}^{\nu}g_sF_{\mu\nu}^{s}in_{+}\partial\chi_{c} +
\frac{1}{2}i\chi_{c}^{\dagger}n_{+}^{\mu}n_{-}^{\nu}g_sF_{\mu\nu}^{s} \chi_c\,.
\end{align}
Finally, we focus on the $\mathcal{L}_{\chi}^{\left(2c\right)}$ term. First,
we use $x_{\perp}^{\mu}x_{\perp\rho}=[x_{\perp}^{\mu}x_{\perp\rho}-\frac{1}{2}x_\perp^2 \delta^\mu_{\perp\rho}] + \frac{1}{2}x_\perp^2 \delta_{\perp\rho}^\mu$ and drop the traceless term in the bracket, which does not contribute when external $\perp$-momentum is set to zero. We can then replace
\begin{align}
    n_-^\nu\left[D_{s}^{\mu},g_sF_{\mu\nu}^{s}\right] \equalhat - n_+^\mu n_-^\nu \frac{1}{2} n_-\partial  g_sF_{\mu\nu}^{s}\,,
\end{align}
where we used the equation of motion for the soft gluon and dropped non-linear terms in $A_s$, which only contribute to multiple-emission. This leads to
\begin{align}
    \mathcal{L}_{\chi}^{\left(2c\right)}&\equalhat
-\frac{1}{8}
\chi^\dagger_{c} x^2_{\perp}n_{+}^{\mu}n_{-}^{\nu}
\left[n_-\partial g_sF_{\mu\nu}^{s}\right]
in_{+}\partial\chi_{c}\,. 
\end{align}
At this point, we integrate-by-parts the $n_-\partial$ derivative and use the equation of motion for the collinear field,
\begin{align}
    n_-\partial \chi_c \equalhat -\frac{\partial^2_\perp}{n_+ \partial} \chi_c\,.
\end{align}
It is justified to use the free equation of motion because we neglect terms that contribute to multiple collinear and soft emissions. In the last step, we integrate-by-parts to remove $\partial_\perp$ acting on the internal line, and we get 
\begin{align}\label{eq:L2c}
    \mathcal{L}_{\chi}^{\left(2c\right)}&\equalhat
-i\frac{1}{2}
\chi^\dagger_{c} n_{+}^{\mu}n_{-}^{\nu}
 g_sF_{\mu\nu}^{s}
 \chi_{c}\,. 
\end{align}
This term cancels the last term in \eqref{eq:L2b} from $\mathcal{L}_{\chi}^{\left(2b\right)}$, and in summary we find 
\begin{equation}\label{eq::SQCD::L2}
\mathcal{L}_{\chi}^{\left(2\right)}\equalhat \mathcal{L}^{(2)}_{\rm orbital} \equiv \frac{1}{2}
\chi^\dagger_{c}\,n_{-}x\,n_{+}^{\mu}n_{-}^{\nu}g_sF_{\mu\nu}^{s}\, in_{+}\partial\chi_{c}\,.
\end{equation}

To identify the subleading term in the Lagrangians derived above with (\ref{eq:lbk-explicit}), we give the form of the angular momentum operator \eqref{eq:sQCD:AMmomspace} explicitly in position space: 
\begin{align}\label{eq:Jx}
    L^{\mu\nu}	=x^{[\mu}\partial^{\nu]}
	=\frac{1}{2}x^{[\mu}n_{-}^{\nu]}n_{+}\cdot\partial+\ldots
	=\underbrace{\frac{1}{4}n_{-}\cdot xn_{+}\cdot\partial\;n_{+}^{[\mu}n_{-}^{\nu]}}_{L_{+-}^{\mu\nu}}
	+\underbrace{\frac{1}{2}x_{\perp}^{[\mu}n_{-}^{\nu]}n_{+}\cdot\partial}_{L_{\perp+}^{\mu\nu}} \,.
\end{align}
The dots after the first equality represent terms that vanish for $p^\mu_\perp=0$. 
We can therefore express the subleading Lagrangians \eqref{eq::SQCD::L1}, \eqref{eq::SQCD::L2} in terms of the orbital angular momentum operator as
\begin{align}
    \mathcal{L}^{(1)}_\chi &= ig_sF_{\mu\nu}^{s \;a} \chi_{c}^\dagger t^a\, L^{\mu\nu}_{\perp+}\,\chi_{c}\,,\\
    \mathcal{L}^{(2)}_{\rm orbital} &= ig_sF_{\mu\nu}^{s \;a}
\chi^\dagger_{c} t^a\,L_{+-}^{\mu\nu}\,\chi_{c}
\,.
\end{align}
The term in $\mathcal{L}^{(1)}_{\chi}$ contributes in the time-ordered product with $\hat{\mathcal{A}}^{(1)}$ given in \eqref{eq:A1operator}, as the $\partial_\perp$ of the $A1$ operator acts on the $x_\perp$ in the Lagrangian.
This, in combination with the form of the $A1$ coefficient given in \eqref{eq:CA1} shows that the matrix element yields the $p_\perp$-derivative of the non-radiative amplitude $\mathcal{A}$, namely
\begin{align}
  i\int d^{4}x\;T\left\{\hat{\mathcal{A}}^{(1)}, \mathcal{L}_{\chi}^{\left(1\right)}\right\} & =    i\int d^{4}x\;T\left\{ \int dt \,\widetilde{C}_{\,\mu}^{A1}(t)  \:  i\partial_\perp^\mu\chi^\dagger_{c},  \chi_{c}^\dagger\left[x_{\perp}^{\mu}n_{-}^{\nu}g_sF_{\mu\nu}^{s}\right]in_{+}\partial\chi_{c}\right\}\nn  \\
  &= -g_s t^a_i \frac{k_\mu \varepsilon^a_\nu(k)}{\nm k}
    \nm^{[\nu}\frac{\partial}{\partial p_{\perp\mu]}}\mathcal{A}^{(1)}\,.
\end{align}
This contribution reproduces the last term in~(\ref{eq:lbk-explicit}).

In summary, we find an operatorial version of the next-to-soft term in the LBK theorem 
\begin{align}
 \mathcal{A}^{(2)}_{\rm rad}  & \equalhat \sum_{i} i\int d^{4}x\;T\left\{\hat{\mathcal{A}}^{(1)}, \mathcal{L}_{i,\,{\chi}}^{\left(1\right)}\right\}+i\int d^{4}x\;T\left\{\hat{\mathcal{A}}^{(0)}, \mathcal{L}_{i,\,{\rm orbital}}^{\left(2\right)}\right\}
 \nn\\
   & =    i\sum_i\int d^{4}x\;T\left\{ \hat{\mathcal{A}},\chi_{c_i}^\dagger t_i^a\, L_i^{\mu \nu} ig_sF_{\mu\nu}^{s\;a} \chi_{c_i}
    \right\}   ,
    \label{eq:LBK-SQCD}
\end{align}
where in the second line, we combined the source and Lagrangian terms 
using (\ref{eq:A0L1}) and $\int d^{4}x\;T\left\{\hat{\mathcal{A}}^{(1)}, \mathcal{L}_{\chi}^{\left(2\right)}\right\} \equalhat 0$ . The sum $\hat{\mathcal{A}}^{(0)}+\hat{\mathcal{A}}^{(1)}=\hat{\mathcal{A}}+\mathcal{O}(\lambda^2)$ represents the non-radiative amplitude expanded up to order $\lambda$.\footnote{Note that here $x$ has typical SCET-scaling, i.e. $x_\perp\sim 1/(\lambda Q)$ and $n_-x \sim 1/Q$. 
However, since the $x_\perp$ term is vanishing when acting on the $\hat{\mathcal A}^{(0)}$, there is no power-enhancement. Thus, \emph{effectively}, in expression (\ref{eq:LBK-SQCD}), $x$  has a hard $\mathcal{O}(\frac{1}{Q})$ scaling, like in the LBK theorem.} We can view \eqref{eq:LBK-SQCD} as a new and fully equivalent representation of the content of the soft theorem: in the absence of collinear radiation, the single soft-emission amplitude follows directly from the time-ordered product of the non-radiative amplitude and an interaction vertex containing the angular momentum operator.

\subsection{Recovering the LBK amplitude}
\label{sec:sQCD:conventional}

To verify the operatorial form and recover the LBK amplitude (\ref{eq:LBK-QCD}), we transform the expression to momentum space and evaluate the time-ordered product in the matrix element. 
We can obtain the expansion of the non-radiative amplitude in a generic reference frame using (\ref{eq:CA0}) and (\ref{eq:CA1}) to be
\begin{equation}
\mathcal{A}=C^{A0}(n_+ p )+p^{\mu}_{i\perp}C_{i,\,\mu}^{A1}(n_+ p )+\mathcal{O}(\lambda^2)\,,
\end{equation}
where $C^{A1}$ can either be computed from the generic non-radiative matching, or from the leading order coefficient and the RPI relation \eqref{eq:RPIs}. With $p^\mu_{i\perp} = 0$, the entire non-radiative amplitude is given by $\mathcal{A}^{(0)}$, and the suppressed terms are only relevant for the angular momentum term, where the derivative $\frac{\partial}{\partial p_{i\perp}}$ can act on left-over $p_{i\perp}$.

Evaluating the leading-order expression explicitly, we find
\begin{align}
      \left\langle p, k \right\rvert i\int d^{4}x\;T\left\{\mathcal{\hat{A}}^{(0)}, \mathcal{L}_{\rm eikonal}^{\left(0\right)}\right\}
   \left\lvert 0 \right\rangle
   &= -g_s t^a\frac{\nm \varepsilon^a(k)}{\nm k}\mathcal{A}^{(0)},
\end{align}
where we used the universal contraction (\ref{eq:eikonal}) and (\ref{eq:CA0}), which relates the leading-power matching coefficient to the non-radiative amplitude. This verifies the operatorial form of the leading LBK theorem (\ref{eq:OP-LP}).
For the subleading term, we evaluate
\begin{eqnarray}
\label{eq:CA}
&& \left\langle p \right\rvert  i\int d^{4}x\:
 T\left\{ \hat{\mathcal{A}},ig_s F_{\mu\nu}^{s\;a}(x_-)\chi_{c}^\dagger t^a L^{\mu \nu} \chi_{c}
    \right\}\left\lvert 0 \right\rangle 
\nn\\
&& \hspace*{0.5cm} = \,
   -g_s \,k_{\mu}\varepsilon_{\nu}^a\,t^a\,\frac{1}{p\cdot k} p^{[\mu} \frac{\partial}{\partial p_{ \nu]}} \left[ {C}^{A0}(n_+p)+p^\mu _{\perp } {C}_\mu^{A1}(n_+p)\right] \nn\\
&& \hspace*{0.5cm}=\, 
   -g_s \,k_{\mu}\varepsilon_{\nu}^a\,t^a\frac{1}{p\cdot k} L^{\mu \nu} \mathcal{A}(p)\,.
\end{eqnarray} 
In the first line of (\ref{eq:CA}), the operator $L^{\mu \nu}$ is assumed to be represented in position space (\ref{eq:Jx}), while in the second line, we assume momentum space representation for $L^{\mu \nu}$, see (\ref{eq:J}). 
Finally, we sum over all collinear directions and recover the radiative amplitude \eqref{eq:LBK-QCD}.
Since we already showed that the operatorial statement \eqref{eq:LBK-SQCD} is universal, and we computed the relation for an arbitrary non-radiative amplitude $\mathcal{A}$, this proves the LBK theorem.
As we can see, working directly within the EFT allows for a short and simple derivation of the LBK theorem, since due to the multipole expansion, the effective Lagrangian already contains all the elements of the angular momentum operator. The derivation gives an intuition for the two universal terms based on the effective gauge-symmetry, where the first one stems from the gauge-covariant derivative and is thus not manifestly gauge-invariant. The second term originates from the subleading interactions that are expressed in terms of the field-strength tensor.
We can also immediately see that a third term would no longer be universal. Here, soft building blocks make an appearance in the operator basis, and the computation becomes process-dependent.

\vskip0.2cm\noindent
In view of the subsequent 
discussion of the gravitational soft theorem, we emphasise 
the following connection between the structure of the 
soft-collinear effective Lagrangian and the soft theorem 
for gauge theory: The effective Lagrangian contains the 
covariant derivative $n_- D_s(x_-)$ of the background 
gauge field $n_- A_s(x_-)$ only at leading power. All 
subleading interactions are expressed in terms of the 
gauge-invariant field strength tensor, multiplied by 
explicit factors of position $x^\mu$ from the multipole 
expansion of soft fields around the classical trajectories 
of the energetic emitters. The covariant derivative 
interaction is related to the LP eikonal term in the soft 
theorem, which is gauge-invariant only after summing over 
all emitter directions, assuming charge conservation 
of the non-radiative process. All subleading terms 
are gauge-invariant, since the interactions involve 
only $F^s_{\mu\nu}$. However, universality ends at the 
NLP, since in higher powers there exist source operators 
containing soft field products invariant under the soft 
gauge symmetry, which have coefficient functions 
unrelated to those of the non-radiative process. The 
coupling to the angular momentum operator arises 
naturally from the dipole terms in the multipole 
expansion, when the full non-linear Lagrangian is 
restricted to single emission at tree level.

\vskip0.2cm\noindent
The following two sections explain how the universal spin 
term arises for soft gauge-boson emission. Readers mostly 
interested in the gravitational soft theorem 
may jump directly to Section~\ref{sec:GR} from here. 


\section{Fermionic QCD}
\label{sec:fQCD}

The derivation of the soft theorem in the fermionic case is very similar to the scalar case. The main difference lies in the subleading term, where the angular momentum operator contains an additional spin contribution. We briefly discuss the derivation of the orbital momentum part and then focus on the spin part. 

In the effective theory, one splits the full-theory spinor field $\psi_c = \xi_c + \eta_c$ and works with the leading component $\xi_c$, integrating out the subleading component $\eta_c$.
The leading component satisfies the projection property
\begin{align}\label{eq:fQCD:projection}
    \frac{\slashed n_{-} \slashed n_{+}}{4}\xi_c =\xi_c\,, 
\end{align}
which implies $\slashed  n_{-}\xi_c=0$.
This is similar to non-relativistic spinors, where one only keeps the leading two components of the spinor field.
This also means that in the amplitude, we have to Taylor-expand the external spinors $\overline{u}(p_i)$, as explained below.

In the $N$-jet operator \eqref{eq:GenericNJet}, the building block is now the gauge-invariant fermionic field $\chi_c$.
As the fields now carry a spinor index, which must be contracted to form Lorentz scalars, the matching coefficients $\widetilde{C}^{A0}(t_1,\ldots, t_N) $ and $\widetilde{C}^{A1\,\mu}(t_1,\ldots, t_N) $ become tensors in spinor space. 
For the sake of a simpler notation, we omit the spinor indices. 
Besides this change, the notation set up in \cref{sec:sQCD} is still valid.
This is one advantage of the SCET formalism: it works for collinear matter fields regardless of their specific spin or gauge representation.

\subsection{Non-radiative matching}

We proceed with the non-radiative matching, following the outline in \cref{sec:sQCD}. The LP matching 
(\ref{eq:CA0}) now reads
\begin{align}\label{eq:CA0f}
 \overline{\xi}_{c_1}(p_1) \ldots \overline{\xi}_{c_N} (p_N) \mathcal{A}^{\left(0\right)} &=
 \langle p_1,\ldots, p_N  | \hat{\mathcal{A}}^{(0)} | 0 \rangle  \nn \\ &=
\int  \left[ dt \right]_N \overline{\xi}_{c_1}(p_1) \ldots \overline{\xi}_{c_N} (p_N) \widetilde{C}^{A0}({t_1,\ldots t_N}) e^{i\sum_i  n_{i+}p_i \; t_i}\\ \nn
  &\equiv  \overline{\xi}_{c_1}(p_1) \ldots \overline{\xi}_{c_N} (p_N) {C}^{A0}(n_{1+}p_1,\ldots, n_{N+}p_N)\,.
\end{align}
Thus the ${C}^{A0}(n_{1+}p_1,\ldots, n_{N+}p_N)$ can be understood as the LP amplitude with the external spinors stripped off,
\begin{align}\label{eq:CA0A}
{C}^{A0}(n_{1+}p_1,\ldots, n_{N+}p_N)= \mathcal{A}^{(0)}\,.
\end{align}
Here $\overline{\xi}_{c_i}(p_i)$ is an $i$-collinear spinor obtained from the expansion of the full QCD spinor
\begin{align}\label{eq:fQCD:FullSpinorRelation}
    \overline{u}(p_i)= \overline{\xi}_{c_i}(p_i)\left(1 - \frac{\slashed p_{i\perp}}{n_{i+}p_i} \frac{\slashed n_{i+}}{2} \right) \,.
\end{align}

The subleading current can be matched like in the scalar case (\ref{eq:CA1}). However, now we take into account the subleading term in the external QCD spinor expansion \eqref{eq:fQCD:FullSpinorRelation}.
This additional term is also universal and follows from reparametrisation invariance. Consequently, for spinors, the relation (\ref{eq:RPIs}) reads
\begin{eqnarray}
\label{eq:subleadingSpinor}
    C^{A1\,\mu}_i(n_{1+}p_1,\ldots,n_{N+}p_N)&=&\bigg[-\frac{\gamma^{\mu}_{i\perp}}{n_{i+}p_i}\frac{\slashed{n}_{i+}}{2}-\sum_{j\neq \,i}\frac{2n^{\mu}_{j-}}{n_{i-}\cdot n_{j-}}\frac{\partial}{\partial n_{i+}p_{i}}\bigg]C^{A0}(n_{1+}p_1,\ldots,n_{N+}p_N)\nn \\
&&\hspace*{-2cm}    \equiv \,C^{A1\mu}_{i,\, {\rm spin}}(n_{1+}p_1,\ldots,n_{N+}p_N)+C^{A1\mu}_{i,\,{\rm orbit}}(n_{1+}p_1,\ldots,n_{N+}p_N)\,.
\end{eqnarray}
Following the split performed in the last line, we also split the NLP operator into the orbital and spin parts proportional to $C^{A1\mu}_{i,{\rm orbit}}(n_{1+}p_1,\ldots,n_{N+}p_N)$ and $C^{A1\mu}_{i, {\rm spin}}(n_{1+}p_1,\ldots,n_{N+}p_N)$, respectively, such that 
\begin{align}\label{eq::fQCD::A1spinorbit}
    \hat{\mathcal{A}}^{(1)}= \hat{\mathcal{A}}_{\rm orbit}^{(1)}+\hat{\mathcal{A}}^{(1)}_{\rm spin}\,.
\end{align}
    
As for the scalar case, we find the universal contraction (\ref{eq:eikonal}), which now reads 
\begin{align}\label{eq:eikonalF}
\wick{
\int d^4x e^{i\frac{1}{2}n_- k\, n_+ x }  \,\langle \c1{p} \vert \;  \c2{\overline{\xi}}_c(0),\;  \c1{\overline{\xi}}_c(x)\;   \frac{\slashed n_+}{2} \c2 \xi_c (x)\;   \vert 0 \rangle  = \overline{\xi}_c(p)\frac{i n_+ p }{2p\cdot k}= \overline{\xi}_c(p)\frac{\slashed n_+ \slashed n_-}{4}\frac{i}{n_-k}\,.
  }
\end{align}
Thanks to the different normalisation of fermionic fields, we do not need to include $n_+\partial$ derivative to achieve (\ref{eq:nmx}) and (\ref{eq:xperp}). 
    
\subsection{Soft theorem}
    
Like in the scalar case, there are no soft gluon building blocks available for the $N$-jet operators, and all contributions must stem from time-ordered products. As in \cref{sec:SoftTheoremSQCD}, we focus on a single external line and choose the kinematics where all $p_i$ are aligned with their reference vectors $\nim$, i.e., $p^\mu_{i\perp} = 0$.

The Lagrangian for fermionic SCET is \cite{Beneke:2002ni}
\begin{align}
\label{eq:Lxi1}
{\cal L}^{(1)}_{\xi} &=
\overline{\xi}_c \left( x_\perp^\mu n_-^\nu \, W_c \,g_s F_{\mu\nu}^s W_c^\dagger 
\right) \frac{\slashed n_+}{2} \xi_c\,,
\\[0.0cm]
{\cal L}^{(2)}_{\xi} &=
  \frac{1}{2} \, \overline \xi_c \left(
  n_-x \: n_+^\mu n_-^\nu \, W_c \,g_sF_{\mu\nu}^{s}  W_c^\dagger
  + x_\perp^\mu x_{\perp\rho} n_-^\nu W_c \big[D^\rho_{s}, 
   g_s F_{\mu\nu}^{s}\big] W_c^\dagger  \right) 
   \frac{\slashed n_+}{2} \xi_c
\nn \\
& + \,\frac{1}{2} \, \overline \xi_c \left(
    i \slashed D_{\perp }  \,
 \frac{1}{i n_+ D} \, x_\perp^\mu \gamma_\perp^\nu \,
 W_c \,g_s F_{\mu\nu}^{s}W_c^\dagger  +   x_\perp^\mu \gamma_\perp^\nu \,
 W_c \,g_s F_{\mu\nu}^{s}W_c^\dagger  \,
 \frac{1}{i n_+ D} \, i \slashed D_{\perp }
 \right) \frac{\slashed n_+}{2}\xi_c\,.
\label{eq:Lxi2}
\end{align}
As before, we neglect non-contributing terms and split the Lagrangian into several parts that we discuss one by one
\begin{align}\label{eq:LFQCD}
{\cal L}^{(1)}_{\xi} &\equalhat
\overline{\xi}_c \left( x_\perp^\mu n_-^\nu  \,g_s F_{\mu\nu}^s 
\right) \frac{\slashed n_+}{2} \xi_c\,, \\
    {\cal L}^{(2)}_{\xi} &\equalhat {\cal L}^{(2a)}_{\xi}+{\cal L}^{(2b)}_{\xi}+{\cal L}^{(2c)}_{\xi}+{\cal L}^{(2s)}_{\xi}\,,
\label{eq:L2FQCD}\end{align}
with analogues of the scalar counterparts (\ref{eq:LSQCD})
\begin{equation}
\begin{aligned}
 {\cal L}^{(2a)}_{\xi} &=   \, \overline \xi_c  x_\perp^\mu g_s F_{\mu\nu}^s \frac{i\partial_\perp^\nu}{in_+\partial}\frac{\slashed  n_+}{2}\xi_c\,,
 \\
    {\cal L}^{(2b)}_{\xi} &=
  \frac{1}{2} \, \overline \xi_c 
  (n_-x) \, n_+^\mu n_-^\nu \, \,g_sF_{\mu\nu}^{s}    \frac{\slashed  n_+}{2}\xi_c\,, \label{eq::fQCD::L2b}  \\
   {\cal L}^{(2c)}_{\xi} &=  \frac{1}{2}\, \overline \xi_c  x_\perp^\mu x_{\perp\rho} n_-^\nu  \big[D^\rho_{s}, 
   g_s F_{\mu\nu}^{s}\big]  
   \frac{\slashed  n_+}{2} \xi_c\,,
\end{aligned}
\end{equation}
and a new, spin-dependent part
\begin{align}\label{eq:FQCD:LSpin}
 {\cal L}^{(2s)}_{\xi} &=   \, \overline \xi_c  g_s \Sigma_\perp^{\mu\nu} iF_{\mu\nu}^s \frac{1}{in_+\partial}\frac{\slashed  n_+}{2}\xi_c \,.
\end{align}
The spin operator $\Sigma_{\mu\nu}$ is expressed in terms of the light-cone components
\begin{align}\label{eq:Sigma}
    \Sigma^{\mu\nu} = \frac{1}{4}[\gamma^\mu, \gamma^\nu]
    \equalhat \underbrace{\frac{1}{4}\lc \gamma_\perp^\mu, \gamma_\perp^\nu\rc}_{\Sigma_\perp^{\mu\nu}} + \underbrace{\left(-\frac{1}{2} \frac{\slashed n_+}{2}\gamma_\perp^{[\mu} n_-^{\nu]}\right)}_{\Sigma_{\perp+}^{\mu\nu}}
    +
    \underbrace{\left(-\frac{1}{4} n_+^{[\mu} n_-^{\nu]}\right)}_{\Sigma_{+-}^{\mu\nu}}
    \,,
\end{align}
where, after $\equalhat$, we dropped the terms that vanish due to the projection property \eqref{eq:fQCD:projection} of collinear spinors and our choice $p^\mu_\perp=0$.

The derivation of the orbital angular momentum term is almost the same as in the scalar case. As before, we find that 
$\mathcal{L}_{\xi}^{\left(2a\right)}(x)  \equalhat 0\,$. Furthermore, the form of $\mathcal{L}_{\xi}^{\left(2b\right)}(x) $ agrees with the first term in (\ref{eq:L2b}), however, the second term is absent, which implies that the contribution from $\mathcal{L}_{\xi}^{\left(2c\right)}(x)$ does not cancel. Instead, $\mathcal{L}_{\xi}^{\left(2c\right)}(x)$ supplies the longitudinal components of the spin as shown below. Thus, we find the orbital angular momentum as in \eqref{eq:LBK-SQCD},
\begin{align}
& i\int d^{4}x\;T\left\{\hat{\mathcal{A}}_{\rm orbit}^{(1)}, \mathcal{L}_{\xi}^{\left(1\right)}\right\}+i\int d^{4}x\;T\left\{\hat{\mathcal{A}}^{(0)}, \mathcal{L}_{\xi}^{\left(2a\right)}+\mathcal{L}_{\xi}^{\left(2b\right)}\right\}
 \nn\\   & =    i\int d^{4}x\;T\,\bigg\{ \hat{\mathcal{A}}_{\rm orbit},\overline{\xi}_c \frac{\slashed n_+}{2} L^{\mu \nu} ig_sF_{\mu\nu}^{s} \xi_c
    \bigg\}   \,.
    \label{eq:LBK-FQCD}
\end{align}

Let us focus on the spin part in the soft theorem. We see that the first term in (\ref{eq:Sigma}) is directly given by $
    T\,\{\hat{\mathcal{A}}^{(0)}, \mathcal{L}_{\xi}^{(2s)}\}\,,
$
while the last term is obtained from the remaining part $\mathcal{L}_{\xi}^{\left(2c\right)}$. Following the same manipulations, which in the scalar case led us to (\ref{eq:L2c}), gives
\begin{align}\label{eq::fQCD::L2cspin}
    \mathcal{L}_{\xi}^{\left(2c\right)}&\equalhat
-i\frac{1}{2}
\overline{\xi}_c n_{+}^{\mu}n_{-}^{\nu}
 g_sF_{\mu\nu}^{s} \frac{1}{i n_+ \partial}\frac{\slashed n_+}{2}
 \xi_c\,. 
\end{align}

In summary, the second-order Lagrangian takes the form
\begin{equation}
    \mathcal{L}^{(2)}_\xi \equalhat \mathcal{L}^{(2)}_{\rm orbit} + \mathcal{L}^{(2)}_{\rm spin}\,,
\end{equation}
where the orbit part \eqref{eq::fQCD::L2b} and spin parts \eqref{eq:FQCD:LSpin}, \eqref{eq::fQCD::L2cspin} are manifest 
\begin{align}
    \mathcal{L}^{(2)}_{\rm orbit} &= g_s iF_{\mu\nu}^{s} \overline\xi_c 
  L^{\mu\nu}_{+-}\frac{\slashed  n_+}{2}\xi_c\,,\label{eq::fQCD::L2orbit}\\
  \mathcal{L}^{(2)}_{\rm spin} &= g_s iF_{\mu\nu}^{s} \overline\xi_c \lp 
  \Sigma_\perp^{\mu\nu} + \Sigma_{+-}^{\mu\nu}
  \rp 
  \frac{1}{i\np\partial}\frac{\slashed n_+}{2}\xi_c\,.\label{eq::fQCD::L2spin}
\end{align}
Note that here, in both the orbital and the spin part, the mixed transverse-longitudinal terms are missing.
Like in the scalar case, the missing orbital term is reproduced by the contribution from $T\{\hat{\mathcal{A}}^{(1)}_{\rm orbit}\,,\mathcal{L}^{(1)}_\xi\}$.
For the spin term, notice that the $\hat{\mathcal{A}}^{(0)}$ operator contains the projection 
\begin{equation}
    \overline\xi_c \mathcal{A}^{(0)} = \overline\xi_c \frac{\slashed n_+ \slashed n_-}{4}\mathcal{A}^{(0)}\,.
\end{equation}
This projection eliminates the mixed transverse-longitudinal contribution, as
\begin{equation}\label{eq:fQCD:projectionSigma}
    \Sigma_{\perp+}^{\mu\nu}\frac{\slashed n_+\slashed n_-}{4} \propto \frac{\slashed n_+}{2}\frac{\slashed n_+\slashed n_-}{4} = 0\,.
\end{equation}
This missing mixed term is related to the new contribution due to $T\,\{\hat{\mathcal{A}}^{(1)}_{\rm spin}, \mathcal{L}_{\xi}^{(1)}\}\,.$
It is important here that the operator $\hat{\mathcal{A}}^{(1)}_{\rm spin}$ is completely fixed by the RPI relation~(\ref{eq:subleadingSpinor}) and thus determined solely by the non-radiative amplitude.
Using translation-invariance of the propagator and the fact that the SCET fermion propagator anti-commutes with $\gamma_\perp^\mu$, we derive
\begin{align}
T\,\bigg\{\mathcal{L}_\xi^{(1)},- \overline{\xi}_c \frac{i \overset{\leftarrow}{\slashed\partial}_\perp}{in_+\overset{\leftarrow}{\partial}}\frac{\slashed n_+}{2} C^{A0}\bigg\} 
&\equalhat\wick{T\left\{
\overline{\xi}_c \left( x_\perp^\mu n_-^\nu  \,g_s F_{\mu\nu}^s 
\right) \frac{\slashed n_+}{2}  \frac{i \slashed \partial_\perp}{in_+\partial}\c3 \xi_c\,, 
\c3 {\overline{\xi}_c} \frac{\slashed n_+}{2}C^{A0}\right\} }
\nn\\
&\equalhat
\wick{T\left\{
g_s i F_{\mu\nu}^s \overline{\xi}_c\,\Sigma^{\mu \nu}_{\perp+} \,
\frac{1}{in_+\partial}\c3 \xi_c\,,
\c3 {\overline{\xi}_c} \frac{\slashed n_+}{2}C^{A0}\right\} }\,.
\label{eq:S2}
\end{align}

Summing the contributions from $T\,\{\hat{\mathcal{A}}^{(0)}\,,\mathcal{L}^{(2)}_{\rm spin}\}$ using \eqref{eq::fQCD::L2spin} and $T\,\{\hat{\mathcal{A}}_{\rm spin}^{(1)}\,,\mathcal{L}^{(1)}_\xi\}$ given in (\ref{eq:S2}), we recover the spin term in the LBK theorem,
\begin{align}
 &  i\int d^{4}x\;T\left\{\hat{\mathcal{A}}_{\rm spin}^{(1)}, \mathcal{L}_{\xi}^{\left(1\right)}\right\}+
 i\int d^{4}x\;T\left\{\hat{\mathcal{A}}^{(0)}, \mathcal{L}_{\rm spin}^{\left(2\right)}\right\}\,
\nn  \\
&\equalhat i \int d^4x\overline{\xi}_c(x)   \,g_s F_{\mu\nu}^s\,  \left(\wick{
  \Sigma^{\mu \nu}_{\perp+}\frac{1}{in_+\partial}\c3 \xi_c\: \c3 {\overline{\xi}_c} \frac{\slashed n_+}{2} +  (\Sigma^{\mu \nu}_{+-} + \Sigma^{\mu \nu}_\perp)\frac{1}{in_+\partial}\frac{\slashed n_+}{2} \c3 \xi_c\: \c3 {\overline{\xi}_c}  }\right)  \, C^{A0}(n_+p)
\nn \\
&=i \int d^4x\overline{\xi}_c(x)   \,g_s i F_{\mu\nu}^s \Sigma^{\mu \nu}\,\left(\wick{
  \frac{1}{in_+\partial}\c3 \xi_c\: \c3 {\overline{\xi}_c} \frac{\slashed n_+}{2} +  \frac{1}{in_+\partial}\frac{\slashed n_+}{2} \c3 \xi_c\: \c3 {\overline{\xi}_c}  }\right) \, \mathcal{A}\,.\label{eq:fQCD:SoftTheoremOperator}
\end{align}
We replaced the matching coefficient $C^{A0}$ by the stripped non-radiative amplitude according to (\ref{eq:CA0A}), and our choice of kinematics implies $\mathcal{A}=\mathcal{A}^{(0)}$ for the spin-dependent term, since it does not contain any $x$-dependent terms. 
In addition, we used the projection properties \eqref{eq:fQCD:projection}, \eqref{eq:fQCD:projectionSigma} to pull out the full $\Sigma^{\mu\nu}$.
After evaluating the matrix element, the term inside the bracket becomes equal to the eikonal factor, similarly to our universal contraction (\ref{eq:eikonalF}).


\section{Vectorial QCD}
\label{sec:vQCD}

We now extend the treatment to the case of vector matter and show in particular how the spin-1 term arises. 
While the expressions look quite different at first, the situation is remarkably similar to the fermionic case.
We consider a complex\footnote{This toy model is just for convenience, as we can immediately transfer most of the complex scalar results to the complex vector case. We do not claim that this construction defines a UV-complete non-linear quantum field theory.} vector field $V_{c}^\mu$ in some representation of $SU(N)$, but in principle the discussion also holds for collinear gluons as emitting particles.
The vector components scale as
\begin{equation}\label{eq::VQCD::VCounting}
    \np V_c \sim 1\,,\quad V_{c\perp} \sim \lambda\,,\quad \nm V_c \sim \lambda^2
\end{equation}
in the SCET power-counting parameter. 
The vector field must come with its own gauge symmetry, which we call $V_c$-gauge, to define a consistent theory.
However, the details of this gauge symmetry as well as the collinear gauge symmetry are irrelevant for the soft theorem, and either by explicit gauge-fixing, or by Wilson-line constructions similar to the case of the gluon, we can define a manifestly $V_c$-gauge invariant field.
It is only necessary that this vector transforms like a matter field under the soft gauge transformation. 
To stress this, we define the gauge-invariant field $\mathcal{V}_{c}^\mu$, which satisfies $\np \mathcal{V}_c = 0$ and only transforms under the soft gauge symmetry.
It is advantageous to work with this gauge-invariant vector field.
First, since $\np \mathcal{V}_c = 0$, there is no $\mathcal{O}(1)$ building block, and the first possible collinear vector operator scales as $\mathcal{V}_{c\perp}\sim \mathcal{O}(\lambda)$.
Second, the $\nm \mathcal{V}_c$ component is subleading. We can express this component in terms of the leading $\mathcal{V}_{c\perp}$ using the equation of motion as
\begin{equation}\label{eq::VQCD::SubleadingVector}
    \nm \mathcal{V}_c = -\frac{2}{i\np \partial}i\partial_{\perp\alpha} \mathcal{V}^{\alpha}_{c\perp} + \mathcal{O}(\mathcal{V}_c^2)\,.
\end{equation}
Thus $\mathcal{V}_{c\perp}$ is the analogue of the spinor $\xi_c$ in \cref{sec:fQCD}, and \eqref{eq::VQCD::SubleadingVector} is the analogue of the spinor-relation \eqref{eq:fQCD:FullSpinorRelation}.
The crucial difference to the fermionic theory is that we write the Lagrangian in terms of the original field $V_{c\mu}$.
For the actual computations, this means that $\mathcal{V}_{c\mu}$ should be expressed in terms of the original field $V_{c\mu}$.
To linear order, they are related as
\begin{equation}\label{eq::VQCD::GaugeInvVRelation}
    \mathcal{V}_{c\mu} = V_{c\mu} - \partial_\mu\frac{\np V_c}{\np \partial} + \dots\,.
\end{equation}

\subsection{Non-radiative matching}

Again, we consider first the non-radiative matching, following the outline in \cref{sec:sQCD,sec:fQCD}.
In the operator basis, we only have $\mathcal{V}_{c\perp}$ as a building block, which enters at $\mathcal{O}(\lambda)$.
The LP matching 
(\ref{eq:CA0}) now reads
\begin{align}\label{eq:CA0V}
 \mathcal{A}^{(0)} &= \varepsilon^*_{\alpha_{1}}(p_1) \ldots \varepsilon^*_{\alpha_{N}} (p_N) \mathcal{A}^{\left(0\right)\alpha_{1}\dots\alpha_{N}}\nn\\
 &=
 \langle p_1,\ldots, p_N  | \hat{\mathcal{A}}^{(0)} | 0 \rangle  \nn \\ &=
 \tilde{\varepsilon}^*_{c_1\alpha_{1\perp}}(p_1) \ldots \tilde{\varepsilon}^*_{c_N\alpha_{N\perp}}(p_N)
\int  \left[ dt \right]_N
 \eta_\perp^{\alpha_1\beta_1}\dots \eta_\perp^{\alpha_N\beta_N} \lp \widetilde{C}^{A0}\rp_{\beta_1\dots\beta_N}
e^{i\sum_i  n_{i+}p_i \; t_i}\nn\\
  &\equiv  \tilde{\varepsilon}^*_{c_1\alpha_{1\perp}}(p_1) \ldots \tilde{\varepsilon}^*_{c_N\alpha_{N\perp}}(p_N)\eta_\perp^{\alpha_1\beta_1}\dots\eta_\perp^{\alpha_N\beta_N} \lp C^{A0}\rp_{\beta_1\dots\beta_N}\,.
\end{align}
In the vectorial case, there are a few subtleties to note:
First, the amplitude $\mathcal{A}^{(0)}$ is written in terms of the full-theory polarisation vectors $\varepsilon_\alpha$, which are not necessarily restricted to transverse components, whereas the $N$-jet operator contains the transverse building block $\mathcal{V}_{c\perp}$ and the corresponding polarisation tensor $\tilde{\varepsilon}_{c_i\alpha_{i\perp}}$, where 
the first index refers to the collinear direction, and the second is the Lorentz index.
These vectors are related via \eqref{eq::VQCD::GaugeInvVRelation} as
\begin{equation}
    \Tilde{\varepsilon}_{c_i\mu_\perp}(k) = \varepsilon_{\mu_\perp}(k) - k_{\mu_\perp} \frac{\nip\varepsilon(k)}{\nip k}\,.
\end{equation}
Next, the full-theory amplitude $\mathcal{A}^{\alpha_1\dots\alpha_n}$ is a rank $N$ tensor in Minkowski space, with one index per external polarisation tensor.
Thus, also the matching coefficient $C^{A0}$ is a tensor, indicated by the brackets.
At leading order, only the $\perp$ components are relevant.
At subleading order, the other components are also relevant, as will be see in the $C^{A1\mu}$ relation \eqref{eq::VQCD::VectorA1} below.
Hence, we do not restrict the indices of $C^{A0}$ to be transverse, but rather write an explicit $\eta_\perp^{\alpha\beta}$ to indicate this projection.
This is essentially the analogue of the projection properties \eqref{eq:fQCD:projection} of the spinor $\xi_c$ in \cref{sec:fQCD}.
Just as in the fermionic case \eqref{eq:fQCD:FullSpinorRelation}, the subleading component of the polarisation vector is related to the leading one via \eqref{eq::VQCD::SubleadingVector} as
\begin{equation}
    \nim \tilde\varepsilon_{c_i}(k) = -\frac{2}{\nip k}\,k_\perp^\alpha \tilde{\varepsilon}_{c_i\alpha\perp}\,.
\end{equation}
Thus, as in \cref{sec:fQCD}, the $\lp {C}^{A0}
\rp_{\beta_1\dots\beta_N}$ corresponds to the LP amplitude with external polarisation vectors stripped off,
\begin{align}\label{eq:CA0AVec}
\lp {C}^{A0}
\rp_{\beta_1\dots\beta_N}= \mathcal{A}^{(0)}_{\beta_1\dots\beta_N}\,.
\end{align}

The subleading current is related to the leading current in a similar fashion as in the fermionic case presented in \eqref{eq:subleadingSpinor}.
Here, we again find the spin-independent contribution present already in the scalar case \eqref{eq:RPIs}, as well as the contribution from the subleading $\nm\mathcal{V}_c$ component, similar to the spin part in \eqref{eq:subleadingSpinor}.
The relation reads
 \begin{equation}
    \begin{split}\label{eq::VQCD::VectorA1}
    \lp C_i^{A1\,\mu}\rp_{\beta_1\dots\beta_i\dots\beta_N}
    &=\bigg[-\frac{ \eta_{\perp\beta_i}^{\mu}}{n_{i+}p_i}\np^{\rho_i} 
    -\sum_{j\neq \,i}\eta_{\perp\beta_i}^{\rho_i}\frac{2n^{\mu}_{j-}}{n_{i-}\cdot n_{j-}}\frac{\partial}{\partial n_{i+}p_{i}}\bigg]
    \lp C^{A0}\rp_{\beta_1\dots\rho_i\dots\beta_N}
    \\
    &\equiv \lp C^{A1\mu}_{i,\,{\rm spin}}\rp_{\beta_1\dots\beta_i\dots\beta_N} + \lp C^{A1\mu}_{i,\,{\rm orbit}}\rp_{\beta_1\dots\beta_i\dots\beta_N}\,.
    \end{split}
    \end{equation}
Here the coefficients of the $A1$ current are contracted as $\lp C^{A1\mu}\rp_{\dots\alpha_i\dots} i\partial_{\perp\mu} \mathcal{V}_{c_i}^{\alpha_{i\perp}}$.
Note the similarity of the first term to the spin-term in \eqref{eq:subleadingSpinor}.
We define the corresponding $\hat{\mathcal{A}}^{(1)}_{\rm orbit}$ $\hat{\mathcal{A}}^{(1)}_{\rm spin}$ accordingly as in \eqref{eq::fQCD::A1spinorbit}.

Just as in the scalar case, the universal contraction (\ref{eq:eikonal}) of the original $V_{c\mu}$ fields should be defined with $\np\partial$ acting on $V_{c\mu}$, and it is now given by 
\begin{align}\label{eq:eikonalV}
 \wick{
\int d^4x e^{i\frac{1}{2}n_- k\, n_+ x }  \langle \c1 p \vert \;  \c2 V_{c\alpha}^\dagger(0),\;  \c1 V^\dagger_{c\nu}(x)\;   \c2 {\np\partial V_{c\beta}} (x)\;   \vert 0 \rangle = \varepsilon_\nu^*(p)\frac{-i \eta_{\alpha\beta}}{\nm k}\,,
  }
\end{align}
where the soft momentum is denoted by $k$ and we use the propagator in Feynman gauge.

With this understanding we simplify the cluttered notation by dropping the Lorentz indices due to contractions with polarisation tensors in the following. 
We keep the ones which are related to contractions with derivatives, e.g. in $C^{A1\mu}$, essentially using the same notation as in \cref{sec:fQCD}, always keeping in mind the fact that $C^{A0}$ is actually a rank-$N$ tensor.

\subsection{Soft theorem}

It is now straightforward to derive the soft theorem. Again, following the discussion in \cref{sec:SoftTheoremSQCD}, we make use of the fact that there are no soft gluon building blocks available, that we can consider a single leg and sum over the individual contributions, and we can choose $p_i^\mu =  \nip p_i\,\frac{\nim^\mu}{2}$.
The relevant part of the soft-collinear Lagrangian for the complex vector field can be conveniently expressed in terms of the linear current, just as in the scalar case. We define the current
\begin{align}
    j_{\mu}^a &\equiv  \lc i \partial_\mu  V^\dagger_{c\alpha}\rc t^a  V_c^\alpha - V^\dagger_{c\alpha} t^a i\partial_\mu V_c^\alpha
     + 2 V^\dagger_{c\alpha} t^a i\partial^\alpha  V_{c\mu} - 2 \lc i\partial^\alpha  V^{\dagger}_{c\mu}\rc t^a   V_{c\alpha}\,.
\end{align}
The first two terms look very reminiscent of the scalar Noether current \eqref{eq:NoetherScalar}, while the last two terms are new, and are relevant for the contributions to the spin operator. Note that these terms also contain the linear terms of the interaction $F_{A}^{\mu\nu}V^\dagger_{c\mu}V_{c\nu}$ by virtue of integration by parts.
Expressed in terms of the current, the Lagrangian contains the same structures as in the scalar and fermionic cases \eqref{eq:LSQCD}, \eqref{eq:L2SQCD} and \eqref{eq:LFQCD}, \eqref{eq:L2FQCD}, respectively, and is given by
\begin{equation}
    \mathcal{L}_V = \mathcal{L}_V^{(0)} + \mathcal{L}^{(1)}_V + \mathcal{L}^{(2)}_V\,.
\end{equation}
The leading term reads
\begin{equation}
    \mathcal{L}^{(0)}_V = -\frac 12 \,{\rm tr}\lp (\partial_\mu V_{c\nu}-\partial_\nu V_{c\mu})^\dagger (\partial^\mu V_c^\nu-\partial^\nu V_c^\mu)\rp + \frac 12 g_s \nm A_s^{a} \np j^a + \mathcal{O}(g^2)\,,
\end{equation}
and the subleading linear interactions are
\begin{align}\label{eq:LVQCD}
    \mathcal{L}_{V}^{\left(1\right)}&	= 
\frac{1}{2} x_{\perp}^{\mu}\, n_{+}j_a\, n_{-}^{\nu}g_sF_{\mu\nu}^{s\;a} \\
\mathcal{L}_{V}^{\left(2\right)}	&\equalhat \mathcal{L}_{V}^{\left(2a\right)}+\mathcal{L}_{V}^{\left(2b\right)}+\mathcal{L}_{V}^{\left(2c\right)}\,,
\end{align}
with
\begin{align}
\mathcal{L}_{V}^{\left(2a\right)} &=
\frac{1}{2} x_{\perp}^{\nu}j^{\mu_\perp}_a g_sF_{\nu\mu}^{s\;a}\,, \\
\mathcal{L}_{V}^{\left(2b\right)}&=
\frac{1}{4}
n_{-}x\,n_{+}^{\mu}\, n_{+} j_a \, n_{-}^{\nu}g_sF_{\mu\nu}^{s\;a}\,, \\
\mathcal{L}_{V}^{\left(2c\right)}&=
\frac{1}{4}  x_{\perp}^{\mu}x_{\perp\rho}\,n_{+} j_a\,n_{-}^{\nu}\,{\rm tr}\left(\left[D_{s}^{\rho},g_sF_{\mu\nu}^{s}\right] t^a \right)
\,.
\end{align}
In this form, we see the complete analogy to the scalar case \eqref{eq:LSQCD}.
However, we can also rearrange the terms to bring them into a form similar to the fermionic Lagrangian.
To achieve this, we have to abandon the compact notation in terms of the Noether current and write the terms explicitly.

To find the analogue of $\mathcal{L}^{(2s)}_\xi$ in \eqref{eq:FQCD:LSpin}, we manipulate $\mathcal{L}^{(2a)}_V$ to obtain
\begin{equation}
    \mathcal{L}^{(2a)}_V \equalhat -ig_s F_{\mu\nu}^{s\;a} V^\dagger_{c\alpha} t^a V_{c\beta} \lp \eta_{\perp}^{\alpha\mu}\eta_\perp^{\beta\nu} - \eta_\perp^{\alpha\nu}\eta_\perp^{\beta\mu}\rp
    = -ig_s F_{\mu\nu}^{s\;a} V_{c\alpha}^\dagger t^a V_{c\beta} (\Sigma_\perp^{\mu\nu})^{\alpha\beta}\,.
\end{equation}
We used  integration by parts, dropping terms proportional to $p_\perp^\mu$, the equation of motion, and $\partial_\mu V^\mu \equalhat 0$, and introduced the spin-1 operator $(\Sigma^{\mu\nu})^{\alpha\beta} = \lp \eta^{\mu\alpha}\eta^{\nu\beta} - \eta^{\mu\beta}\eta^{\nu\alpha}\rp$, decomposed as
\begin{equation}
    (\Sigma^{\mu\nu})^{\alpha\beta}
    \equalhat \underbrace{\eta_{\perp}^{\alpha\mu}\eta_\perp^{\beta\nu} - \eta_\perp^{\alpha\nu}\eta_\perp^{\beta\mu}}_{(\Sigma_\perp^{\mu\nu})^{\alpha\beta}}
      + \underbrace{\frac 12 \eta_\perp^{\alpha[\mu}\nm^{\nu]}\np^\beta - \eta_\perp^{\beta[\mu}\nm^{\nu]}\np^\alpha}_{(\Sigma_{\perp+}^{\mu\nu})^{\alpha\beta}}
    +
    \underbrace{\frac{n_+^{[\mu} \nm^{\nu]}}{4}(n_-^\alpha n_+^\beta - n_+^\alpha n_-^\beta)}_{(\Sigma_{+-}^{\mu\nu})^{\alpha\beta}}
    \,,
\end{equation}
where we already neglected the components that do not contribute in the following.
Just as for the fermionic case, we see that we can explicitly read-off the transverse contribution to the spin angular momentum.

For $\mathcal{L}^{(2c)}_V$, we find, using the same manipulations as for the scalar case \eqref{eq:L2c}
\begin{equation}\label{eq::Vec::L2c}
    \mathcal{L}^{(2c)}_V \equalhat \frac 12 ig_s F_{+-}^{s\;a} V^{\dagger\alpha}_{c} t^a V_{c\alpha}\,,
\end{equation}
where we defined the short-hand notation $F_{+-}^{s\;a}=\np^\mu\nm^\nu F^{s\;a}_{\mu\nu}$.
This term does not have any interpretation, and it does indeed cancel out with some parts of $\mathcal{L}^{(2b)}_V$.
Together, we find
\begin{equation}\label{eq::Vec::L2b}
    \mathcal{L}^{(2b)}_V + \mathcal{L}^{(2c)}_V \equalhat -\frac 12 \nm x g_s F_{+-}^{s\;a} \lp V^\dagger_{c\alpha} t^a i\np\partial V_c^\alpha \rp - \frac 12 ig_sF_{+-}^{s\;a} \lp V^\dagger_{c-} t^a V_{c+} - V^\dagger_{c+} t^a V_{c-}\rp\,.
\end{equation}
The first term generates the orbital angular momentum component $L_{+-}$. The second can be rewritten as 
\begin{align}
    -\frac 12 ig_s F_{+-}^{s\;a} \lp \nm V^\dagger_c t^a \np V_c - \np V^\dagger_c t^a \nm V_c\rp &= 
    -\frac14 \np^{[\mu}\nm^{\nu]}\lp \nm^\alpha\np^\beta - \np^\alpha\nm^\beta\rp V_{c\alpha}^\dagger t^a V_{c\beta} ig_s F_{\mu\nu}^{s\;a}\nn\\
        &= -(\Sigma_{+-}^{\mu\nu})^{\alpha\beta} ig_sF^{s\;a}_{\mu\nu} V^\dagger_{c\alpha} t^a V_{c\beta}\,,
\end{align}
and is related to the longitudinal part of the spin term. We now observe that we can rewrite $\mathcal{L}^{(2)}_V$ in the same form as in the fermionic case: $\mathcal{L}^{(2b)}_V$ generates the orbital angular momentum, and we have an explicit spin term $\mathcal{L}^{(2a)}_V$.
In summary, we find after these manipulations
\begin{align}
    \mathcal{L}^{(2)}_V &\equalhat -\frac 12 \nm x \np^\mu \nm^\nu g_s F^{s\;a}_{\mu\nu} \lp V^\dagger_{c\alpha} t^a i\np\partial V_c^\alpha \rp
    - ig_s F_{\mu\nu}^{s\;a} V_{c\alpha}^\dagger t^a V_{c\beta} \lp (\Sigma_\perp^{\mu\nu})^{\alpha\beta} + (\Sigma_{+-}^{\mu\nu})^{\alpha\beta} \rp\,,
\end{align}
and we see that the first term generates the term in the orbital angular momentum proportional to $\np\partial \nm x$, just as in the scalar \eqref{eq::SQCD::L2} and fermionic \eqref{eq::fQCD::L2orbit} case. 
The second term contains two of the three spin terms, like for the fermion \eqref{eq::fQCD::L2spin}.
The two missing terms are the two mixed transverse-longitudinal terms, namely the orbital angular momentum piece proportional to $\np\partial x_\perp^\mu$ and the spin piece proportional to $\Sigma^{\mu\nu}_{\perp+}$.
Just as in the scalar and fermionic cases, these missing pieces stem from the time-ordered product $T\{\hat{\mathcal{A}}^{(1)},\mathcal{L}^{(1)}_V\}$. Indeed, we can rearrange $\mathcal{L}^{(1)}_V$ as follows, 
\begin{align}
    \mathcal{L}^{(1)}_V &\equalhat
    -x_\perp^\mu \nm^\nu g_sF^{s\;a}_{\mu\nu} V^{\dagger}_{c\alpha} t^a i\np\partial V_c^\alpha
    -
    ig_s \nm^\nu F_{\mu_\perp\nu}^{s\;a} \lp V^{\dagger\mu_\perp}_c t^a \np V_c - \np V^\dagger_c t^a V^{\mu_\perp}_c\rp\nn\\
    &\equalhat -x_\perp^\mu \nm^\nu g_sF^{s\;a}_{\mu\nu} V^{\dagger}_{c\alpha} t^a i\np\partial V_c^\alpha \,,\label{eq::VQCD::L1comp}
\end{align}
which gives a non-vanishing contribution only in the time-ordered product with the $A1$ current.
The second term in the first line of \eqref{eq::VQCD::L1comp} does not contribute at $\mathcal{O}(\lambda)$ in $T\{\hat{\mathcal{A}}^{(0)},\mathcal{L}^{(1)}_V\}$, since it gets projected out by the transverse $\eta_\perp$ (cf. \eqref{eq:CA0V}) and can only give a non-vanishing contribution with the $A1$ current. In this case, however, it vanishes after setting $p_\perp^\mu = 0$.
The first term is essentially the same as the scalar $\mathcal{L}^{(1)}_\chi$ \eqref{eq:LSQCD} and fermionic $\mathcal{L}^{(1)}_\xi$ \eqref{eq:LFQCD}, and it is this term that generates the orbital piece time-ordered product with $\hat{\mathcal{A}}^{(1)}_{\rm orbit}$.
Just as in the fermionic case \eqref{eq:S2}, the time-ordered product with $\hat{\mathcal{A}}^{(1)}_{\rm spin}$ then yields the missing transverse-longitudinal spin term.

In summary, the terms contributing to single-soft emission in the Lagrangian are cast in the form
\begin{equation}
    \mathcal{L}_V \equalhat \mathcal{L}^{(0)}_{\rm kinetic} + \mathcal{L}^{(0)}_{\rm eikonal} + \mathcal{L}^{(1)}_V
    + \mathcal{L}^{(2)}_{\rm orbit} + \mathcal{L}^{(2)}_{\rm spin}\,,
\end{equation}
with the soft-collinear interaction Lagrangians given by 
\begin{align}
    \mathcal{L}^{(0)}_{\rm eikonal} &= -g_s \nm A^a \np V^{\dagger}_{c\alpha} t^a i\np\partial V_c^\alpha\,,\\[0.1cm]
    \mathcal{L}^{(1)}_V
    &= -ig_sF^{s\;a}_{\mu\nu} V^\dagger_{c\alpha} t^a L_{\perp+}^{\mu\nu} V_c^\alpha \\
    \mathcal{L}^{(2)}_{\rm orbit} &= -ig_s F_{\mu\nu}^{s\;a} V^\dagger_{c\alpha} t^a L_{+-}^{\mu\nu} V_c^\alpha\\
    \mathcal{L}^{(2)}_{\rm spin} &= -ig_s F_{\mu\nu}^{s\;a} V_{c\alpha}^\dagger t^a V_{c\beta} \lp (\Sigma_\perp^{\mu\nu})^{\alpha\beta} + (\Sigma_{+-}^{\mu\nu})^{\alpha\beta} \rp\,.
\end{align}
We note the similarity to the fermionic interactions \eqref{eq::fQCD::L2orbit}, \eqref{eq::fQCD::L2spin}.

Combining all contributions, we the subleading next-to-soft term in the soft theorem is
\begin{align}
& i\int d^{4}x\:T\left\{\hat{\mathcal{A}}^{(1)}_{\rm orbit} + \hat{\mathcal{A}}^{(1)}_{\rm spin} , \mathcal{L}^{\left(1\right)}_V\right\}
+ i\int d^{4}x\:T\left\{\hat{\mathcal{A}}^{(0)}, \mathcal{L}_{\rm orbit}^{\left(2\right)} + \mathcal{L}^{(2)}_{\rm spin}\right\}
 \nn\\   & =    i\int d^{4}x\;T\left\{ \hat{\mathcal{A}},-V^\dagger_{c\alpha} \lp \eta^{\alpha\beta} L^{\mu \nu} + (\Sigma^{\mu\nu})^{\alpha\beta}\rp ig_sF_{\mu\nu}^{s\;a} \,V_\beta
    \right\}   \,,
    \label{eq:LBK-VQCD}
\end{align}
and we recover the same structure as for the scalar \eqref{eq:LBK-SQCD} and fermionic \eqref{eq:fQCD:SoftTheoremOperator} case. Although we shall not pursue this further here, the 
striking similarity of the structure of the derivation 
of the fermionic and vectorial case suggests that 
this procedure and the corresponding operatorial statement 
can now be generalised in a straightforward fashion to 
matter fields of arbitrary spin.


\section{Soft theorem in gravity}
\label{sec:GR}

Now that we thoroughly discussed the gauge-theory case, we proceed to the gravitational soft theorem. In gravity, there is an additional next-to-next-to-soft term. Hence, when deriving the gravitational soft theorem, we go to $\mathcal{O}(\lambda^4)$ to obtain the three universal terms. 

In the following, we focus on the gauge principles underlying the terms contributing to single soft emission, i.e. terms linear in the soft graviton field $s_{\mu\nu}$.
This greatly simplifies the discussion, as the complicated structure of the non-linear interactions, as well as interactions between collinear and soft gravitons are avoided. 
For the general, non-linear soft graviton interactions, we refer to the companion work \cite{Beneke:2021aip}.

One major difference to gauge theory is that in gravity the gauge transformations are inherently inhomogeneous in $\lambda$.
This is due to the fact that the components of collinear momenta scale differently in $\lambda$, and the momentum generates the gravitational gauge symmetry, local translations.
Whereas the homogeneous gauge symmetry was the guiding principle in the construction of soft-collinear gauge-theory beyond leading power, in the gravitational case we have to relax this constraint and find a gauge symmetry that respects the soft multipole expansion.
In the following, we briefly review the salient features of SCET gravity \cite{Beneke:2021aip}.
We consider a minimally-coupled complex scalar field to make the connections to the previous sections transparent. In 
this section, we use the short-hand notation 
$n_\pm^\alpha A_{\alpha\beta\ldots} = A_{\pm\beta\cdots}$ 
for the contractions with the collinear reference vectors.

\subsection{Soft gravity}

In the effective theory, the infinitesimal emergent soft gauge transformation consists of two parts \cite{Beneke:2021aip}, a local translation and a local Lorentz transformation. Under these, the collinear field transforms as
\begin{equation}
    \phi_c(x) \to \phi_c(x) - \varepsilon_s^\alpha(x_-)\partial_\alpha\phi_c(x) - \omega_s^{\alpha\beta}(x-x_-)_\alpha \partial_\beta \phi_c + \mathcal{O}(\varepsilon_s^2,\omega_s^{2},\omega_s\varepsilon_s)\,,
\end{equation}
where $\omega_s^{\alpha\beta}(x_-)$ is related to the derivative of the full-theory parameter $\varepsilon_s(x)$ as
\begin{equation}
    \omega_s^{\alpha\beta}(x_-) = \frac 12 \lp\lc\partial^\alpha \varepsilon^{\beta}_s\rc(x_-) - \lc \partial^\beta \varepsilon^{\alpha}_s\rc(x_-)\rp\,,
\end{equation}
and $x_-^\mu$ is defined in \eqref{eq:xminusdef}.
Let us stress that, since the soft gauge symmetry lives only on the classical trajectory $x_-^\mu$ of the energetic particle, the parameters $\varepsilon_s$ and $\omega_s$ must be viewed as independent objects, as the latter is evaluated at $x_-$ only after taking the derivatives. 
Hence, we already anticipate that there are two independent gauge fields.
These fields can be used to define a soft-covariant derivative $\nm D_s$, which is non-linear in the soft graviton field $s_{\mu\nu}$.
To first order in $s_{\mu\nu}$, this derivative is given by
\begin{equation}
    \nm D_s
    = \nm\partial - \frac \kappa2 s_{\alpha-}\partial^\alpha - \frac \kappa2 \lc \partial_\alpha s_{\beta-}\rc J^{\alpha\beta} + \mathcal{O}(s^2)\,, 
\end{equation}
where we introduced the angular momentum
\begin{equation}
    J^{\alpha\beta} = (x-x_-)^\alpha \partial^\beta - (x-x_-)^\beta \partial^\alpha\,.
\end{equation}
Just as their corresponding gauge parameters, the soft field $s_{\mu-}(x_-)$ and its derivative ${\lc \partial_\alpha s_{\mu-}\rc(x_-)}$ are independent objects in the effective theory.
We can thus truly interpret them as independent gauge fields, even though they stem from the same full-theory field, which couple to momentum and angular momentum, respectively.
It is remarkable that the soft sector provides a soft-covariant derivative quite naturally, even though the minimally-coupled scalar field does not contain a gravitational-covariant derivative in the full theory.
We can appreciate the strong similarities to the gauge-theory case outlined in \cref{sec:prelim:soft}.

Besides this soft-covariant derivative, the subleading interactions are expressed entirely in terms of the Riemann tensor
\begin{equation}
    R^s_{\mu\nu\alpha\beta} = \frac \kappa2 \lp \partial_\mu \partial_\beta s_{\nu\alpha} + \partial_\nu \partial_\alpha s_{\mu\beta} - \partial_\mu \partial_\alpha s_{\nu\beta} - \partial_\nu \partial_\beta s_{\mu\alpha}\rp + \mathcal{O}(s^2)\,,
\end{equation}
and its derivatives, in analogy to the gauge-theory case \eqref{eq:LSQCD}, where the subleading interactions are expressed in terms of the field-strength tensor $F_{\mu\nu}^s$.

The full Lagrangian containing all relevant terms up to $\mathcal{O}(\lambda^4)$ is given in Appendix~\ref{sec::appGR::Lagrangian}. We focus on the terms up to $\mathcal{O}(\lambda^2)$ here for brevity.
The Lagrangian for a complex gauge-invariant scalar field $\chi_c$ can be conveniently expressed as
\begin{equation}\label{eq::GR::StructureLagrangian}
\begin{aligned}
    \mathcal{L} &= \frac 12 \lc\np\partial\chi_c\rc^\dagger\nm\partial\chi_c + \frac 12 \lc \nm\partial\chi_c\rc^\dagger \np\partial\chi_c
    + \lc \partial_{\mu_\perp}\chi_c^\dagger\rc\partial^{\mu_\perp}\chi_c\\
    &\quad
    - \frac{\kappa}{4}s_{-\mu}T^{\mu+}
     - \frac{\kappa}{4} \lc \partial_{[\mu} s_{\nu]-}\rc\,(x-x_-)^\mu T^{\nu+} - \frac 18 x_\perp^\alpha x_\perp^\beta  R^s_{\alpha-\beta-} T_{++} + \mathcal{O}(x^3)\,,
     \end{aligned}
\end{equation}
where we introduced the energy-momentum tensor
\begin{equation}
    T^{\mu\nu} = \lc \partial^\mu\chi_c\rc^\dagger \partial^\nu \chi_c + \lc \partial^\nu \chi_c\rc^\dagger \partial^\mu\chi_c - \eta^{\mu\nu} \lc\partial_\alpha\chi_c\rc^\dagger\partial^\alpha\chi_c\,.
\end{equation}
In this form, we see quite transparently the coupling of $s_{\mu-}$ to the energy-momentum tensor $T^{\mu\nu}$, as well as the coupling of its derivative $\partial_{[\alpha} s_{\beta]-}$, which is an independent gauge field in the effective theory, to the angular momentum density 
\begin{equation}
    \mathcal{J}^{\alpha\beta\mu} = (x-x_-)^\alpha T^{\beta\mu} - (x-x_-)^\beta T^{\alpha\mu}\,.
\end{equation}
However, the Lagrangian \eqref{eq::GR::StructureLagrangian} is not homogeneous in $\lambda$, as the scaling of the collinear momenta leads to inhomogeneous contractions between soft fields and collinear derivatives.
Expanding in powers of $\lambda$, we find
\begin{equation}
    \mathcal{L} = \mathcal{L}^{(0)} + \mathcal{L}^{(1)} + \mathcal{L}^{(2)} + \mathcal{O}(\lambda^3)\,,
\end{equation}
where
\begin{align}
\mathcal{L}^{(0)} &= \frac 12 \lc\np\partial\chi_c\rc^\dagger \nm\partial\chi_c + \frac 12 \lc\nm\partial\chi_c\rc^\dagger\np\partial\chi_c + \lc\partial_{\mu_\perp}\chi_c\rc^\dagger\partial^{\mu_\perp}\chi_c -\frac \kappa 8 s_{--} T_{++}\,,\\
    \mathcal{L}^{(1)} &= -\frac \kappa 4 s_{-\mu_\perp}\tensor{T}{^{\mu_\perp}_+}
    - \frac \kappa 8 \lc \partial_{[\mu}s_{-]-}\rc\, x_\perp^\mu T_{++}\,,
\label{eq:GRSCETL1}\\
    \mathcal{L}^{(2)} &= - \frac \kappa 8 s_{+-} T_{+-}
    - \frac{\kappa}{4} \lc \partial_{[\mu_\perp}s_{\nu_\perp]-}\rc\, x_\perp^\mu \tensor{T}{^{\nu_\perp}_+}
    - \frac{\kappa}{16} \lc \partial_{[+}s_{-]-}\rc\, \nm x T_{++}\nn\\
    &\quad
    -\frac  18 x_\perp^\alpha x_\perp^\beta R^s_{\alpha-\beta-} T_{++}\,.
\end{align}

In the soft theorem, the structure of the Lagrangian \eqref{eq::GR::StructureLagrangian} manifests itself as follows:
The leading interaction generates the eikonal term
\begin{equation}
\varepsilon_{\mu-}\,p^\mu\,\frac{\np p}{p\cdot k}\,,    
\end{equation} 
where the first factor $p^\mu$ is due to the coupling, and the second factor is the eikonal propagator. 
$\varepsilon_{\mu\nu}$ denotes the polarisation tensor of the 
emitted graviton. 
The next interaction in \eqref{eq::GR::StructureLagrangian} generates the subleading term 
\begin{equation}
    k_\rho \varepsilon_{\mu-} J^{\rho\mu}\,\frac{\np p}{p\cdot k}\,,
\end{equation}
where we can see that the coupling to the angular momentum in the Lagrangian already provides the correct form.
Here, just as in the gauge-theory case, the term counts as $\mathcal{O}(\lambda^2)$, and the terms in $\mathcal{L}^{(1)}$ only contribute in time-ordered products with suppressed $N$-jet operators, like $T\{\hat{\mathcal{A}}^{(1)},\mathcal{L}^{(1)}\}$.
These first two terms are the exact analogues of the leading eikonal term in QCD, and they also stem from the gravitational covariant derivative.
Finally, the Riemann tensor terms which are present in $\mathcal{L}^{(2)},\mathcal{L}^{(3)}$ and $\mathcal{L}^{(4)}$ generate the sub-subleading term
\begin{equation}
    \frac{1}{2}\varepsilon_{\mu\nu} k_\rho k_\sigma J^{\rho\mu}\frac{J^{\sigma\nu}}{p\cdot k}\,,
\end{equation}
which starts to contribute at $\mathcal{O}(\lambda^4)$.
Here, one factor of $J$ is due to the coupling, while the second factor is from the eikonal propagator in combination with the explicit $x_\perp$, just as in the second term in QCD.
In the following, we make these ideas explicit and derive the soft theorem to $\mathcal{O}(\lambda^4)$, following closely the derivation in the gauge-theory case.

\subsection{$N$-jet operators}

Most of the concepts introduced in \cref{sec:sQCD} can be carried over to the gravitational case, but there are some differences.

The first difference to the gauge-theory case concerns the $N$-jet operator 
\eqref{eq:GenericNJet}.
For gravity, these operators are defined in a translationally-invariant fashion as
\begin{equation}
    \hat{\mathcal{A}} = \int d^4x\: \hat{\mathcal{A}}(x) = \int d^4x\: T_{x} \hat{\mathcal{A}}(0) T^{-1}_{x}\,,
\end{equation}
where $\hat{\mathcal{A}}(0)$ is the same $N$-jet operator as defined in \eqref{eq:GenericNJet},
\begin{align}
 \hat{ \mathcal{A}}(0) = \sum_X \int [dt]_N\: \widetilde{C}^X({t_1,\ldots t_N}) \left(\prod_{i=1}^N J^X_i(t_i)\right)J^X_s(0) \,,
\label{eq:Njetagain}
  \end{align}
  and $T_x$ denotes a translation to the point $x$.
Once we evaluate the matrix element, the integral over $x$ turns into the momentum-conserving $\delta$-function. We can therefore adopt the convention that we drop the integral over $x$, work with the unintegrated $N$-jet operators \eqref{eq:Njetagain} as in the previous sections, and impose momentum conservation by hand in the final amplitude.
This simplifies the notation greatly.
The operator basis can be constructed analogously to the QCD case, and the generic building blocks are given by the analogues of \eqref{eq:JA0Blocks}.
Note that this time, the $A2$ building block
\begin{equation}
    J^{A2\mu\nu}_{\partial^2\chi_i^\dagger} = i\partial_{i\perp}^\mu i\partial_{i\perp}^\nu\chi_i^\dagger(t_i\nip)
\end{equation}
is relevant for the $\mathcal{O}(\lambda^4)$ contributions, as it is related to the second derivative
\begin{equation}
     \mathcal{A}^{(2)}= 
 p_{i \perp }^\mu p_{i\perp}^\nu
 \left.\lp\frac{\partial^2}{\partial p^{\mu}_{i\perp}\partial p^{\nu}_{i\perp}} \mathcal{A}\rp\right\rvert_{p_i^\mu = n_{i+}p_i\, n^\mu_{i-}/2 } \,,
\end{equation}
 of the non-radiative amplitude,\footnote{Note that we can eliminate $\nm p$ via $\nm p = - \frac{p_\perp^2}{\np p}$, so we can expand the amplitude using only $p_\perp$.} and the amplitude is expanded as
\begin{equation}
\mathcal{A}=C^{A0}(n_+ p )+p^{\mu}_{i\perp}C_{i,\,\mu}^{A1}(n_+ p )+
p^{\mu}_{i\perp}p^{\nu}_{i\perp}C_{i,\,\mu\nu}^{A2}(n_+ p )
+ \mathcal{O}(\lambda^3)\,.
\end{equation}

The soft building blocks differ slightly from the gauge-theory case.
In gravity, the covariant derivative $\nm D_s$ contains two independent gauge fields, one linked to local translations and the other to local Lorentz transformations.
As in the gauge-theory case, this covariant derivative can be eliminated using the equations of motion.
The next allowed soft building block is then the Riemann tensor $R^s_{\mu\nu\alpha\beta}$, the analogue of the field-strength tensor $F^s_{\mu\nu}$.
However, the Riemann tensor contains two derivatives of the soft field, and thus counts as $\mathcal{O}(\lambda^6)$.
Hence, in gravity, \emph{there are no soft graviton building blocks in the operator basis until $\mathcal{O}(\lambda^6)$.}
In other words, the first \emph{three} terms in gravity are universal for all processes, and only at $\mathcal{O}(\lambda^6)$, process-dependent building blocks can enter.
This already proves that the gravitational soft theorem contains three universal pieces, including a next-to-next-to soft term, and we determine them in the same fashion as the gauge-theory case presented in \cref{sec:SoftTheoremSQCD}.

To sum up, the entire soft emission process up to $\mathcal{O}(\lambda^4)$ can be described with purely collinear building blocks \eqref{eq::NJet::RelevantOperators}
and time-ordered products with the Lagrangian.
The non-radiative matching works exactly as discussed in \cref{sec:sQCD}, and there is no adaptation needed.

We can now proceed with the derivation of the soft theorem.
This derivation is completely analogous to the one presented in \cref{sec:SoftTheoremSQCD}, but we extend the discussion and go to $\mathcal{O}(\lambda^4)$ to also find the universal sub-subleading (next-to-next-to-soft) term.
For the scalar field, the gravitational soft theorem \eqref{eq:SoftTheorem} is given by
    \begin{equation}\label{eq::GR::SoftTheoremAmplitude}
        \mathcal{A}_{\mathrm{rad}} = \frac{\kappa}{2}\sum_i \left( \frac{\varepsilon_{\mu\nu}(k)p_i^\mu p_i^\nu}{p_i\cdot k} + \frac{\varepsilon_{\mu\nu}(k)p_i^\mu k_\rho L_i^{\nu\rho}}{p_i\cdot k} 
        +\frac{1}{2}\frac{\varepsilon_{\mu\nu}(k) k_\rho k_\sigma L_i^{\rho\mu}L_i^{\sigma\nu}}{p_i\cdot k}
        \right)\mathcal{A}\,.
\end{equation}
We make use of the same manipulations and choices as we did in the previous sections, explained in detail in \cref{sec:sQCD:setup}.
Most importantly, we choose a reference frame where the external collinear momenta satisfy $p^\mu_\perp = 0$ and make use of $\equalhat$ to drop terms that do not contribute to the single soft-emission matrix element.
Due to the absence of soft graviton building blocks, we can consider the emission off a single leg.
In the following, we present the terms contributing to the soft emission and skip most of the computations, which are essentially the same as performed in detail in \cref{sec:sQCD:subleading}. For the interested reader, the full Lagrangian contributing to single soft emission up to $\mathcal{O}(\lambda^4)$ as well as details regarding the computation can be found in \cref{sec::appGR}.

\subsection{Leading-power term}\label{sec:GR:leading}

Just as in the gauge-theory case, the leading contribution can already be read-off from the Lagrangian.
It is given by the time-ordered product of the leading current $\hat{\mathcal{A}}^{(0)}$ with the leading-power Lagrangian,
which we rewrite as
\begin{align}
    \mathcal{L}^{(0)}_{\rm eikonal} &= \frac{\kappa}{4} s_{--}\, \chi_c^\dagger \np\partial\, \np\partial\chi_c\,.
\end{align}
 Here, the first $\np\partial$ is the analogue of the colour-generator $t^a$ in QCD, and the second $\np\partial$ generates the eikonal propagator, as in \cref{sec:SoftTheoremSQCD}, 
\eqref{eq:sQCD:eikonal}.
We obtain the leading contribution to the radiative amplitude as
\begin{align}
      \mathcal{A}_{\rm rad}^{(0)} &\equalhat i\int d^{4}x\;T\left\{\mathcal{\hat{A}}^{(0)}, \mathcal{L}^{\left(0\right)}_{\rm eikonal}\right\}\\
      &= i \int d^4x\: T\left\{\mathcal{\hat{A}}^{(0)}, -\frac{\kappa}{4} \chi_c^\dagger \,s_{--}i\np \partial\, i\np\partial \chi_c \right\}\,,\label{eq:gr:leadingOp}
 \end{align}
 where we can immediately read off the eikonal term proportional to $\np\partial\lp s_{--} \np\partial\chi_c\rp$. 
Note that $s_{--}$ depends only on $x_-$, and hence $\np\partial$ commutes with $s_{--}$.

\subsection{Subleading-power / next-to-soft term}

Just as in the gauge-theory case, with $p_{\perp}^\mu=0$, there is no contribution at $\mathcal{O}(\lambda)$ in \eqref{eq::GR::SoftTheoremAmplitude}, and the subleading term, given by
 \begin{equation}\label{eq::GR::SoftTheoremSubleadingAmplitude}
        \frac{\kappa}{2}\frac{\varepsilon_{\mu\nu}(k)\,p^\mu k_\rho L^{\nu\rho}}{p\cdot k} \,
       \mathcal{A}\,,
\end{equation}
enters at $\mathcal{O}(\lambda^2)$.
The angular momentum is given by the two terms in \eqref{eq:Jx}
and these two terms are reproduced analogous to the scalar QCD case, the first one stemming from $T\{\hat{\mathcal{A}}^{(0)}\,,\mathcal{L}^{(2)}\}$ and the second one from $T\{\hat{\mathcal{A}}^{(1)}\,,\mathcal{L}^{(1)}\}$.

First, we check that there is no contribution from $T\{\hat{\mathcal{A}}^{(0)}\,,\mathcal{L}^{(1)}\}$:
 In $\mathcal{L}^{(1)}$, \eqref{eq:GRSCETL1}, we integrate by parts to write it as
 \begin{align}
     \mathcal{L}^{(1)} &= -\frac \kappa 2 s_{-\mu_\perp} \lc \partial^{\mu}_\perp\chi_c\rc^\dagger \np\partial\chi_c - \frac \kappa4 \lc\partial_{[\mu}s_{\nu]-}\rc x_\perp^\mu\nm^\nu \lc \np\partial\chi_c\rc^\dagger \np\partial\chi_c\,,
 \end{align}
 and we see that there is no contribution at $\mathcal{O}(\lambda)$, as
 \begin{align}
    \int d^{4}x\;T\left\{\hat{\mathcal{A}}^{(0)}, \mathcal{L}^{\left(1\right)}\right\} \equalhat 0\,.
\end{align}
However, just as in the gauge-theory case, there is a non-vanishing contribution of the second term from the time-ordered product with $\hat{\mathcal{A}}^{(1)}$.

We can split $\mathcal{L}^{(2)}$ in a similar fashion as in the gauge-theory case \eqref{eq:LSQCD} as
\begin{align}
   \mathcal{L}^{(2a)} &=
    - \frac{\kappa}{4} \lc \partial_{\mu}s_{\nu-} - \partial_\nu s_{\mu-}\rc x_\perp^\mu \lp \lc \partial_\perp^\nu \chi_c\rc^\dagger \np\partial \chi_c + \lc \np\partial\chi_c\rc^\dagger \partial_\perp^\nu\chi_c\rp\,,\\
    \mathcal{L}^{(2b)} &= 
    -\frac{\kappa}{8} \lc \partial_\mu s_{\nu-}\rc \np^{[\mu}\nm^{\nu]} \nm x \lc\np\partial\chi_c\rc^\dagger \np\partial\chi_c \,,\\
    \mathcal{L}^{(2c)} &= 
    - \frac{1}{4}x_\perp^\alpha x_\perp^\beta R^s_{\alpha-\beta-}\lc\np\partial\chi_c\rc^\dagger \np\partial\chi_c\,,\\
    \mathcal{L}^{(2d)} &=
    \frac{\kappa}{4}s_{+-} \lc\partial_{\perp}^\alpha\chi_c\rc^\dagger \partial_{\perp\alpha}\chi_c\,.
\end{align}
Using integrations by parts and $p^\mu_\perp = 0$, we find for the first contribution
\begin{equation}
    \mathcal{L}^{(2a)}\equalhat - \frac{\kappa}{4} \lc \partial_{[\mu}s_{\nu]-}\rc \eta_{\perp}^{\mu\nu}\chi_c^\dagger \np\partial\chi_c\equalhat 0\,,
\end{equation}
which vanishes by symmetry.
For $\mathcal{L}^{(2b)}$, no additional manipulations are needed and we find  the $L^{\mu\nu}_{+-}$ orbital angular momentum term
\begin{equation}
    \mathcal{L}^{(2b)} \equalhat \frac{\kappa}{8} \lc \partial_\mu s_{\nu-}\rc \np^{[\mu}\nm^{\nu]}\chi_c^\dagger\: \lc\np\partial \nm x \np\partial\chi_c\rc\,.
\end{equation}
Next, we have $\mathcal{L}^{(2c)}$ with the two $x_\perp$ factors. Again, we write $x_\perp^\alpha x_\perp^\beta$ in terms of a trace and a traceless part to get 
\begin{equation}
    \mathcal{L}^{(2c)} \equalhat 
    - \frac{1}{8}x_\perp^2 \eta_\perp^{\alpha\beta}\tensor{R}{^{s}_{\alpha_\perp-\beta_\perp-}}\lc\np\partial\chi_c^\dagger\rc \np\partial\chi_c\,,
\end{equation}
where we dropped non-linear terms in $s_{\mu\nu}$.
With $\eta_\perp^{\alpha\beta}\tensor{R}{^{s}_{\alpha_\perp-\beta_\perp-}} = \eta^{\alpha\beta}\tensor{R}{^{s}_{\alpha-\beta-}}$ and using the source-less equation of motion $R^s_{--}=0$, we find that this term vanishes.
At the linear order in $s_{\mu\nu}$, this is equivalent to working with the transverse-traceless external polarisation tensor.
Finally, $\mathcal{L}^{(2d)}\equalhat 0$ because $p^\mu_\perp = 0$.

In summary, the subleading Lagrangian is
expressed as
\begin{align}\label{eq::GR::L1orbital}
    \mathcal{L}^{(1)} &\equalhat \frac{\kappa}{2} \lc\partial_\mu s_{\nu-}\rc \chi_c^\dagger \overset{\leftarrow}{L}\!\phantom{\,}_{+\perp}^{\mu\nu} \np \partial\chi_c\,,\\
    \mathcal{L}^{(2)}_{\rm orbital} &\equalhat \frac{\kappa}{2}  \lc\partial_\mu s_{\nu-}\rc \chi_c^\dagger \overset{\leftarrow}{L}\!\phantom{\,}_{+-}^{\mu\nu} \np \partial\chi_c\,,
\end{align}
where we identified the angular momentum using \eqref{eq:Jx} and defined the short-hand notation
\begin{equation}
    \overset{\leftarrow}{L}_{\mu\nu} = \overset{\leftarrow}{\partial}_{[\mu} x_{\nu]}\,,
\end{equation}
where $\chi\overset{\leftarrow}{\partial}_\mu = -\lc\partial_\mu\chi\rc$.
In this form, we see the universal contraction and the coupling to the angular momentum.
Alternatively, we can write e.g. \eqref{eq::GR::L1orbital} as
\begin{equation}
    \mathcal{L}^{(1)} \equalhat \frac{\kappa}{2} \lc\partial_\mu s_{\nu-}\rc \chi_c^\dagger \np \partial L_{+\perp}^{\mu\nu}\chi_c\,,
\end{equation}
where we immediately recognise the structure of \eqref{eq::GR::SoftTheoremSubleadingAmplitude}.

The operatorial version of the subleading term is then given by
\begin{align}
    \hat{\mathcal{A}}^{(2)}_{\rm rad} & \equalhat i\int d^{4}x\;T\left\{\hat{\mathcal{A}}^{(1)}, \mathcal{L}^{\left(1\right)}\right\}+i\int d^{4}x\;T\left\{\hat{\mathcal{A}}^{(0)}, \mathcal{L}_{\rm orbital}^{\left(2\right)}\right\}
 \nn\\
 & =    \int d^{4}x\;T\left\{ \hat{\mathcal{A}},\frac\kappa 2\chi_{c}^\dagger \overset{\leftarrow}{L}\!\phantom{\,}^{\mu \nu} \lc \partial_\mu s_{\nu-}\rc i\np \partial\chi_{c}
    \right\}\,.\label{eq:gr:subleadingOp}
\end{align}
Upon evaluating the matrix element, we find the eikonal propagator and a coupling to the angular momentum.
Unlike the gauge theory case \eqref{eq:LBK-SQCD}, where the field-strength tensor $F_{\mu\nu}^s$ appeared, this time the subleading term takes the form of an \emph{eikonal term}, just as the leading term, where the analogue of the charge is given by the orbital angular momentum $L^{\mu\nu}$, and the gauge field corresponds to $\partial_{[\mu} s_{\nu]-}$.
Indeed, this term is linked to the second group of terms in the Lagrangian, which are coupled to the angular momentum tensor, and is only gauge-invariant once we impose angular momentum conservation. {\em Hence the two-fold soft gauge symmetry immediately implies two eikonal terms in the soft theorem.} Only the third term is then expressed in a manifestly gauge-invariant fashion, via the Riemann tensor.

\subsection{Sub-subleading / next-to-next-to-soft term}

We proceed with the derivation of the next-to-next-to-soft term. 
We highlight only the main points of its derivation and relegate the detailed computation to \cref{sec::app::subsubleading}.
There is no contribution at $\mathcal{O}(\lambda^3)$ for 
$p_{\perp}^\mu=0$, so we turn to the sub-subleading term from \eqref{eq::GR::SoftTheoremAmplitude},
    \begin{equation}
        \frac{\kappa}{4}\frac{\varepsilon_{\mu\nu}(k) k_\rho k_\sigma L^{\rho\mu}L^{\sigma\nu}}{p\cdot k}
        \mathcal{A}\,.
\end{equation}
When evaluating an on-shell amplitude, the two angular momenta $L^{\mu\nu}$ can be taken to act only on the amplitude \cite{Bern:2014vva}.
In position space, at the Lagrangian level, it is convenient to thus define the combination
\begin{equation}
    \overset{\leftarrow}{L}_{\mu\rho}\overset{\rightarrow}{L}_{\nu\sigma} \equiv \lp \overset{\leftarrow}{\partial}_{[\mu} x_{\rho]}\rp\lp x_{[\nu}\overset{\rightarrow}{\partial}_{\sigma]}\rp\,.
\end{equation}
Expanding the product of angular momenta, we find four terms given by
\begin{align}
    \overset{\leftarrow}{L}\!\phantom{\,}^{\mu\rho}\overset{\rightarrow}{L}\!\phantom{\,}^{\nu\sigma} &= \frac{1}{16} \np^{[\mu}\nm^{\rho]}\np^{[\nu}\nm^{\sigma]}\, \np\overset{\leftarrow}{\partial}\nm x \nm x\np \partial
    + \frac 14 x_\perp^{[\mu}\nm^{\rho]} x_\perp^{[\nu}\nm^{\sigma]} \np\overset{\leftarrow}{\partial}\np\partial
    \nn\\
    &\quad + \frac 18 \np^{[\nu}\nm^{\sigma]}\, x_\perp^{[\mu} \nm^{\rho]} \np\overset{\leftarrow}{\partial} \nm x \np\partial
    + \frac 18 \np^{[\mu}\nm^{\rho]}\, x_\perp^{[\nu} \nm^{\sigma]} \np\overset{\leftarrow}{\partial} \nm x \np\partial\,.
\end{align}
Identifying the linear single-emission terms in the Riemann tensor as
\begin{equation}
    R^s_{\mu\alpha\nu\beta} = -\frac \kappa2 \lp k_\nu k_\alpha \varepsilon_{\mu\beta} + k_{\mu}k_\beta \varepsilon_{\nu\alpha} - k_\alpha k_\beta \varepsilon_{\mu\nu} - k_\mu k_\nu \varepsilon_{\alpha\beta} \rp + \mathcal{O}(\varepsilon^2)\,,
\end{equation}
we can write 
\begin{align}
    \kappa\varepsilon^{\mu\nu}(k) k^\rho k^\sigma {\overset{\leftarrow}{L}}_{\rho\mu}\overset{\rightarrow}{L}_{\sigma\nu} &= -\frac 18 R^s_{+-+-} \np \partial (\nm x)^2 \np \partial
    - \frac 12 R^s_{\mu-+-} \np \partial x_\perp^\mu \nm x \np\partial\nn  \\
    &\quad
    - \frac 12 R^s_{\mu-\nu-} \np \partial x_\perp^\mu x_\perp^\nu \np \partial\,.
\end{align}
These three terms originate from three different time-ordered products, which we recognise from their explicit $x$-dependence. 
The first term, containing $(\nm x)^2$, contributes in the time-ordered product $T\{\hat{\mathcal{A}}^{(0)}\,,\mathcal{L}^{(4)}\}$, so this term has to be identified inside $\mathcal{L}^{(4)}$.
The second term contains one $x_\perp$, so it yields a non-vanishing contribution only with at least one $\partial_\perp$ inside the current, i.e. it contributes inside $T\{\hat{\mathcal{A}}^{(1)}\,,\mathcal{L}^{(3)}\}.$
Finally, the last term contains two factors of $x_\perp$, so it needs to act on $\partial_\perp^2$ to give a contribution. It appears in the product $T\{\hat{\mathcal{A}}^{(2)}\,,\mathcal{L}^{(2)}\}$.
We can immediately identify the relevant terms in the respective Lagrangians.
After some manipulations (details in Appendix~\ref{sec::app::subsubleading}), we find 
\begin{align}\label{eq::GR::L4contribution}
    \mathcal{L}^{(4)} &\equalhat  \frac{1}{16} R^s_{+-+-} \chi_c^\dagger \np\partial (\nm x)^2 \np\partial\chi_c = \frac{1}{4} R^s_{\mu\alpha\nu\beta} \,\chi_c^\dagger \overset{\leftarrow}{L}\!\phantom{\,}^{\mu\alpha}_{+-}\overset{\rightarrow}{L}\!\phantom{\,}^{\nu\beta}_{+-} \chi_c\,,\\
    \mathcal{L}^{(3)} &\equalhat \frac{1}{4}  R^s_{\alpha-\beta-} \chi_c^\dagger \np\partial x_\perp^\alpha \nm x \np^\beta \np\partial \chi_c = \frac 14 R^s_{\mu\alpha\nu\beta}\,\chi_c^\dagger( \overset{\leftarrow}{L}\!\phantom{\,}_{+-}^{\mu\alpha} \overset{\rightarrow}{L}\!\phantom{\,}_{+\perp}^{\nu\beta} + \overset{\leftarrow}{L}\!\phantom{\,}_{+\perp}^{\mu\alpha}\overset{\rightarrow}{L}\!\phantom{\,}_{+-}^{\nu\beta})\chi_c\,,\\
    \mathcal{L}^{(2)} &\equalhat \frac 14 R^s_{\alpha-\beta-} \chi_c^\dagger \np\partial x_\perp^\alpha x_\perp^\beta \np\partial \chi_c
    = \frac 14 R^s_{\mu\alpha\nu\beta}\,\chi_c^\dagger \overset{\leftarrow}{L}\!\phantom{\,}_{+\perp}^{\mu\alpha} \overset{\rightarrow}{L}\!\phantom{\,}_{+\perp}^{\nu\beta} \chi_c\,,\label{eq::GR::L2contribution}
\end{align}
and the sub-subleading-power term of the soft theorem can 
be cast into the operatorial statement
\begin{align}
    \hat{\mathcal{A}}^{(4)} & \equalhat i\int d^{4}x\;T\left\{\hat{\mathcal{A}}^{(2)}, \mathcal{L}_{\xi}^{\left(2\right)}\right\}+i\int d^{4}x\;T\left\{\hat{\mathcal{A}}^{(1)}, \mathcal{L}_{\xi}^{\left(3\right)}\right\}
    + i\int d^{4}x\;T\left\{\hat{\mathcal{A}}^{(0)}, \mathcal{L}_{\xi}^{\left(4\right)}\right\}
 \nn\\
 & =   i\int d^{4}x\;T\left\{ \hat{\mathcal{A}},\frac14\chi_{c}^\dagger \overset{\leftarrow}{L}\!\phantom{\,}^{\mu \nu} \overset{\rightarrow}{L}\!\phantom{\,}^{\alpha\beta} R^s_{\mu\alpha\nu\beta}\chi_{c}
    \right\}\nn\\
& \equalhat  i\int d^{4}x\;T\left\{ \hat{\mathcal{A}},\frac14\chi_{c}^\dagger L^{\mu \nu} L^{\alpha\beta} R^s_{\mu\alpha\nu\beta}\chi_{c}
    \right\}\,,
\label{eq:gr:subsubleadingOp}
\end{align}
which exhibits the desired form. In the last line, we transformed the left-right angular momenta $\overset{\leftarrow}{L}\!\phantom{\,}^{\mu \nu} \overset{\rightarrow}{L}\!\phantom{\,}^{\alpha\beta}$ into the standard form $L^{\mu\nu}L^{\alpha\beta}$ using on-shell properties and equations of motion similar to the proof given in \cite{Bern:2014vva}.

In summary, we see that all three terms of the soft theorem can be cast into an operatorial statement. In view of the derivation provided here, the individual terms acquire a new interpretation. The first {\em two} terms \eqref{eq:gr:leadingOp}, \eqref{eq:gr:subleadingOp} take the form of an eikonal term and generalise the leading term \eqref{eq:OP-LP} in gauge theory. This follows because the  effective theory for soft-collinear gravitational interactions contains two soft background-gauge fields, one, $s_{\alpha-}(x_-)$, coupled to the momentum, and another, $\partial_{[\alpha}s_{\beta]-}(x_-)$ to the angular momentum. These two gauge fields appear in the soft-covariant derivative of soft-collinear gravity, and in turn these interactions determine the first two terms of the soft theorem. This explains why in the gravitational soft theorem, the first two terms are gauge invariant only after summing over all emitters, and assuming the conservation of the corresponding charges, momentum and angular momentum, in the non-radiative process. All further subleading soft-collinear 
interactions can be expressed in terms of the 
gauge-invariant Riemann tensor at $x_-$. However, 
unlike in gauge theory, the Riemann tensor contains 
two derivatives and arises only at second (quadrupole) 
order in the multipole expansion, that is, at sub-subleading 
order. (In gravity, the dipole terms are related to the 
gauge field coupling to angular momentum.) The universality of soft emission ends at this 
sub-subleading order, since in higher powers there exist source operators 
containing soft field products involving the soft 
Riemann tensor invariant under the soft 
gauge symmetries, which have coefficient functions 
unrelated to those of the non-radiative process. 
The two factors of angular momenta in the sub-subleading 
soft theorem are seen to have different interpretations. 
One factor is related to the charge of the soft theorem, 
similar to the one in the subleading term, while the 
second relates to the coupling to the Riemann tensor, 
similar to the appearance of $J_{\mu\nu}$ from 
the coupling to $F^s_{\mu\nu}$ in the gauge-theory 
soft theorem. In this way, the gravitational soft 
theorem is directly linked to the structure of the 
soft-collinear gravity Lagrangian, restricted to 
single-emission at tree level.

\subsection{Loop corrections to the soft theorem}

Both the gauge-theory and the gravitational soft theorem are 
modified by loop corrections \cite{Bern:2014oka,Larkoski:2014bxa}.
However, in gravity, the structure of these modifications is 
quite different, as can easily be seen from power-counting 
in the EFT perspective.

In SCET, loop contributions arise from three different loop 
momentum regions, the hard, the collinear and the soft region, 
corresponding to the hard, collinear and soft modes in the 
effective theory. The hard modes are integrated out, thus the contributions of the hard loops are inside the matching coefficients $\widetilde{C}^{X}(t_i)$ and part of the non-radiative amplitude.
Hence, hard loops never affect the soft theorem, insofar as they modify the underlying non-radiative process.

In the following, we therefore focus on collinear and soft loops. Gravity differs from gauge theory in two important aspects in the soft and collinear sector \cite{Beneke:2021aip}, ultimately due to the dimensionful coupling:
\begin{itemize}
\item[$i)$] In the purely-collinear sector, that is in the Lagrangian terms containing only collinear but no soft fields, there are no leading power interactions. The $\lambda$ expansion corresponds to the weak-field expansion, and the first collinear interaction appears in $\mathcal{O}(\lambda)$.
	Roughly speaking, we expand collinear gravity in collinear momenta $p_\perp\sim\lambda$.
	\item[$ii$)] In the purely-soft sector,  that is in the Lagrangian terms containing only soft but no collinear fields, there are also no leading power interactions. Here, the weak-field expansion agrees with the $\lambda^2$ expansion, corresponding to an expansion in soft momenta $k\sim\lambda^2$. Purely-soft interaction vertices thus start at $\mathcal{O}(\lambda^2)$.
\end{itemize}
Hence, whenever a purely-collinear or a purely-soft interaction takes place, the contribution is already suppressed by at least one order of $\lambda$ or $\lambda^2$, respectively. In gravity, only soft-collinear interactions exist at leading power.

This has a drastic impact on the loop corrections to the three 
universal terms in the soft theorem. In the remainder of this 
subsection, we show within the EFT framework that the leading-power eikonal term is not modified, the subleading term is only corrected at one-loop, and the sub-subleading term by one- and two-loop contributions. These conclusions agree with \cite{Bern:2014oka} and sharpen the all-order power-counting of soft and collinear loop corrections.

\begin{figure}
	\centering
	\includegraphics[width=0.22\textwidth]{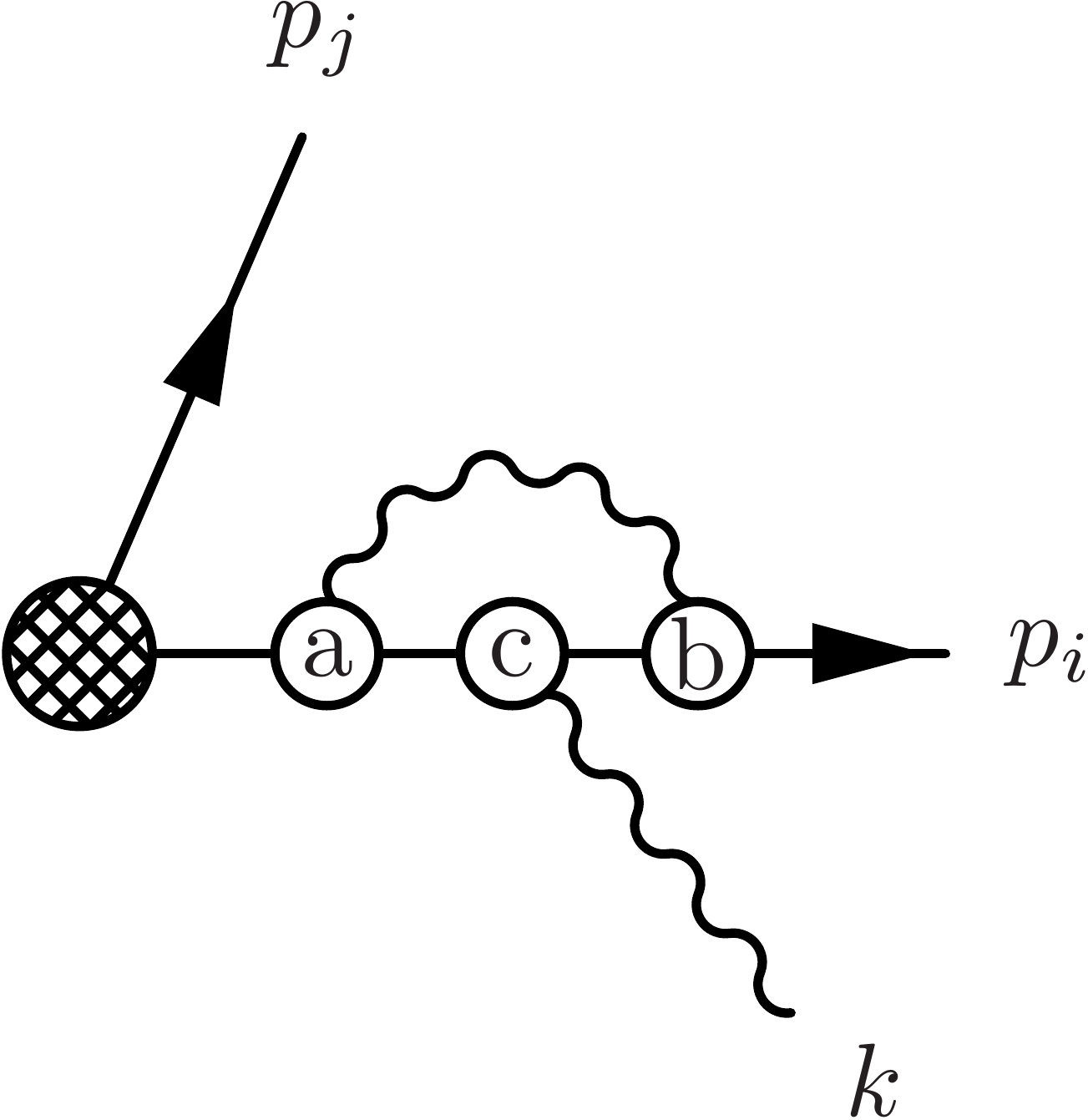}
	\qquad
	\includegraphics[width=0.22\textwidth]{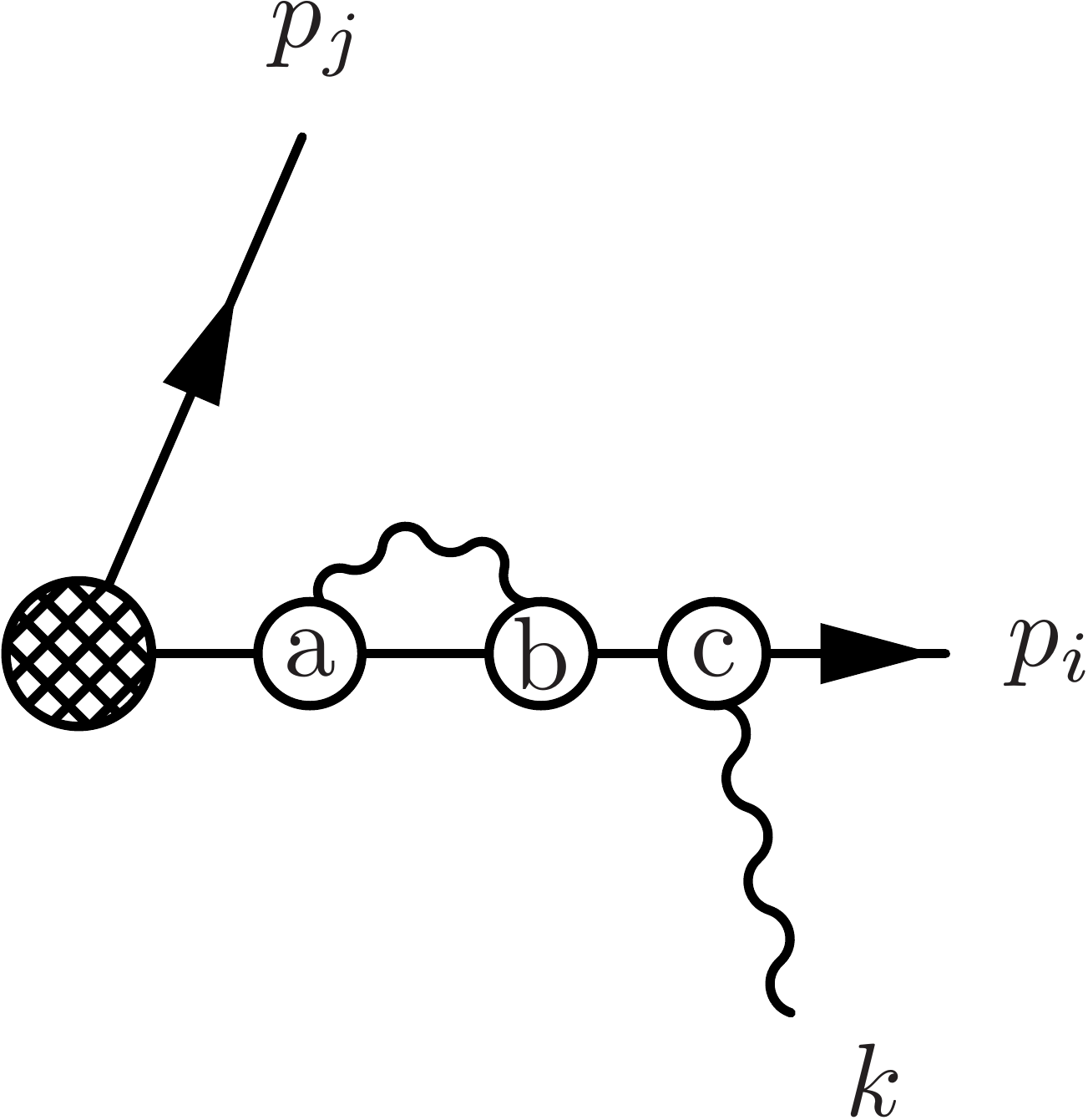}
	\qquad
	\includegraphics[width=0.22\textwidth]{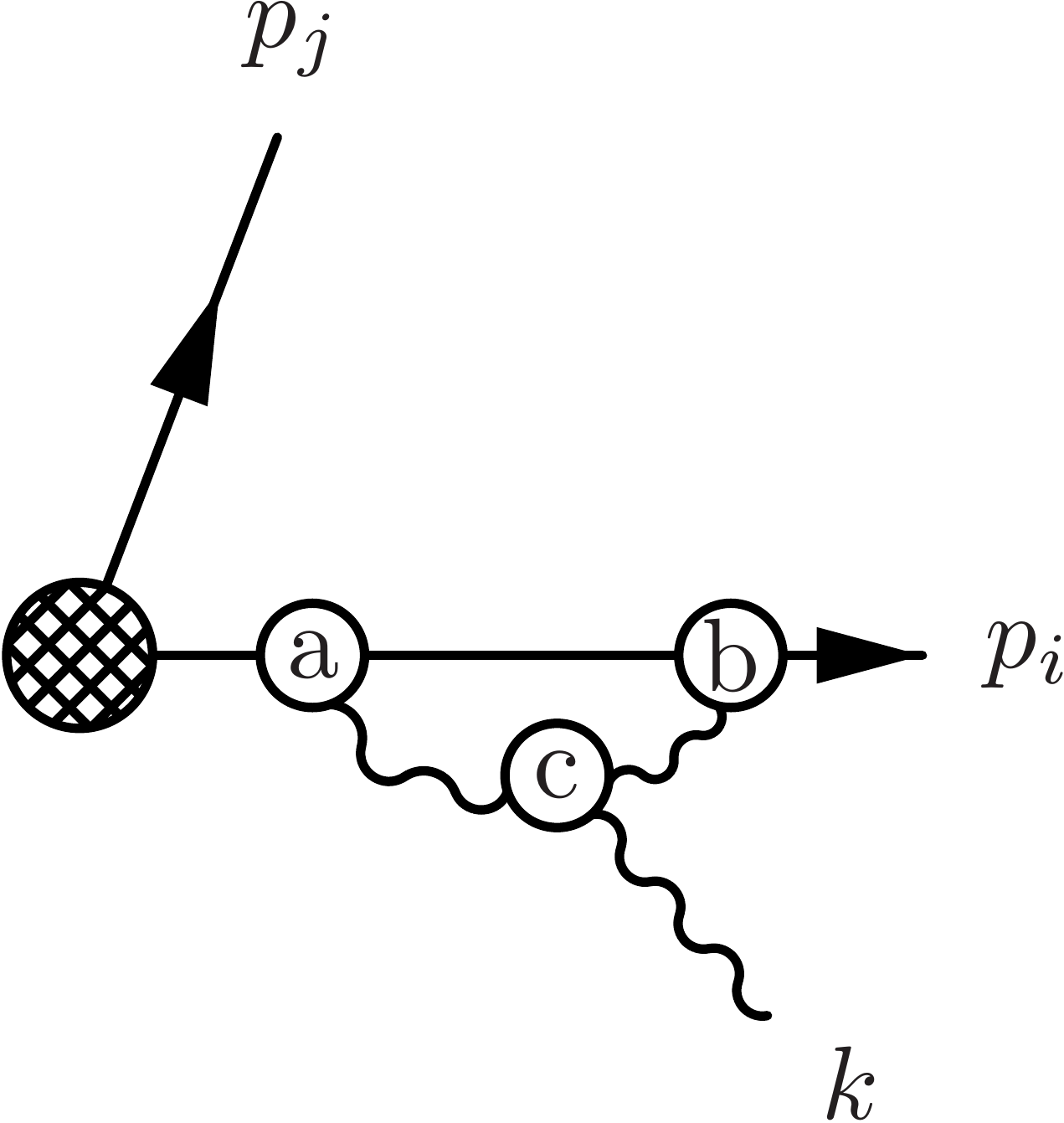}
	\quad
	\caption{Diagram classes modifying the soft emission process with loop correction to a single collinear leg. Collinear interaction vertices are power-suppressed by at least one order in $\lambda$, so $a,b\geq1$, suppressing the collinear loop by $\lambda^2$. For soft loops, one can fix soft light-cone gauge in the direction of the collinear leg, effectively decoupling the soft modes. This means that $a,b\geq2$ and the soft loop is suppressed by $\lambda^4$. In addition, the soft loop is scaleless unless the graviton is emitted via a purely-soft vertex, as depicted in the last diagram. This process is further suppressed by $c\geq2$, yielding a total suppression by $\lambda^6$.
}
\label{fig:singleleg}
\end{figure}

Let us first add one collinear loop to the emission process.
The first possibility is to connect the $i$-collinear loop only to the $i$-collinear leg (or legs, if one considers multiple $i$-collinear particles), as depicted in \cref{fig:singleleg}.
Due to the aforementioned point $i)$, these attachments must stem from the subleading purely-collinear Lagrangian $\mathcal{L}_i^{(k)}$, $k\geq 1$.
As we need to attach the loop twice, this yields a suppression of at least $\mathcal{O}(\lambda^2)$.
Alternatively, one can form a tadpole using a vertex containing two gravitons. These vertices start at $\mathcal{O}(\lambda^2)$.
The second possibility is to attach the loop to the hard scattering, by adding an additional $i$-collinear building block to the $N$-jet operator, and then connecting it to the $i$-collinear leg.
However, these building blocks are also suppressed by another power in $\lambda$, so this contribution also counts at least as $\mathcal{O}(\lambda^2)$.
In summary, the collinear one-loop contribution is suppressed at least by $\mathcal{O}(\lambda^2)$, thus it cannot affect the leading term of the soft theorem.

\begin{figure}[t]
	\centering
	\includegraphics[align=t,width=0.17\textwidth]{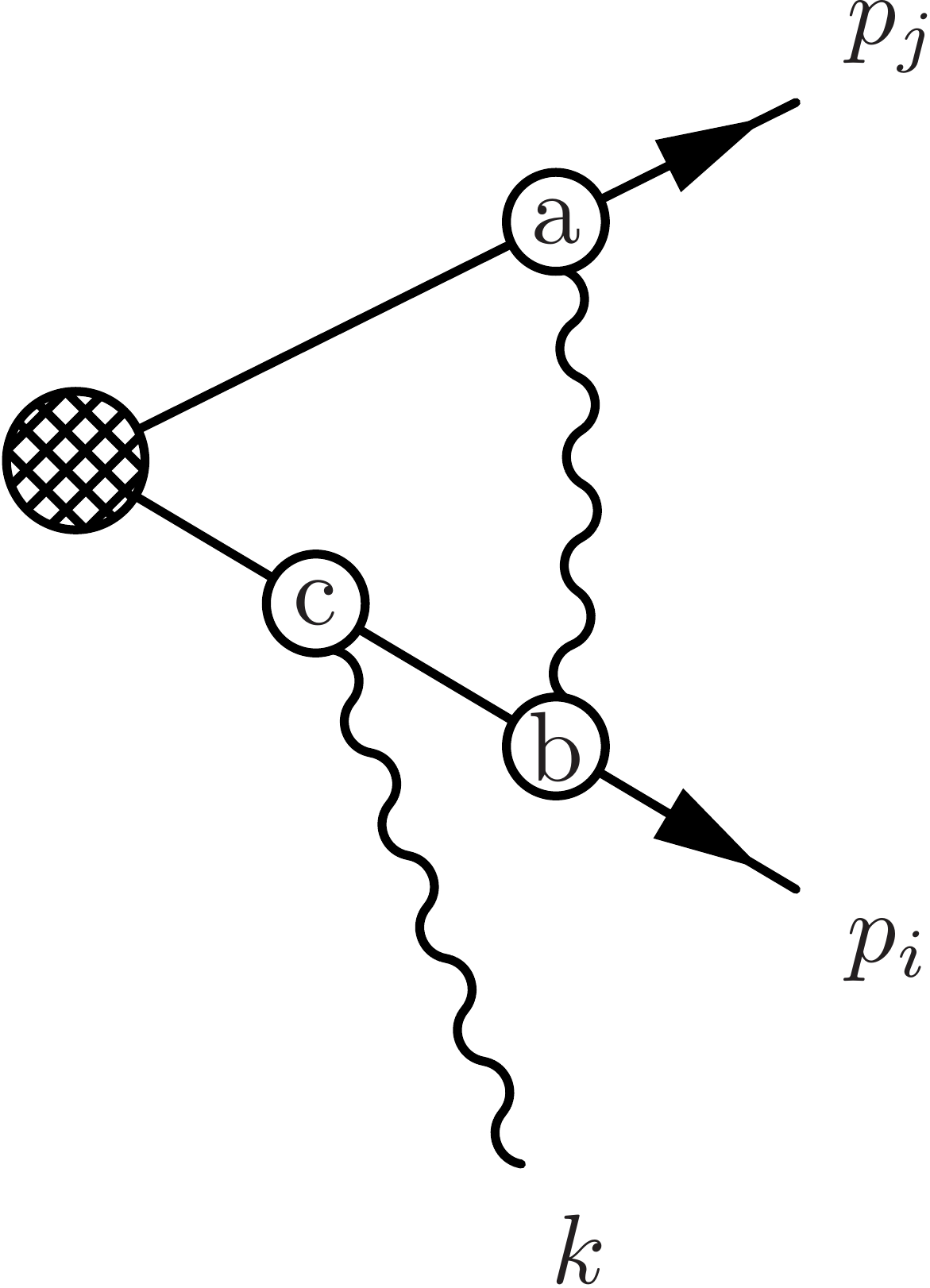}
	\qquad
	\includegraphics[align=t,width=0.17\textwidth]{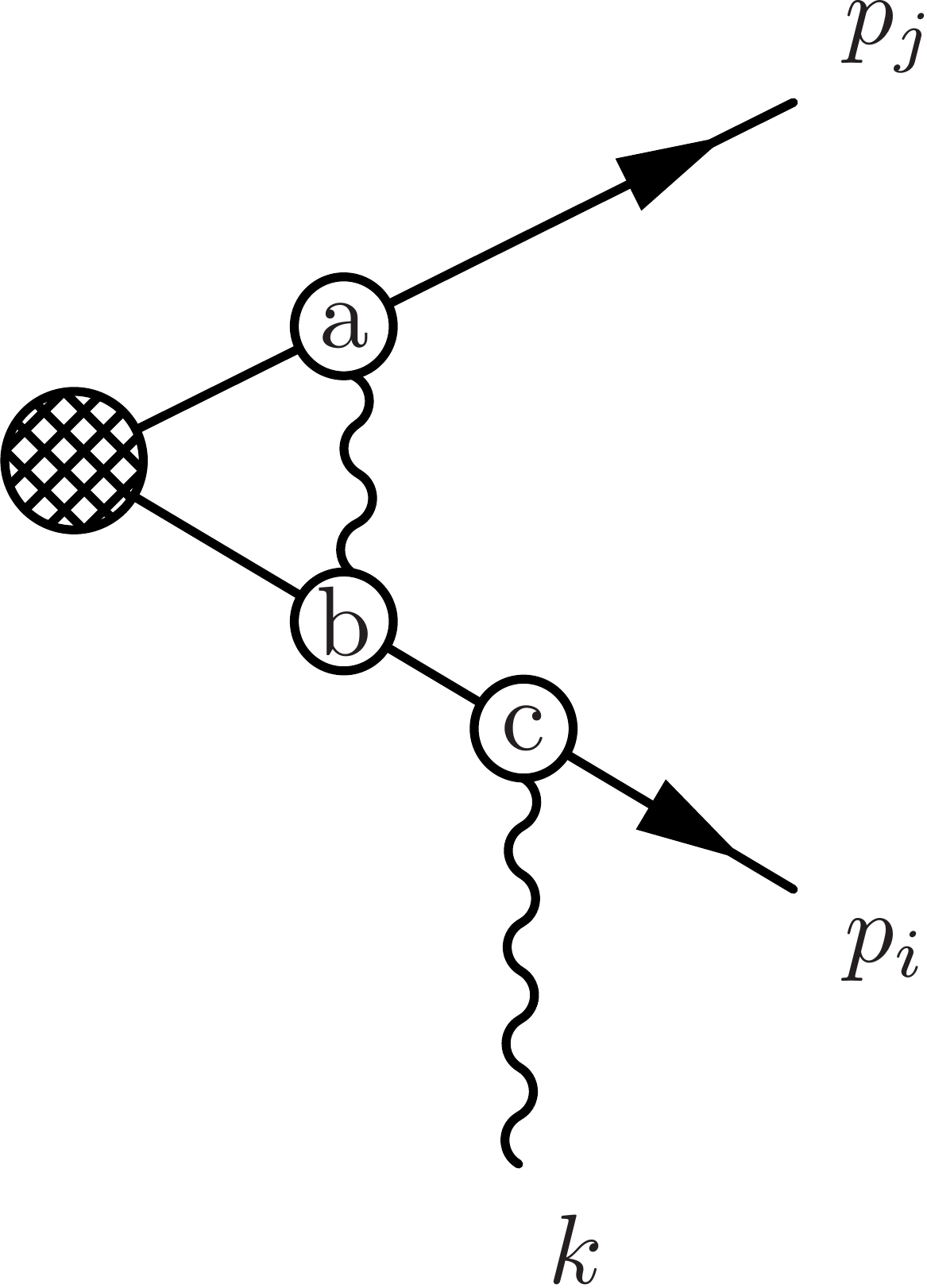}
	\qquad
	\includegraphics[align=t,width=0.17\textwidth]{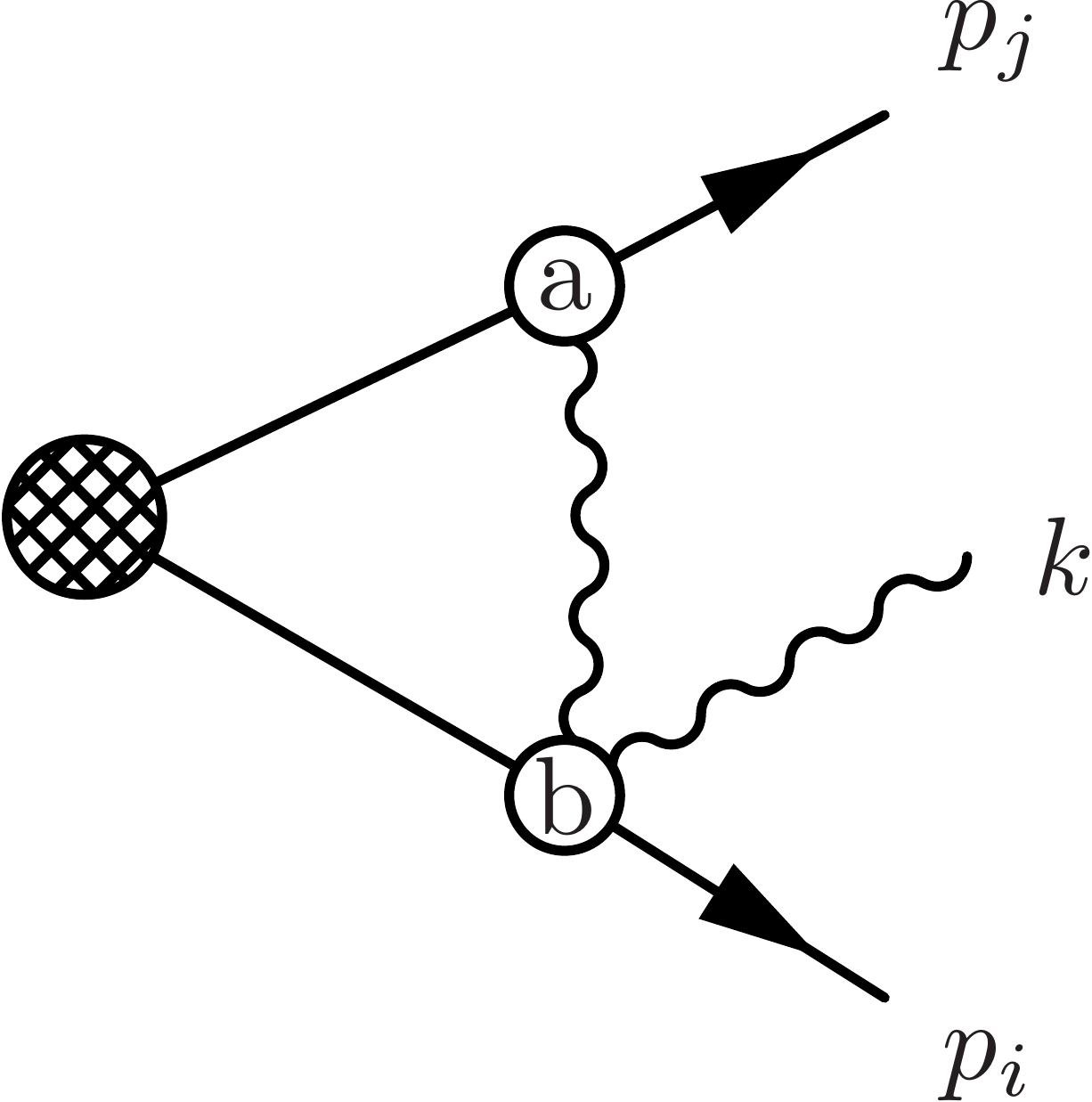}
	\qquad
	\includegraphics[align=t,width=0.17\textwidth]{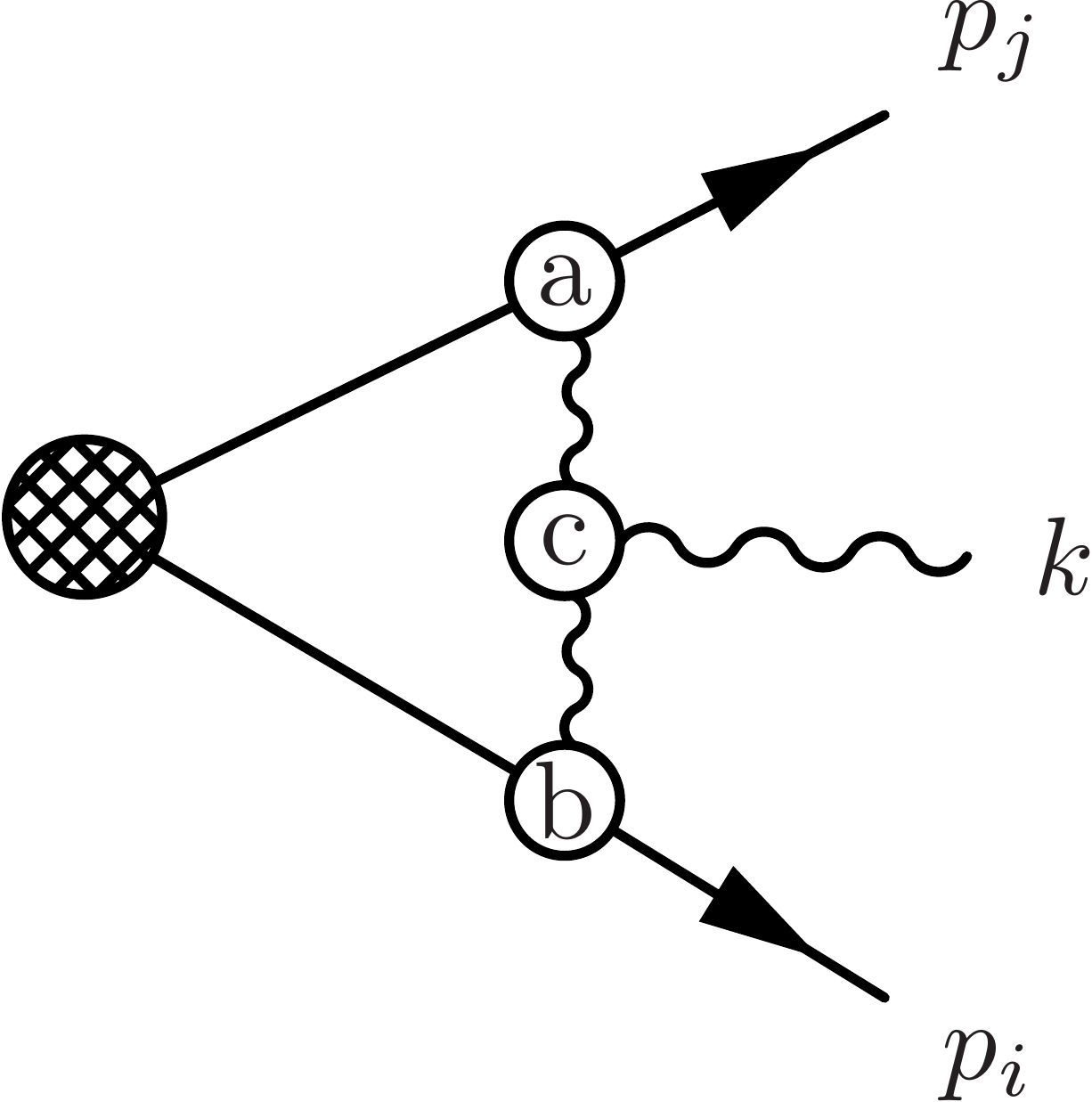}
	\caption{Diagram classes modifying the soft emission process, where the soft loop connects two legs of different directions.
	The soft-collinear interactions are present starting from $a=b=0$.
	Due to multipole expansion, soft-collinear vertices are only sensitive to the $n_{i-}k$ component.
	For on-shell legs, the soft loop vanishes unless a soft scale, provided by the injection of the full soft momentum $k$, is present.
	This can only happen by a purely-soft interaction vertex.
	Hence, only the last diagram has a non-vanishing contribution. Here, $c\geq2$ as it is a purely-soft interaction, and the loop is suppressed by $\lambda^2$ in total.
	}
	\label{fig:twoleg}
\end{figure}

Next, let us consider the soft one-loop contributions.
Here, it is important to note two major simplifications:
First, if the soft loop connects only a single $i$-collinear leg, as depicted in \cref{fig:singleleg}, one can fix soft light-cone gauge in the $n_{i-}$ direction, effectively decoupling the soft-covariant derivative.
For gravity, this implies that the first possible soft-collinear vertex is the Riemann-tensor term in $\mathcal{L}_i^{(2)}$, thus suppressed by at least $\mathcal{O}(\lambda^2)$.
To attach the loop to the leg, one needs two such vertices, so this loop is already suppressed by $\mathcal{O}(\lambda^4)$, without considering the soft emission itself.
Therefore, these types of diagrams are too power-suppressed and not important for the discussion.
The only relevant soft-loop contribution arises from a loop connecting two legs of different collinear sectors, depicted in \cref{fig:twoleg}, which can already appear at $\mathcal{O}(1)$ using the leading-power interactions.\footnote{See Appendix of 
\cite{Beneke:2019oqx} for a similar discussion for soft interactions in gauge theory at next-to-leading power.}  However, we shall now argue that these contributions vanish unless the external soft graviton is connected to the loop through a purely-soft interaction, as shown in the last diagram in the figure. The loop depicted in \cref{fig:twoleg}, with the soft graviton emission removed, is given by the (dimensionally regulated) integral
\begin{align}
A \sim \frac{i\kappa^2}{16}\int \frac{d^d l}{(2\pi)^d} \frac{p_{i+}^2}{p_i^2 + n_{i+}p_i n_{i-}l + i0} \frac{p_{j+}^2}{p_j^2 - n_{j+}p_j n_{j-}l + i0} \frac{(n_{i-}n_{j-})^2}{l^2+i0}\,.
\end{align}
For on-shell external particles, $p_i^2 = p_j^2 = 0$, this simplifies to
\begin{equation}\label{eq::1loopfinal}
A \sim -\frac{i\kappa^2}{16} \,p_{i+}p_{j+} (n_{i-}n_{j-})^2 \int \frac{d^d l}{(2\pi)^d} \frac{1}{l^2+i0} \frac{1}{n_{i-}l +i0} \frac{1}{n_{j-}l + i0}\,.
\end{equation}
The integral is scaleless, and vanishes. If one now attaches 
the soft graviton to one of the collinear lines (see first three 
diagrams in \cref{fig:twoleg}), one can always route the 
$k$-momentum such that it appears in the eikonal 
propagators of only one of the legs, say $i$. In this way, the loop 
integral \eqref{eq::1loopfinal} may be modified to include 
eikonal propagators of the form $(n_{i-}(l+k) +i0)^{-1}$. 
Since only the $n_{i-} k$ component of the soft momentum can 
ever appear in the denominator, one cannot form a soft 
invariant and the soft loop integral will remain scaleless 
and vanishing.\footnote{From the Lagrangian perspective, this 
follows from the multipole expansion of the soft-collinear Lagrangian, after which soft fields are evaluated at $x_{i-}$, 
which implies that only $n_{i-}k \,n_{i+}^\mu/2$ enters 
the momentum-conservation delta-function at a soft-collinear vertex.} In order for soft loops to yield a non-zero contribution, one needs to bring the full external soft momentum $k$ 
into the loop integral. This requires the external soft graviton to couple to the loop through a purely-soft interaction (as in the last diagram in the figure). Such interaction vertices involve the full momentum conservation delta function and lead to 
propagators $1/(l+k)^2$. However, by point $ii)$ above, such a 
purely-soft vertex comes at the cost of power-suppression 
by at least $\lambda^2$. Hence, soft one-loop corrections can 
also not affect the leading term in the soft theorem.

These considerations easily generalise to any loop order. 
The key results from the above discussion can be summarised 
as follows:
\begin{itemize}
	\item[$i)$] A collinear loop can only be connected by purely-collinear vertices, which are power-suppressed in gravity.
	Thus, adding a collinear loop always brings suppression of at least $\mathcal{O}(\lambda^2)$.
	\item[$ii)$] A soft loop is scaleless, unless it is directly connected to the external soft graviton by a \emph{purely-soft interaction vertex}, due to the multipole expansion in soft-collinear vertices.
	Since purely-soft interactions are power-suppressed in gravity by a factor of $\lambda^2$, adding a soft loop yields a suppression of at least $\mathcal{O}(\lambda^2)$.
\end{itemize}

\begin{figure}
\centering	
\includegraphics[width=0.2\textwidth]{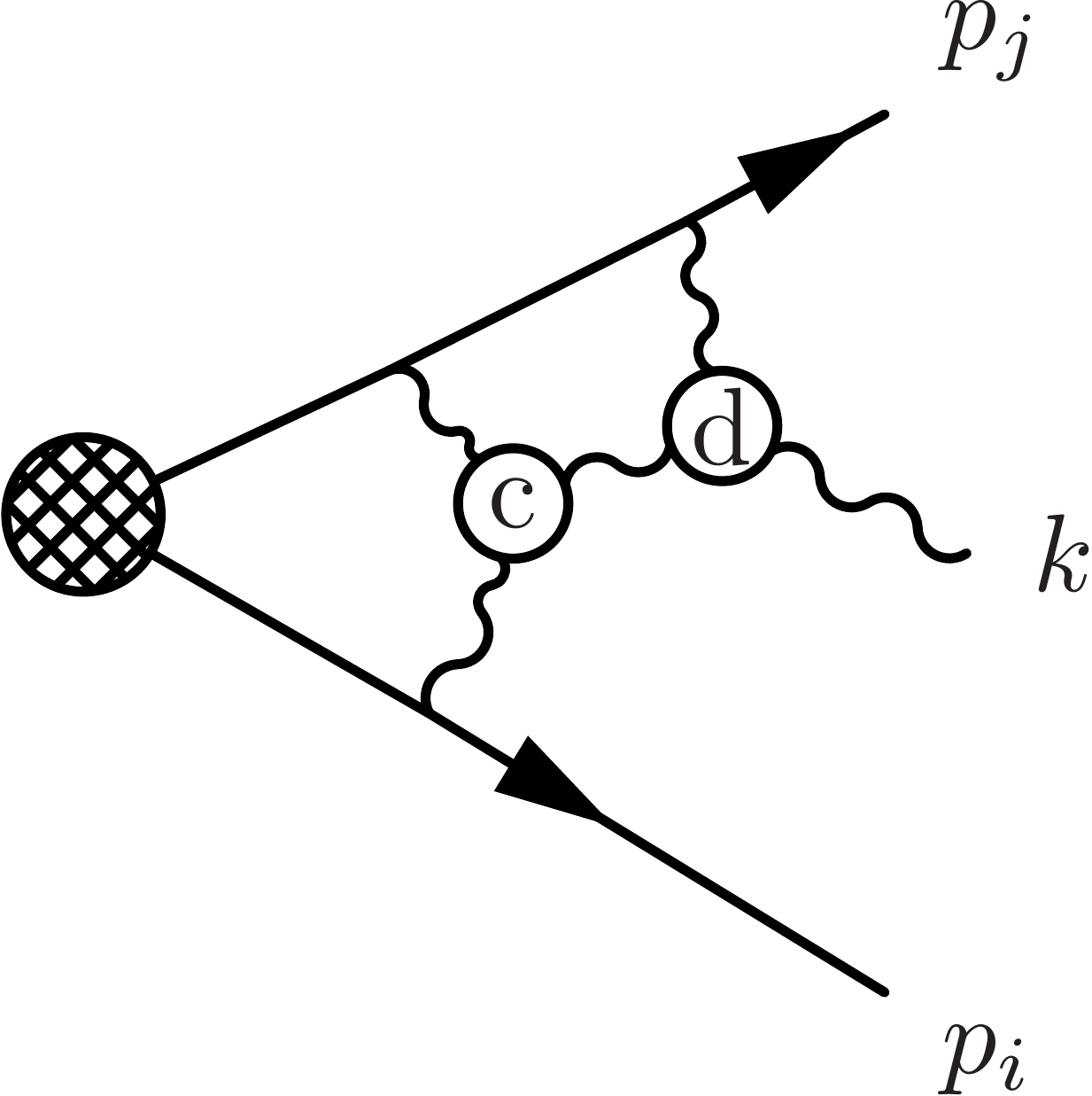}
\caption{An example of a non-vanishing soft two-loop contribution. In order for the soft scale $k$ to be present in both loop momenta, the loops must be attached via purely-soft interactions to the emitted graviton, so $c,d\geq 2$. Any soft loop yields an additional $\lambda^2$ suppression, and the two-loop contribution therefore starts at sub-subleading order $\lambda^4$. }
\label{fig:twoloop}
\end{figure}

For purely-collinear loops, the suppression by $\mathcal{O}(\lambda^2)$ per loop immediately allows us to conclude that the subleading soft factor can only be modified at the one-loop order but not by collinear two- or higher-loop contributions.
Similarly, the sub-subleading factor can only be modified by collinear one- and two-loop processes, since three-loop is suppressed by at least $\lambda^6$.

For purely soft, as well as mixed soft and collinear loops, one again needs to inject the full external soft momentum $k$ into the loops, otherwise they are scaleless and vanish.
For multiple soft loops, this is only possible if the soft emission is attached to the soft particle in the loop, and each additional soft loop is also directly connected to this loop via purely-soft emission vertices, as depicted in \cref{fig:twoloop}.
If a loop is not directly connected (via purely-soft vertices) to the soft emission, it is scaleless and vanishes.
Thus, effectively, as was the case for collinear loops, each additional soft loop comes with at least $\mathcal{O}(\lambda^2)$ from the purely-soft vertex.
We can again conclude that the subleading term can only be affected by one-loop processes, and the sub-subleading term is modified by one- and two-loop diagrams.

In summary, the power-counting and multipole expansion of soft-collinear effective theory, combined with the necessity of a soft scale in soft loops for them to not vanish, immediately imply the following result, already obtained in \cite{Bern:2014oka}: The leading soft factor is never modified by loops, the subleading factor is only corrected by one-loop, and the sub-subleading factor is only modified by one- and two-loop contributions. Higher loop-corrections cannot affect the terms of the gravitational soft theorem.


\section{Summary and outlook}

Despite over 60 years of history, soft theorems are still an active field of research. Adopting the perspective of effective Lagrangians, we connect the structure of the NLP terms in the soft theorems for gauge theories and perturbative 
gravity to the emergent soft gauge theories of the respective Lagrangians. This point of view is especially revealing 
for gravity:
\begin{itemize}
\item It explains without calculation why soft graviton 
emission is universal to sub-sub\-leading order (rather than 
subleading order only as for gauge bosons). Since one 
constructs soft-gauge invariant sources with soft fields 
to these orders, the emission is controlled by Lagrangian 
interactions, which are independent of the hard process.
\item The leading and subleading terms, despite resembling 
the two gauge-theory terms after substituting colour charge 
by momentum, should in fact both be interpreted as 
eikonal terms, stemming from the covariant derivative 
of soft-collinear gravity and representing the coupling 
to the charges of the soft gauge symmetry. This 
explains why, unlike in gauge theory, the first two terms 
in the soft theorem require charge conservation of 
the hard process in order to be gauge invariant.
\item The two angular momentum factors in the 
sub-subleading term, while representing the same 
mathematical expressions, have different interpretations. 
One factor is related to the charge of one of 
the soft gauge symmetries, while the other arises 
from the kinematic multipole expansion, in analogy 
with gauge theory.
\end{itemize}

The EFT formalism allows us to cast the soft theorem into 
an operatorial form, since it follows from simple manipulations of the Lagrangian, which makes the universal nature of the theorem manifest, including the origin of the spin term (as demonstrated here for the case of gauge theories). Power-counting and general 
properties of soft-collinear Lagrangian further allow for a classification of loop corrections to the soft factors. In addition, it can immediately be applied to extensions of QCD, the Einstein-Hilbert theory, and matter theories with non-minimal coupling, as long as the underlying gauge or diffeomorphism symmetry is respected.
In this way, one can immediately judge from power-counting and the Lagrangian  
if a higher-order effective operator, like the chromomagnetic $\overline{\psi} \sigma^{\mu\nu} F_{\mu\nu}\psi$, the gravitational $\sqrt{-g}R^2$ or a non-minimal matter coupling $\sqrt{-g}R^2\phi$ affects the soft theorem. Importantly, since these extensions do not affect the soft gauge symmetry, it follows that there will always be two universal terms in gauge theory, and three universal terms in gravity, as the number of these terms is only linked to the effective soft gauge symmetry, and not to the actual form of the Lagrangian.

 The present work derives from recent progress on the formulation of the SCET for gravity beyond leading power \cite{Beneke:2021aip}. The general construction of this EFT reveals the intricate structure of soft-collinear gravity as a gauge theory, which explains why the soft theorem looks the way it does.
In this respect, it is worth noting that the soft-collinear effective Lagrangian automatically provides rules to generalise soft amplitudes to multiple emissions, including quantum corrections.
This perspective complements the insights obtained from spinor-helicity methods \cite{Cachazo:2014fwa} or the double-copy mapping \cite{Beneke:2021ilf}, and ties them more closely to the properties of (effective) field theories.
It would be interesting to further explore the connection of the EFT formulation of soft physics to asymptotic symmetries \cite{Strominger:2013jfa,Strominger:2013lka,He:2014laa,Campiglia:2015kxa,Strominger:2017zoo}.

\subsubsection*{Acknowledgement}
This work was supported by the Excellence Cluster ORIGINS
funded by the Deutsche Forschungsgemeinschaft 
(DFG)
under Grant No. EXC - 2094 - 390783311.
RS is supported by the United States Department of Energy under Grant Contract DE-SC0012704.


\appendix

\section{Details for gravity}
\label{sec::appGR}

\subsection{SCET gravity Lagrangian}
\label{sec::appGR::Lagrangian}

The Lagrangian for SCET gravity is given by a power series \cite{Beneke:2021aip}
\begin{equation}
    \mathcal{L} = \mathcal{L}_{\rm kinetic} + \mathcal{L}^{(0)} + \mathcal{L}^{(1)} + \mathcal{L}^{(2)} + \mathcal{L}^{(3)} + \mathcal{L}^{(4)}\,.
\end{equation}
The terms linear in the soft graviton can be expressed via the energy-momentum tensor
\begin{equation}
    T_{\mu\nu} = \lc \partial_\mu \chi_c\rc^\dagger \partial_\nu \chi_c + \lc \partial_\nu \chi_c\rc^\dagger \partial_\mu \chi_c - \eta_{\mu\nu} \lc\partial_\alpha\chi_c\rc^\dagger \partial^\alpha \chi_c\,.
\end{equation}
We use the short-hand notation 
$n_\pm^\alpha A_{\alpha\beta\ldots} = A_{\pm\beta\cdots}$ 
for the contractions with the collinear reference vectors.
For the single-emission terms, we find
\begin{align}
    \mathcal{L}^{(0)} &= 
    -\frac \kappa 8 s_{--} T_{++}\,,\\
    \mathcal{L}^{(1)} &= -\frac \kappa 4 s_{-\mu_\perp}\tensor{T}{^{\mu_\perp}_+}
    - \frac \kappa 8 \lc \partial_{[\mu}s_{\nu]-}\rc\nm^\nu x_\perp^\mu \,T_{++}\,,\\
    \mathcal{L}^{(2)} &=
    - \frac{\kappa}{2} \lc \partial_{[\mu}s_{\nu]-}\rc x_\perp^\mu \tensor{T}{^{\nu_\perp}_+}
    - \frac{\kappa}{16} \lc \partial_{[\mu}s_{\nu]-}\rc \np^\mu \nm^\nu \nm x \,T_{++}\nn\\
    &\quad
    -\frac 18 x_\perp^\alpha x_\perp^\beta R^s_{\alpha-\beta-} T_{++}
    - \frac \kappa 8 s_{+-} \,T_{+-}\,,\\
    \mathcal{L}^{(3)} &= -\frac 18 x_\perp^\alpha \nm x n_+^\beta R^s_{\alpha-\beta-} T_{++}
    - \frac{1}{24}x_\perp^\alpha x_\perp^\beta x_\perp^\nu \brac{\partial_\nu R^s_{\alpha-\beta-}}T_{++}\nn \\
    &\quad
    - \frac{1}{8}
    \lc \partial_{[\mu}s_{\nu]-}\rc \np^\nu \nm x\tensor{T}{^{\mu_\perp}_+} 
    -\frac 18 \lc \partial_{[\mu}s_{\nu]-}\rc x_\perp^\mu\np^\nu T_{+-}
    -\frac 13 x_\perp^\alpha x_\perp^\beta R^s_{\alpha\mu_\perp\beta-} \tensor{T}{^{\mu_\perp}_+}\,,
    \\
    \mathcal{L}^{(4)} &= 
    -\frac{1}{32}(\nm x)^2 R^s_{+-+-}T_{++} 
    - \frac{1}{24}x_\perp^\alpha \nm x n_+^\beta x_\perp^\nu \brac{\partial_\nu R^s_{\alpha-\beta-}} T_{++}\nn \\
    &\quad
    -\frac{1}{48}x_\perp^\alpha x_\perp^\beta \nm x n_+^\nu \brac{\partial_\nu R^s_{\alpha-\beta-}}T_{++}
    -\frac{1}{96}x_\perp^\alpha x_\perp^\beta x_\perp^\rho x_\perp^\sigma \brac{\partial_\rho \partial_\sigma R^s_{\alpha-\beta-}}T_{++}\nn \\
    &\quad
    -\frac{1}{6} x_\perp^\alpha \nm x n_+^\beta (R^s_{\alpha\mu_\perp\beta-} + R^s_{\beta\mu_\perp\alpha-})\tensor{T}{^{\mu_\perp}_+}\nn \\
    &\quad 
    -\frac{1}{16}x_\perp^\alpha x_\perp^\beta x_\perp^\nu \brac{\partial_\nu R^s_{\alpha\mu_\perp\beta-}}\tensor{T}{^{\mu_\perp}_+}
    -\frac{1}{6}x_\perp^\alpha x_\perp^\beta R^s_{\alpha\mu_\perp\beta\nu_\perp}T^{\mu_\perp\nu_\perp}
    \nn\\
    &\quad
    + \frac{1}{12}x_\perp^\alpha x_\perp^\beta R^s_{\alpha+\beta-}T_{+-}\,.
\end{align}

 \subsection{Expansion of the soft theorem}
 
The soft theorem \eqref{eq:SoftTheorem} reads
    \begin{equation}\label{eq::app::SoftTheoremAmplitude}
        \mathcal{A}_{\mathrm{rad}} = \frac{\kappa}{2}\sum_i \left( \frac{\varepsilon_{\mu\nu}(k)p_i^\mu p_i^\nu}{p_i\cdot k} + \frac{\varepsilon_{\mu\nu}(k)p_i^\mu k_\rho L_i^{\nu\rho}}{p_i\cdot k} 
        +\frac{1}{2}\frac{\varepsilon_{\mu\nu}(k) k_\rho k_\sigma L_i^{\rho\mu}L_i^{\sigma\nu}}{p_i\cdot k}
        \right)\mathcal{A}\,.
\end{equation}
With our choice of reference vectors $n_{i\pm}$ such that 
$p_i^\mu = p_{i+} \frac{\nim^\mu}{2}$, the first term in the sum is simply given by
 \begin{equation}
\frac{\kappa}{4}\frac{\varepsilon_{--}(p_{i+})^2}{p_{i+} k_-}\mathcal{A}^{(0)}\,,
 \end{equation}
 and since $p^\mu_{i\perp} = 0$, there is no higher-order term generated by its expansion.
 The second term is given by
 \begin{align}
     \frac{\kappa}{2}\frac{\varepsilon_{\mu\nu}(k)p_i^\mu k_\rho L_i^{\nu\rho}}{p_i\cdot k}\mathcal{A} &\equalhat \frac{\kappa}{4}\frac{
     \varepsilon_{--}p_{i+} k_\rho - \varepsilon_{-\rho}p_{i+} k_+}{k_-} 
     \lp \frac{1}{2}\nip^\rho \frac{\partial}{\partial p_{i+}}\mathcal{A}^{(0)} + \frac{\partial}{\partial p_{i\perp\rho}}\mathcal{A}^{(1)}\rp
 \end{align}
\vskip-0.2cm\noindent
 and counts as $\mathcal{O}(\lambda^2)$, as $\frac{\partial}{\partial p_{i\perp}}\sim \lambda^{-1}$ can only act on the suppressed $\mathcal{A}^{(1)}$ piece.
 There is no contribution at $\mathcal{O}(\lambda^3)$. 
 For the third term
 \begin{equation}
   \frac{\kappa}{4}\frac{\varepsilon_{\mu\nu}(k) k_\rho k_\sigma L_i^{\rho\mu}L_i^{\sigma\nu}}{p_i\cdot k}\mathcal{A}\,,
 \end{equation}
 we insert the angular momentum $L_{\mu\nu}$ and find that it
 is given by the sum of the three terms
 \begin{align}\label{eq::app::SoftTheorem3Term1}
     &\frac{\kappa}{32}(k_+^2\varepsilon_{--} -2 k_+ k_- \varepsilon_{+-} + k_-^2\varepsilon_{++})\frac{1}{p_{i+} k_-} p_{i+}^2 \frac{\partial}{\partial p_{i+}}\frac{\partial }{\partial p_{i+}} \mathcal{A}^{(0)}\,,\\
     &\frac{\kappa}{8}(k_\rho k_+ \varepsilon_{--} - k_- k_+\varepsilon_{-\rho} - k_\rho k_- \varepsilon_{+-} + k_-^2 \varepsilon_{\rho+})\frac{1}{p_{i+} k_-} p_{i+}^2\frac{\partial}{\partial p_{i+}}\frac{\partial}{\partial p_{i\perp\rho}}\mathcal{A}^{(1)}\,,\label{eq::app::SoftTheorem3Term2}\\
     &\frac{\kappa}{8}(k_\rho k_\sigma \varepsilon_{--} - 2 k_- k_\rho \varepsilon_{\sigma-} + k_-^2\varepsilon_{\rho\sigma})
     \frac{1}{p_{i+} k_-} p_{i+}^2
    \frac{\partial}{\partial p_{i\perp\rho}}\frac{\partial}{\partial p_{i\perp\sigma}}\mathcal{A}^{(2)}\label{eq::app::SoftTheorem3Term3}\,,
 \end{align}
 which all count as $\mathcal{O}(\lambda^4)$.

\subsection{Details for the sub-subleading-power term}\label{sec::app::subsubleading}

First, it is easy to convince oneself that there is no $\mathcal{O}(\lambda^3)$ contribution from $T\,\{\hat{\mathcal{A}}^{(0)},\mathcal{L}^{(3)}\}$ and $T\,\{\hat{\mathcal{A}}^{(1)},\mathcal{L}^{(2)}\}$ when $p^\mu_{i\perp} = 0$. 
The $\mathcal{O}(\lambda^4)$ term we seek is the operatorial analogue of
\begin{equation}
    \frac{\kappa}{4}\frac{\varepsilon_{\mu\nu}(k) k_\rho k_\sigma J_i^{\rho\mu}J_i^{\sigma\nu}}{p_i\cdot k}\mathcal{A}\,,
\end{equation}
which has three corresponding terms \eqref{eq::app::SoftTheorem3Term1}--\eqref{eq::app::SoftTheorem3Term3}.
The possible contributions stem from the time-ordered products $T\,\{\hat{\mathcal{A}}^{(0)},\mathcal{L}^{(4)}\}$, $T\,\{\hat{\mathcal{A}}^{(1)},\mathcal{L}^{(3)}\}$ and $T\,\{\hat{\mathcal{A}}^{(2)},\mathcal{L}^{(2)}\}$, 
which we compute in the following.
The terms corresponding to \eqref{eq::app::SoftTheorem3Term1} are proportional to $\mathcal{A}^{(0)}$, so they stem from the first time-ordered product.
Similarly for the other two terms \eqref{eq::app::SoftTheorem3Term2}, \eqref{eq::app::SoftTheorem3Term3}, which are given by the second and third time-ordered product, respectively.

\subsubsection{Contribution from $T\,\{\hat{\mathcal{A}}^{(0)}\,, \mathcal{L}^{(4)}\}$}

In $T\,\{\hat{\mathcal{A}}^{(0)}\,, \mathcal{L}^{(4)}\}$, we first have
\begin{equation}\label{eq::GR::subsubcontrib1}
    \mathcal{L}^{(4a)} = -\frac{1}{16}(\nm x)^2 R^s_{+-+-} \lc\np\partial\chi_c\rc^\dagger \np\partial\chi_c\,.
\end{equation}
This term is already in the right form and contributes to the sub-subleading term.
Thus, we show in the following that all other terms in this time-ordered product either cancel against each other, or give a vanishing contribution.

We start with 
\begin{align}
    \mathcal{L}^{(4b)} &= - \frac{1}{12}x_\perp^\alpha \nm x \,n_+^\beta x_\perp^\nu \brac{\partial_\nu R^s_{\alpha-\beta-}}\lc\partial_+\chi_c\rc^\dagger\partial_+\chi_c\nn\\
    &\quad
    -\frac{1}{24}x_\perp^\alpha x_\perp^\beta \nm x n_+^\nu \brac{\partial_\nu R^s_{\alpha-\beta-}}\lc\partial_+\chi_c\rc^\dagger\partial_+\chi_c\,.
\end{align}
We obtain for the first term
\begin{eqnarray}
&&   -\frac{1}{12} \nm x \,n_+^\beta x_\perp^\alpha x_\perp^\nu \brac{\partial_{\nu} R^s_{\alpha-\beta-}} \lc\partial_+\chi_c\rc^\dagger\partial_+\chi_c
= -\frac{1}{24}x_\perp^2\nm x \brac{\partial^{\alpha_\perp}R^s_{\alpha_\perp-+-}}\lc\partial_+\chi_c\rc^\dagger\partial_+\chi_c
\nonumber\\
&&\hspace*{1cm} = \,-\frac{1}{24}\,x_\perp^2 \nm x \Big[\partial^\alpha R^s_{\alpha-+-} - \frac 12 \partial_- R^s_{+-+-}
\Big]\lc\partial_+\chi_c\rc^\dagger\partial_+\chi_c
\nonumber\\
    &&\hspace*{1cm} = \,\frac{1}{48} \nm x \lc \partial_-R_{+-+-}\rc \brac{\partial_+\chi_c}^\dagger \partial_+\chi_c
= -\frac{1}{48} \nm x R_{+-+-} \brac{\partial_+\chi_c}^\dagger  \lc\partial_-\partial_+\chi_c\rc\nonumber\\
    &&\hspace*{1cm}= \,\frac{1}{48} x_\perp^2 \nm x R_{+-+-} \lc\partial_+\chi_c\rc^\dagger\brac{\partial_\perp^2\chi_c}
= \frac{1}{12}\nm x R_{+-+-} \lc\partial_+\chi_c\rc^\dagger\chi_c\,,
\end{eqnarray}
using $\partial^\mu R^s_{\mu-\nu-}=0$ and $x_\perp^2\partial_\perp^2 = 4 + \dots$. For the second term, we find
\begin{align}
     -\frac{1}{24}x_\perp^\alpha x_\perp^\beta \nm x \brac{\partial_+ R^s_{\alpha-\beta-}}\lc\partial_+\chi_c\rc^\dagger\partial_+\chi_c &= -\frac{1}{48}x_\perp^2 \nm x \brac{\partial_+ R^s_{--}}\lc\partial_+\chi_c\rc^\dagger\partial_+\chi_c\label{eq::app::4b2}\,,
\end{align}
which is in fact zero, as the graviton equation of motion reads $R^s_{--}=0$.
In short, we have
\begin{equation}
    \mathcal{L}^{(4b)} \equalhat \frac{1}{12}\nm x R^s_{+-+-} \lc\partial_+\chi_c\rc^\dagger\chi_c \label{eq::nm x term}\,.
\end{equation}
Next, there is 
\begin{equation}
    \mathcal{L}^{(4c)} = -\frac{1}{48}x_\perp^\alpha x_\perp^\beta x_\perp^\rho x_\perp^\sigma \brac{\partial_\rho \partial_\sigma R^s_{\alpha-\beta-}} \lc\np\partial\chi_c\rc^\dagger\np\partial\chi_c\,.
\end{equation}
We use
\begin{equation}
    x_\perp^\alpha x_\perp^\beta x_\perp^\rho x_\perp^\sigma \equalhat \frac{1}{8}x_\perp^4 (\eta_\perp^{\alpha\beta}\eta_\perp^{\rho\sigma} + \eta_\perp^{\alpha\rho}\eta_\perp^{\beta\sigma} + \eta_\perp^{\alpha\sigma}\eta_\perp^{\beta\rho})\,,
\end{equation}
as well as
\begin{equation}
    x_\perp^4 \partial_\perp^4 \equalhat 64 + (\partial_\perp x_\perp\dots)
\end{equation}
and the scalar equation of motion 
\begin{equation}
    n_-\partial \chi_c \equalhat -\frac{\partial^2_\perp}{n_+ \partial} \chi_c\,,
\end{equation}
to find
\begin{equation}
    \mathcal{L}^{(4c)} \equalhat \frac{1}{12} R^s_{+-+-}\, \chi_c^\dagger \chi_c\,.\label{eq::R+-+- term}
\end{equation}
The next contribution stems from
\begin{equation}
    \mathcal{L}^{(4d)} = -\frac{1}{6} x_\perp^\alpha \nm x n_+^\beta (R^s_{\alpha\mu_\perp\beta-} + R^s_{\beta\mu_\perp\alpha-})\lp \lc\partial^{\mu_\perp}\chi_c\rc^\dagger \partial_+\chi_c + \lc \np\partial \chi_c\rc^\dagger \partial^{\mu_\perp}\chi_c\rp.
\end{equation}
We can immediately see that the first term in the bracket vanishes by $p^\mu_\perp = 0$. For the second one we integrate by parts and obtain
\begin{equation}
    \mathcal{L}^{(4d)} = -\frac{1}{12} \nm x R^s_{+-+-}\lc\np\partial\chi_c\rc^\dagger \chi_c\,,
\end{equation}
which cancels with $\eqref{eq::nm x term}$ from $\mathcal{L}^{(4b)}$.
Next, there is
\begin{equation}
    \mathcal{L}^{(4e)} = \frac{1}{12} x_\perp^\alpha x_\perp^\beta \eta_\perp^{\mu\nu}\tensor{R}{^{s}_{\mu_\perp\alpha\nu_\perp\beta}} \lc\np\partial\chi_c\rc^\dagger \nm\partial\chi_c\,,
\end{equation}
which, after using the scalar equation of motion, dropping the traceless term and integration by parts, yields 
\begin{equation}
    \mathcal{L}^{(4e)} \equalhat -\frac{1}{12} R^s_{+-+-} \chi_c^\dagger\chi_c\,.
\end{equation}
This cancels with \eqref{eq::R+-+- term} from $\mathcal{L}^{(4c)}$.
Finally, there are the left-over terms
\begin{align}
    \mathcal{L}^{(4f)} &= -\frac{1}{16}x_\perp^\alpha x_\perp^\beta x_\perp^\nu \brac{\partial_\nu R^s_{\alpha\mu_\perp\beta-}}\lp \lc\partial^{\mu_\perp}\chi_c\rc^\dagger \np \partial\chi_c + \lc \np\partial \chi_c\rc^\dagger \partial^{\mu_\perp}\chi_c\rp\nn\\
    &\quad
    -\frac{1}{6}x_\perp^\alpha x_\perp^\beta R^s_{\alpha\mu_\perp\beta\nu_\perp}\lc\partial^{\mu_\perp}\chi_c\rc^\dagger \partial^{\nu_\perp}\chi_c
    + \frac{1}{6}x_\perp^\alpha x_\perp^\beta R^s_{\alpha+\beta-}\lc\partial_{\mu_\perp}\chi_c^\dagger\rc\partial^{\mu_\perp}\chi_c\nn\\
    &\quad
    + \frac{1}{12}x_\perp^\alpha x_\perp^\beta \tensor{R}{^{s\mu_\perp}_{\alpha\mu_\perp\beta}} \lc\partial_{\nu_\perp}\chi_c\rc^\dagger\partial^{\nu_\perp}\chi_c\,.\label{eq::app:L4leftover}
\end{align}
We can immediately see that the last three terms in \eqref{eq::app:L4leftover} give no contribution, as there is always one external $p^\mu_\perp = 0$. For the first term, we integrate by parts and drop the traceless contribution to see the cancellation. In total we have
\begin{equation}
    \mathcal{L}^{(4f)} \equalhat 0\,.
\end{equation}
To summarise, the entire contribution from $T\,\{\hat{\mathcal{A}}^{(0)}\,, \mathcal{L}^{(4)}\}$ is given by \eqref{eq::GR::subsubcontrib1}, so
\begin{equation}
    \hat{\mathcal{A}}^{(4)} \supset i\int d^4x\:
    T\left\{\hat{\mathcal{A}}^{(0)}\,, \frac{\kappa}{16}\chi_c^\dagger 
    \np \partial \lp (\nm x)^2 R^s_{+-+-} \np\partial \chi_c\rp
    \right\}\,.
\end{equation}

\subsubsection{Contribution from $T\,\{\hat{\mathcal{A}}^{(1)}\,, \mathcal{L}^{(3)}\}$}

In $T\,\{\hat{\mathcal{A}}^{(1)}\,, \mathcal{L}^{(3)}\}$, 
the first term
\begin{equation}\label{eq::GR::subsubcontrib2}
    \mathcal{L}^{(3a)} = -\frac{1}{4} x_\perp^\alpha \nm x \np^\beta R^s_{\alpha-\beta-} \lc \np\partial \chi_c\rc^\dagger \np\partial \chi_c\,,
\end{equation}
already gives the correct orbital momentum contribution.
Again, we show that all other terms cancel out.
First,
\begin{align}
        \mathcal{L}^{(3b)} &= -\frac{1}{12}x_\perp^\alpha x_\perp^\beta x_\perp^\nu \brac{\partial_\nu R^s_{\alpha-\beta-}}\lc\np\partial\chi_c\rc^\dagger\np\partial\chi_c\nn\\
        &\quad
        - \frac{1}{3}x_\perp^\alpha x_\perp^\beta R^s_{\alpha\mu_\perp\beta-}\lp \lc\partial^{\mu_\perp}\chi_c\rc^\dagger \np\partial\chi_c + \lc\np\partial\chi_c\rc^\dagger\partial^{\mu_\perp}\chi_c\rp\,, 
    \end{align}
gives a vanishing contribution, using the same manipulations as before. Concretely, the first term enters as
\begin{equation}
    -\frac{1}{12}x_\perp^\alpha x_\perp^\beta x_\perp^\nu \brac{\partial_\nu R^s_{\alpha-\beta-}}C^{A1\mu_\perp}\lc\np\partial\chi_c\rc^\dagger\brac{\partial_{\mu_\perp}\np\partial\chi_c}
    \equalhat -\frac{1}{6}R^s_{+-\beta_\perp-} C^{A1\beta_\perp}\lc\np\partial\chi_c\rc^\dagger\chi_c\,,
\end{equation}
and the second one yields
\begin{equation}
    -\frac{1}{3} x_\perp^\alpha x_\perp^\beta R^s_{\alpha\mu_\perp\beta-} C^{A1\rho_\perp}\lc\np\partial\chi_c\rc^\dagger \brac{\partial^{\mu_\perp}\partial^{\rho_\perp}\chi_c}
    \equalhat
    \frac{1}{6}R^s_{+-\alpha_\perp-} C^{A1\alpha_\perp}\lc \np\partial\chi_c\rc^\dagger\chi_c\,,
\end{equation}
hence $\mathcal{L}^{(3b)} \equalhat 0$.
For all other terms in $\mathcal{L}^{(3)}$, we immediately see that they do not contribute.
In summary, the only surviving term is \eqref{eq::GR::subsubcontrib2}, so
\begin{equation}
    \hat{\mathcal{A}}^{(4)} \supset i\int d^4x\:
    T\left\{\hat{\mathcal{A}}^{(1)}\,, \frac{1}{4}\chi_c^\dagger 
    \np \partial \big( x_\perp^\alpha \nm x \np^\beta R^s_{\alpha-\beta-} \np \chi_c\big)
    \right\}\,.
\end{equation}

\subsubsection{Contribution from $T\,\{\hat{\mathcal{A}}^{(2)}\,, \mathcal{L}^{(2)}\}$}

Finally, there is only one contribution to  $T\,\{\hat{\mathcal{A}}^{(2)}\,, \mathcal{L}^{(2)}\}$, given by
\begin{equation}
    \mathcal{L}^{(2c)} =
        -\frac{1}{8}x_\perp^\alpha x_\perp^\beta R^s_{\alpha-\beta-}\lc\partial_+\chi_c\rc^\dagger\partial_+\chi_c\,.
\end{equation}
We immediately notice that the $x_\perp^2$ term eliminates $\partial_\perp^2$ of the $A2$ current and we find the second $\perp$-derivative contribution to the sub-subleading-power term.
There are no further contributions.

In summary, the contributing terms to the emission are given by
\begin{align}
    \mathcal{L}^{(2)}_{\rm orbital} &= \frac{1}{4} R^s_{\alpha\mu\beta\nu}\chi_c^\dagger \overset{\leftarrow}{L}\!\phantom{\,}_{\perp+}^{\alpha\mu} L_{\perp+}^{\beta\nu}\chi_c\,,\\
    \mathcal{L}^{(3)}_{\rm orbital} &= \frac{1}{2} R^s_{\alpha\mu\beta\nu} \chi_c^\dagger \overset{\leftarrow}{L}\!\phantom{\,}_{\perp+}^{\alpha\mu}L_{+-}^{\beta\nu}\chi_c\\
    \mathcal{L}^{(4)}_{\rm orbital} &= \frac{1}{4}R^s_{\alpha\mu\beta\nu} \chi_c^\dagger \overset{\leftarrow}{L}\!\phantom{\,}^{\mu\alpha}_{+-}L^{\nu\beta}_{+-} \chi_c\,,
\end{align}
as claimed in \eqref{eq::GR::L4contribution}--\eqref{eq::GR::L2contribution}.


\bibliography{gravity}

\end{document}